\documentclass[twocolumn]{aastex631}

\usepackage{CJKutf8}
\usepackage{amsmath, rotating}
\usepackage{soul}
\usepackage{graphicx}
\usepackage{subfigure}

\newcommand\schi{{\sc  H\,i}}%
\newcommand\schii{{\sc H\,ii}}%
\newcommand\rmht{\mathrm{H_2}}%
\newcommand\idl{{\sc IDL}}%
\newcommand\clumpfind{{\sc CLUMPFIND}}%
\newcommand\kms{\mathrm{km\,s^{-1}}}%
\newcommand\Kkms{\mathrm{K\,km\,s^{-1}}}%
\newcommand\vlsr{v_\mathrm{LSR}}%
\newcommand\msol{\mathrm{M_\odot}}%
\newcommand\Rsun{\mathrm{R_0}}%
\newcommand\thetasun{\mathrm{\Theta_0}}%
\newcommand\Ts{T_\mathrm{s}}%
\newcommand\Nhi{N_\mathrm{HI}}%
\newcommand\Nht{N_\mathrm{H_2}}%
\newcommand\Mhi{M_\mathrm{HI}}%
\newcommand\Mht{M_\mathrm{H_2}}%
\newcommand\Mhtall{M_\mathrm{H_2,all}}%
\newcommand\Mgas{M_\mathrm{HI+H_2}}%
\newcommand\fht{f_{\rmht}}%
\newcommand\SDhi{\Sigma_\mathrm{HI}}%
\newcommand\SDht{\Sigma_\mathrm{H_2}}%
\newcommand\SDSFR{\Sigma_\mathrm{SFR}}%
\newcommand\SDgas{\Sigma_\mathrm{HI+H_2}}%
\newcommand\lc{l_\mathrm{c}}%
\newcommand\bc{b_\mathrm{c}}%
\newcommand\Tgs{\mathcal{G}(v)}%
\newcommand\vc{v_\mathrm{c}}%
\newcommand\vsigma{\sigma_\mathrm{v}}%
\newcommand\vsigmac{\sigma_{v,\mathrm{crit}}}%
\newcommand\vVT{v_\mathrm{VT}}%
\newcommand\rmajor{r_\mathrm{major}}%
\newcommand\rminor{r_\mathrm{minor}}%
\newcommand\rgeo{r_\mathrm{geo}}%
\newcommand\rgeohi{r_\mathrm{geo,HI}}%
\newcommand\rgeoco{r_\mathrm{geo,CO}}%
\newcommand\Rgc{R_\mathrm{GC}}%
\newcommand\Xco{X_\mathrm{CO}}%
\newcommand\alphav{\alpha_\mathrm{vir}}%
\newcommand\W{W}%
\newcommand\wise{{\it WISE}}%
\newcommand\wmap{{\it WMAP}}%

\graphicspath{{./}{figures/}}

\begin{document}
\begin{CJK}{UTF8}{}
 \CJKfamily{mj}

\title{Neutral atomic and molecular clouds and star formation in the outer Carina arm}


\author[0000-0001-8467-3736]{Geumsook Park (박금숙)}
\affil{Korea Astronomy and Space Science Institute, 
776 Daedeokdae-ro, Yuseong-gu, Daejeon 34055, Republic of Korea}
\affil{Research Institute of Natural Sciences, 
Chungnam National University
99 Daehak-ro, Yuseong-gu, Daejeon 34134, Republic of Korea}

\author[0000-0002-2755-1879]{Bon-Chul Koo}
\affil{Department of Physics and Astronomy, Seoul National University, Seoul 08861, Republic of Korea}
\affil{Research Institute of Basic Sciences, Seoul National University, Seoul 08826, Republic of Korea}

\author[0000-0003-2412-7092]{Kee-Tae Kim}
\affil{Korea Astronomy and Space Science Institute, 776 Daedeokdae-ro, Yuseong-gu, Daejeon 34055, Republic of Korea}

\author[0000-0002-1723-6330]{Bruce Elmegreen}
\affil{IBM Thomas J. Watson Research Center, Yorktown Heights, New York 10598, USA}

\begin{abstract}
We present a comprehensive investigation of \schi\ (super)clouds, molecular clouds (MCs), and star formation in the Carina spiral arm of the outer Galaxy.
Utilizing HI4PI and CfA CO survey data, we identify \schi\ clouds and MCs based on the ($l$, $\vlsr$) locations of the Carina arm.
We analyzed 26 \schi\ clouds and 48 MCs.
Most of the identified \schi\ clouds are superclouds, with masses exceeding $10^6~\msol$.
We find that 15 of these superclouds have associated MC(s) with $\Mhi \gtrsim 10^6~\msol$ and $\SDgas \gtrsim$ 50~$\msol \rm pc^{-2}$. 
Our virial equilibrium analysis suggests that these CO-bright \schi\ clouds are gravitationally bound or marginally bound.
We report an anti-correlation between molecular mass fractions and Galactocentric distances, and a correlation with total gas surface densities.
Nine CO-bright \schi\ superclouds are associated with \schii\ regions, indicating ongoing star formation.
We confirm the regular spacing of \schi\ superclouds along the spiral arm, which is likely due to some underlying physical process, such as gravitational instabilities.
We observe a strong spatial correlation between \schii\ regions and MCs, with some offsets between MCs and local \schi\ column density peaks.
Our study reveals that in the context of \schi\ superclouds, the star formation rate surface density is independent of \schi\ and total gas surface densities but positively correlates with molecular gas surface density.
This finding is consistent with both extragalactic studies of the resolved Kennicutt-Schmidt relation and local giant molecular clouds study of \citet{lada2013}, emphasizing the crucial role of molecular gas in regulating star formation processes.

\end{abstract}


\section{Introduction} \label{sec:intro}
\end{CJK}

Stars form within the densest regions of molecular clouds, which are predominately found within giant \schi\ clouds, known as \schi\ superclouds \citep{grabelsky1987, elmegreen1987, lada1988}.
Star complexes, the largest scale of star-forming regions in galaxies, form within giant molecular clouds (GMCs) inside \schi\ superclouds \citep{elmegreen2004}.
Therefore, the formation of \schi\ superclouds marks the onset of star formation.
Understanding the properties of \schi\ superclouds is essential for comprehending global star formation properties in galaxies.

\citet{fukui2010} reviewed the importance of studying the association between GMCs and \schi\ gas in understanding the formation and evolution of GMCs.
They summarized the finding that while GMCs often form in \schi\ filaments, intense \schi\ is not always associated with CO gas.
However, the GMC-\schi\ association has not been well established in the Milky Way, leaving room for further investigation.
This connection suggests a strong interplay between the atomic and molecular phases of the interstellar medium, which is crucial for understanding star formation processes.

\schi\ superclouds typically exhibit masses ranging from $10^6$ to $10^7~\msol$ and sizes up to approximately 1~kpc, as reported by various studies such as those by \citet{elmegreen1987} and \citet{engargiola2003}.
The regular distribution of these superclouds along the spiral arms of galaxies, including the Milky Way, has been well-established through \schi\ observations \citep[e.g.,][]{mcgee1964, elmegreen1983, boulanger1992}.
\citet{mcgee1964} conducted an early study that demonstrated a series of \schi\ superclouds can be observed in the Carina spiral arm in the fourth quadrant of the Milky Way.

According to \citet{grabelsky1987}, there is good agreement between 
distributions of \schi\ superclouds and molecular clouds (MCs) in spiral arms.
However, previous \schi\ surveys with low angular resolution limited 
the ability to reveal detailed structures of \schi\ superclouds in both space and velocity.
The recent all-sky \schi\ $4\pi$ (HI4PI) survey \citep{hi4pi2016}
provides a significantly improved angular resolution, offering 3 to 8 times better resolution compared to previous surveys.
In this paper, we aim to identify and analyze \schi\ superclouds and GMCs in the longitude range from 288\arcdeg\ to 340\arcdeg\ at positive velocities, corresponding to the {\it outer} Carina arm.

The outer Carina arm is a favorite subject for research because it does not have kinematic distance ambiguity, which reduces the likelihood of being influenced by outer spiral arms.
Despite this advantage, the outer Carina arm has received limited attention since the 1980s.
In this study, we use the locations ($l$, $\vlsr$) of the outer Carina arm as determined by \citet[][hereafter referred to as Paper~I]{koo2017} to extract \schi\ emission associated with the spiral arm.
Paper~I traced the spiral arms in the outer Galaxy using the LAB \schi\ data with a beam size of 30\arcmin--36\arcmin\ full-width at half-maximum (FWHM)  \citep{kalberla2005}.
The study identified four spiral arms, including the Sagittarius-Carina, Perseus, Outer, and Scutum-Centaurus arms, using a combination of local peak intensities and radial velocities integrated along latitudes.
Through the use of ($l$, $\vlsr$) diagrams of local peak intensities integrated along latitudes,
Paper~I clearly defined the four spiral arms and prominent interarm features.

Many previous studies have explored the relationships between gas and star formation on both local and extragalactic scales.
For example, \citet{bigiel2008} demonstrated that the star formation rate surface density correlates well with molecular gas surface density but shows little or no correlation with atomic gas surface density in nearby spiral galaxies.
Furthermore, \citet{lada2013} found that star formation rates in local GMCs are independent of total gas surface density, emphasizing the importance of molecular gas in star formation.
These results highlight the importance of understanding the properties and distribution of \schi\ superclouds in relation to molecular gas and star-forming regions.

In this paper, we present a comprehensive study of the outer Carina arm, focusing on \schi\ clouds and their relationships with MCs and star-forming regions.
Section~\ref{sec:data} provides brief descriptions of \schi\ and $^{12}\textrm{CO}$ (hereafter CO) survey data used in this study.
In Section~\ref{sec:clouds}, we outline the process of identifying \schi\ clouds and MCs in the outer Carina arm and determining their physical parameters, such as velocity, distance, and mass. 
We also examine the spatial distribution and virial equilibrium of the \schi\ superclouds identified in this study. 
Section~\ref{sec:HInCO} compares the properties of \schi\ clouds and molecular clouds in the outer Carina arm.
In Section~\ref{sec:sf}, we compare the properties of star-forming regions with the two cloud components and analyze the star formation rates  in the \schi\ superclouds.
Finally, in Section~\ref{sec:summary}, we summarize our results and draw our conclusions.

\section{Data} \label{sec:data}

\subsection{\schi\ Data from the HI4PI Survey} \label{sec:hidata}

In order to obtain images with a higher angular resolution, we utilize the HI4PI survey \citep{hi4pi2016}.
This survey combines the Effelsbeg-Bonn \schi\ Survey (EBHIS) observed with the 100~m Effelsberg radio telescope in the northern hemisphere \citep{kerp2011, winkel2016}, and the Galactic All-Sky Survey (GASS) collected with the 64~m Parkes telescope in the southern hemisphere \citep{mcclure2009, kalberla2010, kalberla2015}.
The HI4PI survey has an angular resolution of 16\farcm2 FWHM and a velocity channel width of 1.29~$\kms$, 
with a brightness temperature noise level of $\sim43$~mK.

During the analysis, we encountered a false absorption-like feature around ($l$, $b$) $\sim (291.5\arcdeg, -0.5\arcdeg)$ when comparing with other data, such as SGPS data (Parkes multibeam data) \citep{mcclure2005}.
To address this issue, the erroneous pixels were replaced with the SGPS data. 
The SGPS data were rebinned and interpolated to match the grid coordinates of the HI4PI data. 
Additionally, the reprocessed SGPS data were multiplied by 0.956 to account for the slightly different brightness temperature levels, which were obtained by averaging brightness temperature ratios from four neighboring peaks. 

\subsection{CO Data from the CfA Survey} \label{sec:codata}

We use the composite Galactic CO survey data of \citet{dame2001}.
The survey is comprised of several large-scale, unbiased surveys using the CfA 1.2~m telescope in the northern hemisphere (with an angular resolution of 8\farcm4) and CfA-Chile 1.2~m telescope in the southern hemisphere (with an angular resolution of 8\farcm8).
The composite CO survey data are grid sampled with a sampling interval of 7\farcm5 and a velocity width of 1.3~$\kms$.
The data for the Galactic 4th quadrant, that are used in this study, provide a main beam temperature noise level of $\sim 0.17$~K. 
To obtain a CO cube with the same angular resolution as the \schi\ data, we employed Gaussian smoothing with a kernel size of 13\farcm8 FWHM.
Throughout this paper, we utilized the smoothed CO data with an angular resolution of 16\farcm2 FWHM for all images and analysis.

\section{Atomic and molecular clouds in the outer Carina arm} \label{sec:clouds}

\subsection{Cloud Identification} \label{sec:clf}

\begin{figure*}
\epsscale{1}
\includegraphics[width=\textwidth]{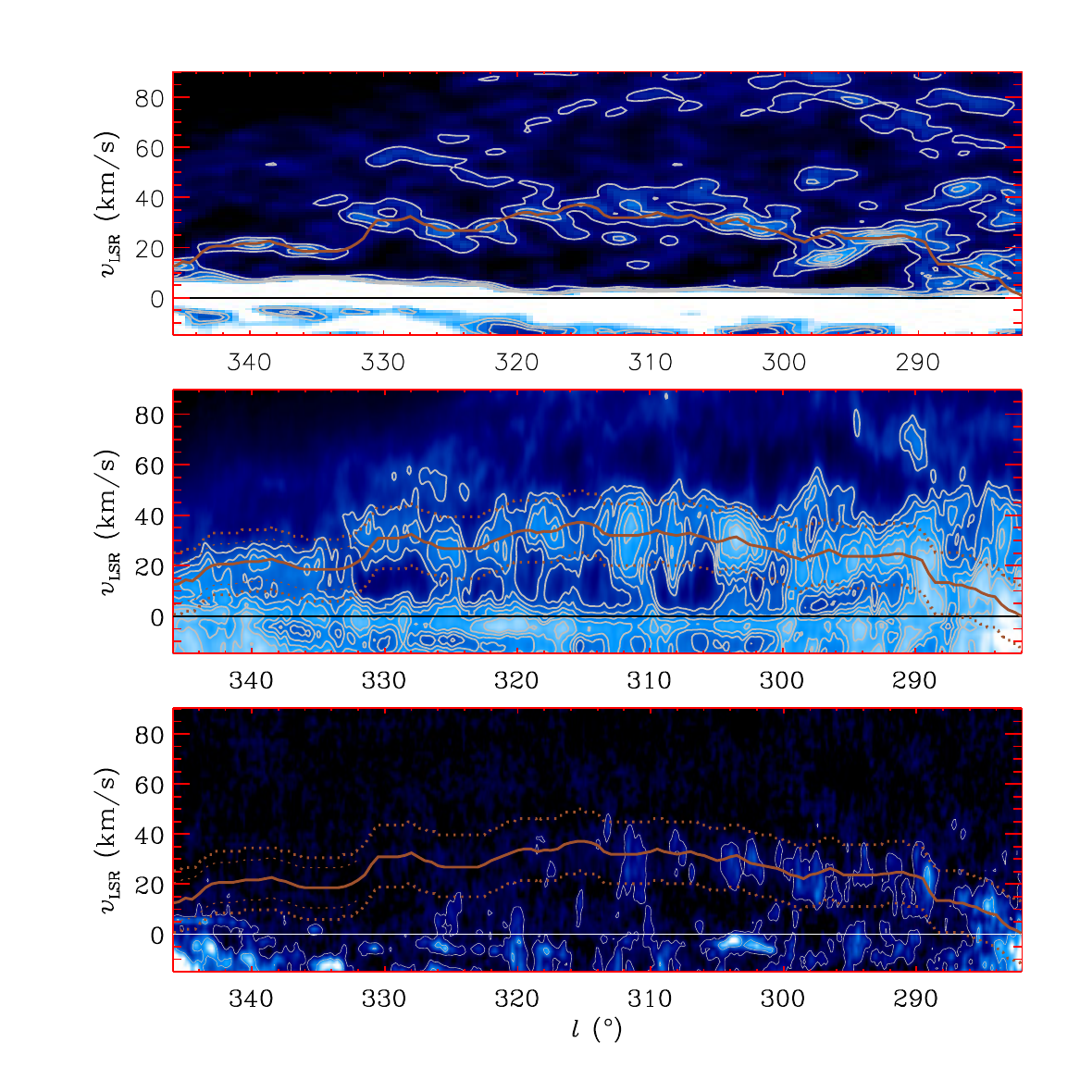}
\caption{
($l$, $\vlsr$) diagrams with the locations of the outer Carina arm.
The outer Carina arm traces obtained by Paper~I are overlaid with a zigzagging solid line in each panel.
{\it Top:}
$b$-integrated local peak intensities in \schi\
($= \sum_{b=-10\arcdeg}^{b=+30\arcdeg} T_{\rm b,max}(l, b, \vlsr)$) from the LAB \schi\ data used in Paper~I.
Contour levels are 20, 40, 60, and 80~K.
{\it Middle:}
$b$-averaged \schi\ intensities from $b=-2\arcdeg$ to +1\arcdeg\ from the HI4PI data.
Contour levels are 30, 40, 50, 60, 70, and 80~K.
{\it Bottom:}
Similar to the middle panel, but showing the CfA CO data.
Contour levels are 0.1 and 0.5~K.
All three images are displayed on a square-root scale.
Two thick dotted lines spaced approximately $25~\kms$ apart indicate the velocity range from which the $v$-integrated maps in Figure~\ref{fig:gas_lb} are obtained.
Two thin dotted lines at $l > 332\arcdeg$ aid in locating an interval of $\pm 8~\kms$  from the solid line mentioned in Sections~\ref{sec:clf_step2}.
A horizon (black or white) line is included to indicate where $\vlsr = 0~\kms$.
\label{fig:gas_lv}}
\end{figure*}

In Paper~I, we presented systemic LSR velocity information along the outer Carina arm.
Figure~\ref{fig:gas_lv} depicts three $b$-integrated ($l$, $\vlsr$) diagrams of \schi\ local peak intensities as well as ordinary \schi\ and CO emissions.
The zigzagging solid lines in the figure denote the locations of the outer Carina arm, which were traced in Paper~I.
A majority of the emission features associated with the outer Carina arm appear to be distributed between two thick dotted lines, approximately $25~\kms$ in width.
The spiral arm exhibits clear visible clumpy structures in both \schi\ and CO emissions.

Due to the ubiquitous nature of \schi\ emission, particularly in the Galactic plane, identifying three-dimensional clumps from cube data is challenging.
Therefore, our clump identification procedure involved two main steps.
In the first step, we determine the projected ($l$, $b$) area from a $v$-integrated image (as discussed in Section~\ref{sec:clf_step1}).
In the second step, we use Gaussian decomposition to measure the mean velocity component(s) of each cloud from the area-averaged spectra (as explained in Section~\ref{sec:clf_step2}).
Although the CO PPV cube is less complex than the \schi\ cube, we employ this method to define MCs for consistency.\footnote{In this study, the identification of clouds is based on the $v$-integrated map and does not take into account the three-dimensional CO distribution. As a result, multiple components along a line of sight may be merged into a single identified MC.} 

\subsubsection{Step~I: Spatial Identification} \label{sec:clf_step1}

\begin{figure*}
\centering
\subfigure[$v$-integrated \schi\ and CO maps of the outer Carina arm]{\includegraphics[width=\textwidth]{%
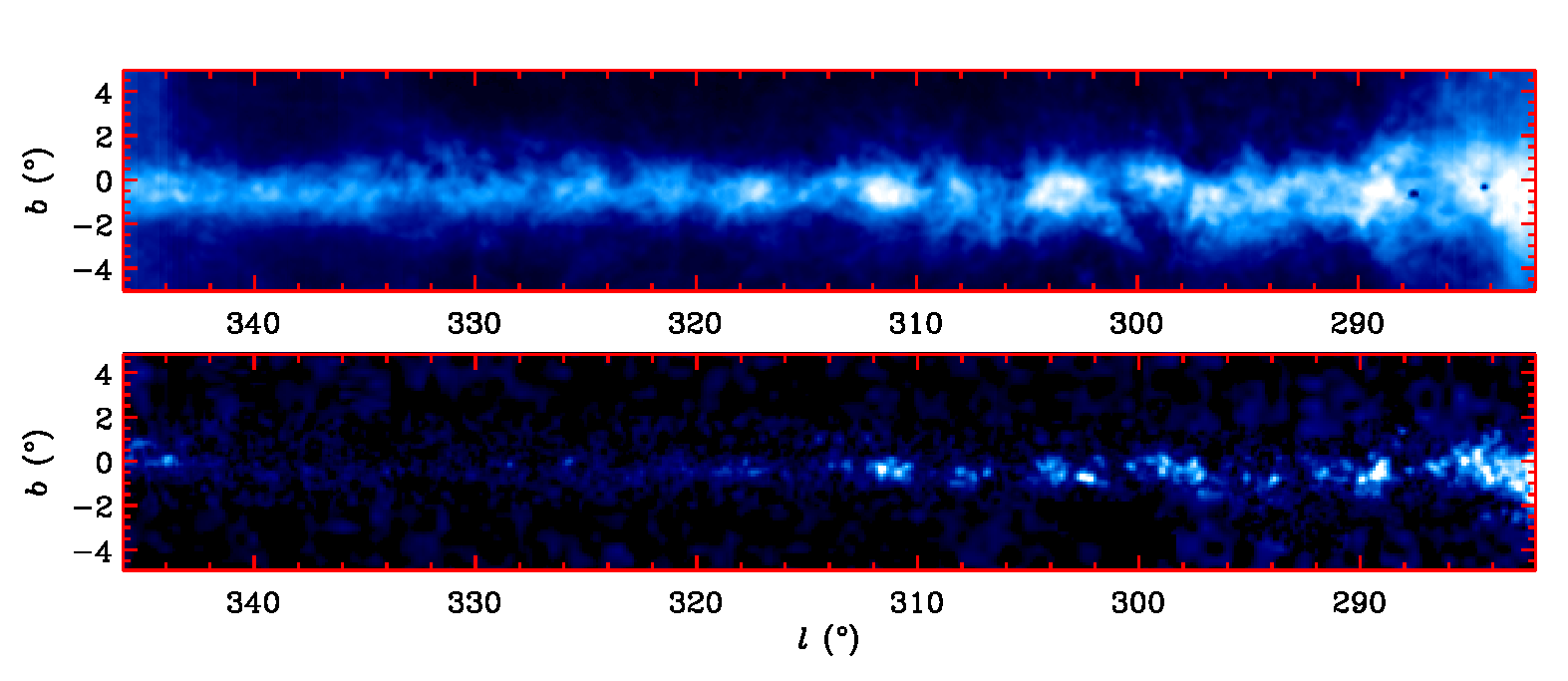}
\label{fig:gas_lb}}
\centering
\subfigure[\schi\ and CO maps of the outer Carina arm with cloud identification]{\includegraphics[width=\textwidth]{%
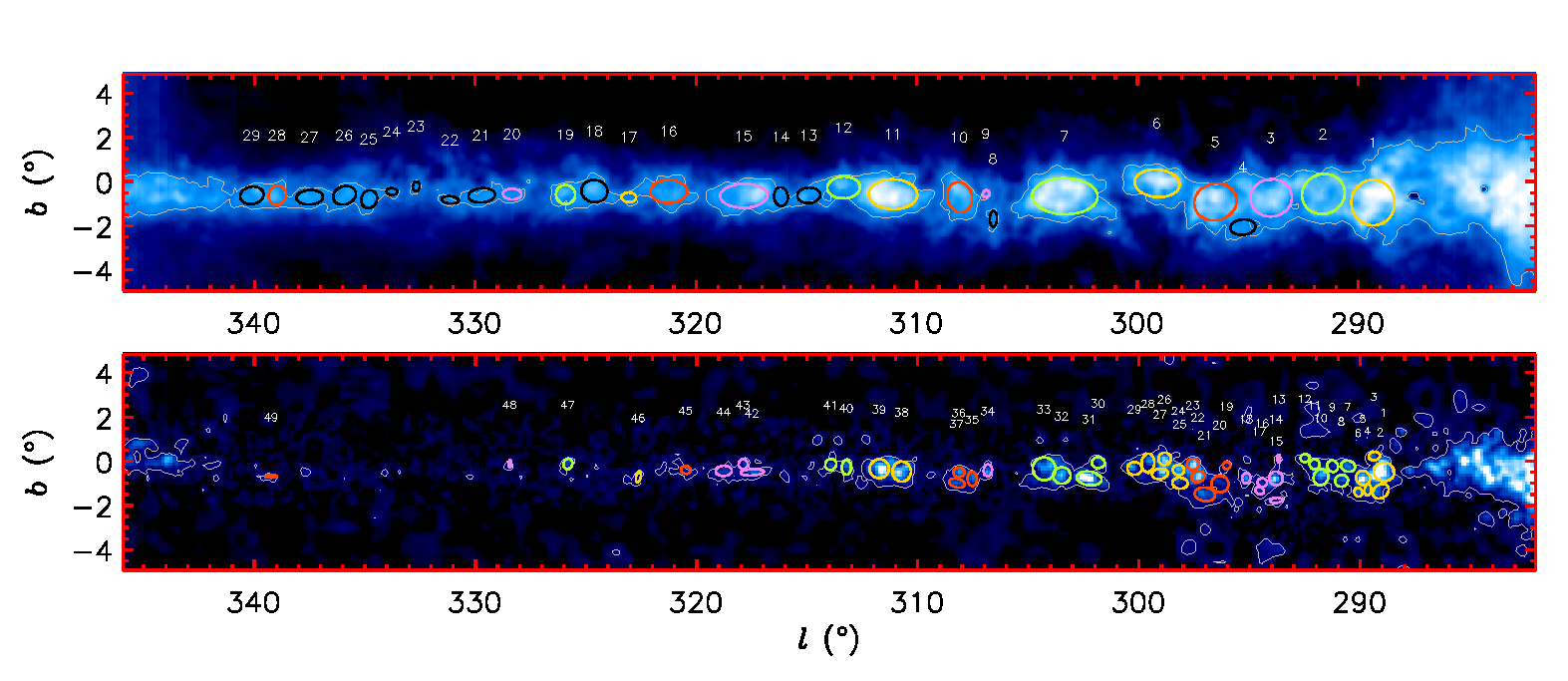}
\label{fig:cloud_clf}}
\caption{
($l$, $b$) maps of the outer Carina arm in \schi\ and CO.
These maps are obtained by integrating spectra within a range of about $25~\kms$ centered around the outer Carina arm traces in Figure~\ref{fig:gas_lv}.
The top panels show a $v$-integrated HI4PI \schi\ map on a square-root scale, with a range of 100 to 2600~$\Kkms$.
The bottom panels show $v$-integrated CfA CO map on the same scale, with a range of 0 to 20~$\Kkms$.
Figure~\ref{fig:cloud_clf} is the same as Figure~\ref{fig:gas_lb}, but with contours and labels indicating \schi\ and CO clouds identified in this study.
A contour in the top panel indicates the lowest \schi\ level of 1210~$\Kkms$ described in Section~\ref{sec:clf_step1}, while a contour in the bottom panel is drawn at the lowest CO level of 2.5~$\Kkms$ described in Section~\ref{sec:clf_step1}.
Best-fit ellipses are drawn to aid in locating the \schi\ clouds in Figure~\ref{fig:cloud_clf}, and, in its bottom panel, the colors of circles are the same as those of associated \schi\ clouds.
The colors of the circles indicate whether they are matched with MCs, and the floating numbers correspond to those assigned to the \schi\ clouds or MCs.
Three \schi\ clouds, H4, H9, and H22, and one MC of M34 are excluded from further analysis and discussion (see Section~\ref{sec:clf_step1} for details).
}
\label{fig:lbmaps}
\end{figure*}

To identify clumpy structures of the arm, we begin by creating a $v$-integrated ($l$, $b$) diagram using data within a width of approximately $25~\kms$ centered at the outer Carina arm traces depicted in Figure~\ref{fig:gas_lv}.
The resulting ($l$, $b$) diagrams in \schi\ and CO, presented in Figure~\ref{fig:gas_lb}, exhibit clumpy structures along Galactic longitudes similar to those shown in earlier studies, such as Figure~6a of \citet{mcgee1964} or Figure~15 of \citet{grabelsky1987},
albeit at a higher resolution. 
To identify \schi\ clouds and MCs from Figure~\ref{fig:gas_lb}, we employ the \idl/\clumpfind\ algorithm \citep{williams1994}.
This algorithm works by contouring an input image from the highest level to the lowest one.
The outcome, such as the number and size of clouds, depends sensitively on the input contour levels.
To obtain appropriate levels of the $v$-integrated intensity, we conduct empirical attempts and determine the following: (1210, 1260, 1320, 1430, 1700, 1810, 1850)~$\Kkms$ for \schi\ and (2.5, 4.5, 7, 10, 15, 20)~$\Kkms$ for CO.
The minimum of the \schi\ contour levels corresponds to $\Nhi \simeq 2.2 \times 10^{21}~\mathrm{cm^{-2}}$, assuming the line to be optically thin, while the minimum CO threshold is comparable to $\Nht \simeq 5 \times 10^{20}~\mathrm{cm^{-2}}$.
We identify twenty-nine \schi\ clouds and forty-nine MCs along the outer Carina arm, spanning from $l=288\arcdeg$ to 340\arcdeg\ 
(see Figure~\ref{fig:cloud_clf}). 
While defining MCs, we try to delineate clumpy structures as individual objects as much as possible, which differs slightly from the identification of \schi\ clouds. 

Some \schi\ clouds were found to have abnormal, discontinuous pixels included in the clumpfind-result area.
These pixels were manually removed.
Additionally, two \schi\ clouds, H2, and H5, were obtained by combining two or three clumpfind-pieces, as they appeared to be connected with each other.  
On the other hand,
we observed that for three \schi\ clouds, H4, H9, and H22, their latitudinal intensity distributions suggest that the identified area is not a significant contributor within the given longitude range. 
These clouds may be part of a larger cloudy structure. 
Therefore, we have excluded them from further analysis and discussion, despite the inferred parameters presented in Table~\ref{tab:hiclouds}. 
As a result, this paper focuses on twenty-six \schi\ clouds.

The process of identifying \schi\ clouds involves defining bright areas against widely-extended diffuse \schi\ emission.
On the other hand, the identification of MCs takes into account most of the detectable CO emissions from the given survey data.
Almost all of the identified MCs listed in this paper are  likely to be associated with the \schi\ clouds mentioned herein.
A few MCs located at $l \sim 288\arcdeg$ to 289\arcdeg\ are not included in this paper, as they are associated with other \schi\ emission features that are not explored here.
Additionally, we limit the minimum pixel number of MCs to 5, which corresponds to a circle area with a radius of approximately $9\farcm5$.

We examined the presence and overlap of MCs within the identified \schi\ cloud areas and determined the possible association between MCs and \schi\ clouds. 
Our analysis revealed that 16 cataloged \schi\ clouds are associated with MCs.
Among the CO-bright \schi\ clouds, six are matched with single MCs, while ten are matched with two or more MCs. 
We also found that H9 is associated with the relevant MC (M34), although it was excluded from the statistical analysis.
Furthermore, M34 was also excluded from further analysis, resulting in a total of 48 MCs that will be discussed in this paper.

\subsubsection{Step~II: Gaussian Decomposition} \label{sec:clf_step2}

In order to understand the kinematic characteristics of clouds, we attempt to isolate the cloud velocity component(s) using multiple Gaussian fitting from a velocity profile averaged within the identified cloud area in ($l$, $b$) space.
However, this task is challenging, particularly for \schi\ profiles, due to contamination by different arm or {\it interarm} components. 
This complexity results in a mixing of many velocities, making it difficult to determine the number of features present.
To efficiently and accurately extract the velocity component(s) of the \schi\ cloud, we obtain an area-averaged velocity profile from a specific region within the cloud.
The cloud area is determined using the \clumpfind\ process, which clips at the minimum velocity-integrated intensity, 1210~$\Kkms$.
We then fit the area-averaged velocity profile to a Gaussian Mixture Model (GMM) using the \texttt{sklearn.mixture.GaussianMixture} function 
in python.\footnote{An example of a one-dimensional GMM can be found at \url{https://www.astroml.org/book_figures/chapter4/fig_GMM_1D.html}. }
This algorithm, which is an unsupervised learning algorithm for clustering, is advantageous in that it does not require an initial setting of the number of Gaussian components and their parameters.
In situations where a profile exhibits complexity, the resulting best-fit values can differ based on the value of the ``\textit{random\_state}" parameter, which controls the random seed for initialization.
To mitigate this, we performed 100 iterations with \textit{random\_state} values ranging from 0 to 99, and utilized a value of \textit{tol} (= 1e-5; the convergence threshold) larger than the default threshold, as determined by manual inspection, 
The average parameters and their standard deviations were computed.
\footnote{Out of the identified \schi\ clouds, convergence issues were encountered in some iterations of the GMM fitting for eight of them. Consequently, statistical values were derived from the iterations that achieved successful convergence and those that did not were excluded from the calculation.}
Smaller standard deviation values suggest that the decomposition results converge to a specific case with greater clarity.
Conversely, larger standard deviation values indicate that the profile for a given velocity range may have multiple possible solutions.

The kinematic selection of the decomposed Gaussian components is based on their central velocities, which are compared to the mean velocity of the outer Carina arm traces with the longitude range of each cloud.
We accept Gaussian components that fall within a velocity range of $\pm 12.5~\kms$ for $l \leq 332\arcdeg$ and $\pm 8~\kms$ for $l > 332\arcdeg$.  
The narrower velocity range for the latter is chosen because the corresponding \schi\ emission features, as seen in the position-velocity diagram (see Figure~\ref{fig:gas_lv}), are distributed over a relatively narrower velocity range than in the former case.

On the other hand, the CO emission features appear to be relatively isolated and exhibit almost no extended diffuse emission feature.
We performed a decomposition process of the velocity component(s) for the CO data, similar to that described above for the \schi\ data, but with the additional criterion that only those Gaussian components with peaks that are $3 \times T_\mathrm{rms}$ or higher were retained.
$T_\mathrm{rms}$ is the noise level, which was determined from regions where no astronomical signal is present.

\subsection{Derivation of Physical Parameters} \label{sec:para}

\subsubsection{Position and Size} \label{sec:size}

The central position ($\lc$, $\bc$) of each cloud is determined by fitting a best-fit ellipse to Galactic coordinates within the clumpfind-result area using the IDL procedure MPFITELLIPSE \citep{markwardt2009}.
This fitting also calculates the semi-major/minor axes ($\rmajor$ and $\rminor$, respectively).
For \schi\ clouds, a position angle (PA) is a free-fitting parameter, whereas it is set to be 0 for MCs.
The presented PA in this paper is a counter-clockwise rotation angle measured from the $+b$-axis.
The best-fit ellipses represent about 50\% of the clumpfind-result area on average.
In estimating other parameters, such as column density and mass, we use a geometrical mean radius ($\rgeo$), which is defined as the radius of an imaginary circle with the same clumpfind-result area ($A$) as given by Equation~\ref{eq:rgeo}.
\begin{equation}
  \rgeo = \sqrt{A/\pi}.
\label{eq:rgeo}
\end{equation}

\subsubsection{Central Velocity and Velocity Dispersion} \label{sec:vsigma}

The central velocity ($\vc$) and velocity dispersion ($\vsigma$) of each cloud are estimated through Gaussian decomposition of the area-averaged velocity ($\overline{T}(v)$) profile, as described in Section~\ref{sec:clf_step2}.
These parameters are intensity-weighted and computed as follows:
\begin{equation}
  \vc = \frac{\sum_{v} v\,\Tgs dv}{\sum_{v}\Tgs dv}, 
\label{eq:vc}
\end{equation}
\begin{equation}
  \vsigma = \sqrt{\frac{\sum_{v} (v-\vc)^2\,\Tgs dv}{\sum_{v}\Tgs dv}}, 
\label{eq:vsigma}
\end{equation}
where $\Tgs ( = \sum_\mathrm{clouds} \overline{T}(v)_\mathrm{Gauss})$ 
represents the sum of Gaussian components that are identified as being associated with each cloud, also explained in Section~\ref{sec:clf_step2}.

\subsubsection{Distance} \label{sec:dist}

Based on the derived central velocity, we calculate the kinematic distance to each cloud by employing the flat Galactic rotation model of \citet{reid2016}, which updates the model presented in \citet{reid2014}.
The model provides the Sun's distance to the Galactic center as $\Rsun = 8.34$~kpc and
the Sun's rotational speed as $\thetasun = 240$~$\kms$.

\subsubsection{Mass and Surface Density} \label{sec:MnSD} 

The mass of each \schi\ cloud is determined using the expression:
\begin{equation}
   \Mhi = \Nhi \times \pi \rgeohi^2 \times m_\mathrm{H} \times 1.36, 
\label{eq:mhi}
\end{equation}
where $\rgeohi$ is the geometrical mean radius of the \schi\ cloud as defined earlier, $m_\mathrm{H}$ is the mass of a hydrogen atom, and 1.36 is mean atomic mass per H atom \citep{allen1973}.

The \schi\ column density ($\Nhi$) is calculated under the assumption of a constant spin temperature ($\Ts$), following  \citet{levine2006apj}, where $\Ts$ is assumed to be 155~K.
The calculation is based on the area-averaged velocity profile, from which the portion associated with a given cloud is extracted.
The equation for $\Nhi$ is given by:
 \begin{equation}
 \Nhi = 1.82 \times 10^{18}\ \sum_{v} -\Ts \mathrm{ln}\left( 1-\frac{\Tgs}{\Ts} \right) dv.
\label{eq:Nhi} 
\end{equation}

Similarly, the mass of each MC ($\Mht$),  which represents the total mass of molecular hydrogen, is calculated using the following equation:
\begin{equation}
   \Mht = \Xco \times \sum_{v} \Tgs dv \times \pi \rgeoco ^2 \times m_\mathrm{H} \times 2.72,
\label{eq:mht}
\end{equation}
where $\Xco$ is the CO-to-$\rmht$\ conversion factor of $2\times 10^{20}$ $(\Kkms)^{-1}\,\mathrm{cm^{-2}}$ \citep{bolatto2013}, and 2.72 is the mean atomic mass per H$_2$ molecule.

Although CO is commonly used as a tracer of molecular hydrogen gas, it is known that molecular gas exists without being detected by CO lines.
This phenomenon has been observed \citep[e.g.,][]{grenier2005, abdo2010, planck2011a24, planck2011a19} and theoretically predicted \citep[e.g.,][]{vanDishoeck1988, wolfire2010}.
To address this issue, alternative tracers such as OH, $\textrm{HCO}^{+}$, and $\textrm{C}^{+}$ lines have been utilized in many observational studies \citep[e.g.,][]{liszt1996,lucas1996,tang2017,park2018}.
Studies by \citet{pineda2013} have revealed the existence of CO-dark molecular gas, which is warm and diffuse, extends over a wider range of Galactocentric distance ($\Rgc$) than the cold and dense molecular gas traced by $^{12}\textrm{CO}$ and $^{13}\textrm{CO}$.
The contribution of CO-dark molecular gas increases with increasing $\Rgc$, resulting in an increase in the $\Xco$ conversion factor as well.
From the relation between $\Xco$ and $\Rgc$ presented in Figure~20 of \citet{pineda2013}, it is evident that $\Xco$ increases by a factor of approximately 1.5 to 2.5 in the $\Rgc$ range of (9, 11)~kpc.
For this study, we have adopted the conventional $\Xco$, which means that the resulting $\Mht$ values may represent a lower limit of the actual total molecular cloud mass.

\subsubsection{Basic Properties} \label{sec:para_res}

Tables~\ref{tab:hiclouds} and \ref{tab:coclouds} provide the physical parameters of the 29 \schi\ clouds and 49 MCs identified in this study.
The geometrical mean radii of 26 \schi\ clouds, except H4, H9, and H22, range from 74~pc to 318~pc with a mean value of 187~pc.
The \schi\ cloud mass ranges from $4.2 \times 10^5~\msol$ to $2.0 \times 10^7~\msol$, with most clouds (23 of 26) being more massive than $10^6~\msol$.
The mean \schi\ cloud mass is calculated to be $5.9 \times 10^6~\msol$. 
According to the classification of \citet[][hereafter, EE87]{elmegreen1987}, clouds with a mass greater than $\sim 10^6~\msol$ are referred to as \schi\ {\it superclouds}, and most clouds in this study can be classified as such.

\begin{deluxetable*}{ccccc lccc|r rrlc}
\centering
\tabletypesize{\scriptsize}
\setlength{\tabcolsep}{1mm}
\tablewidth{0pt}
\tablecaption{\schi\ Superclouds in the Carina Arm ($288\arcdeg \lesssim l \lesssim 340\arcdeg$) \label{tab:hiclouds}}
\tablehead{
 & 
\colhead{$\lc$\tablenotemark{\scriptsize *}} & \colhead{$\bc$\tablenotemark{\scriptsize *}} & \colhead{$\rmajor$\tablenotemark{\scriptsize *}} & \colhead{$\rminor$\tablenotemark{\scriptsize *}} & \colhead{PA\tablenotemark{\scriptsize *}} & 
\colhead{$\rgeo$} & 
\colhead{$\vc$\tablenotemark{\scriptsize \dag}} & \colhead{$\vsigma$ {\scriptsize \dag}} & 
\colhead{$d$} & \colhead{$\Rgc$} & 
\colhead{$\rgeo$} & \colhead{$\SDhi$\tablenotemark{\scriptsize \dag}} & \colhead{$\Mhi$ {\scriptsize \ddag}} \\
\colhead{\#$_{\rm HI}$} & 
\colhead{(\arcdeg)} & \colhead{(\arcdeg)} & \colhead{(\arcdeg)} & \colhead{(\arcdeg)} & \colhead{(\arcdeg)} & 
\colhead{(\arcdeg)}&
\colhead{($\kms$)} & \colhead{($\kms$)} & 
\colhead{(kpc)} & \colhead{(kpc)} & 
\colhead{(pc)} & \colhead{($\msol \rm pc^{-2}$)} & \colhead{($\msol$)}
\vspace{-2mm}\\
\colhead{(1)} &
\colhead{(2)} & \colhead{(3)} & \colhead{(4)} & \colhead{(5)} & \colhead{(6)} & 
\colhead{(7)} &
\colhead{(8)} & \colhead{(9)} & 
\colhead{(10)} & \colhead{(11)} & 
\colhead{(12)} & \colhead{(13)} & \colhead{(14)} 
}
\startdata
\phn H1 & 289.34 (0.01)& $-$0.95 (0.01)&  1.03 (0.01)& 0.98 (0.01)&\phn\phn 8  (9)  &  1.43 & $+$19.3 (2.4)&  8.2 (1.2)    &   7.4 &   9.1 &     184 &   67.4 (14.5) &  7.2e+6 (1.6e+6)   \\
\phn H2 & 291.58 (0.01)& $-$0.56 (0.01)&  0.98 (0.01)& 0.90 (0.01)&       111  (10) &  1.30 & $+$24.9 (2.3)&  8.4 (1.3)    &   8.3 &   9.4 &     189 &   54.0 (9.2)  &  6.0e+6 (1.0e+6)   \\
\phn H3 & 293.95 (0.01)& $-$0.74 (0.01)&  0.94 (0.01)& 0.85 (0.01)&\phn    96  (7)  &  1.27 & $+$24.7 (1.7)&  8.0 (1.0)    &   8.9 &   9.4 &     197 &   58.4 (9.7)  &  7.1e+6 (1.2e+6)   \\
\phn H4 & 295.23 (0.02)& $-$2.06 (0.01)&  0.57 (0.01)& 0.35 (0.01)&\phn    95  (3)  &  0.62 & $+$22.0 (2.1)&  8.2 (1.2)    &   8.9 &   9.3 &      96 &   42.1 (7.7)  &  1.2e+6 (2.2e+5)   \\
\phn H5 & 296.47 (0.01)& $-$0.92 (0.01)&  0.97 (0.01)& 0.83 (0.01)&       104  (3)  &  1.29 & $+$22.3 (1.6)&  8.6 (0.9)    &   9.3 &   9.3 &     209 &   64.8 (8.7)  &  8.9e+6 (1.2e+6)   \\
\phn H6 & 299.11 (0.02)& $-$0.08 (0.01)&  1.02 (0.01)& 0.62 (0.01)&\phn    85  (1)  &  1.18 & $+$24.4 (2.4)&  8.4 (1.1)    &  10.0 &   9.4 &     205 &   53.8 (8.7)  &  7.1e+6 (1.1e+6)   \\
\phn H7 & 303.29 (0.02)& $-$0.66 (0.01)&  1.49 (0.01)& 0.82 (0.01)&\phn    89  (1)  &  1.60 & $+$29.4 (2.1)&  8.0 (1.1)    &  11.4 &   9.8 &     318 &   61.9 (11.9) &  2.0e+7 (3.8e+6)   \\
\phn H8 & 306.53 (0.01)& $-$1.68 (0.02)&  0.36 (0.02)& 0.16 (0.01)&       179  (3)  &  0.34 & $+$33.6 (1.3)&  8.2 (1.0)    &  12.5 &  10.1 & \phn 74 &   44.7 (6.2)  &  7.7e+5 (1.1e+5)   \\
\phn H9 & 306.86 (0.01)& $-$0.55 (0.01)&  0.17 (0.01)& 0.14 (0.01)&       163  (27) &  0.23 & $+$34.2 (2.1)&  9.1 (1.2)    &  12.6 &  10.1 & \phn 50 &   44.5 (8.2)  &  3.4e+5 (6.2e+4)   \\
    H10 & 308.06 (0.01)& $-$0.71 (0.01)&  0.69 (0.01)& 0.56 (0.01)&\phn    15  (5)  &  0.91 & $+$31.7 (2.8)&  7.3 (1.2)    &  12.7 &  10.0 &     202 &   42.2 (10.5) &  5.4e+6 (1.3e+6)   \\
    H11 & 311.07 (0.02)& $-$0.57 (0.01)&  1.10 (0.01)& 0.68 (0.01)&\phn    89  (1)  &  1.28 & $+$33.1 (2.9)&  8.6 (1.3)    &  13.5 &  10.2 &     303 &   61.2 (15.3) &  1.8e+7 (4.5e+6)   \\
    H12 & 313.31 (0.02)& $-$0.27 (0.01)&  0.74 (0.01)& 0.49 (0.01)&\phn    92  (2)  &  0.80 & $+$32.8 (2.1)& 12.5 (5.9)\phn&  14.0 &  10.2 &     195 &   46.8 (9.9)  &  5.6e+6 (1.2e+6)   \\
    H13 & 314.89 (0.02)& $-$0.62 (0.01)&  0.52 (0.02)& 0.35 (0.01)&\phn    94  (4)  &  0.57 & $+$36.2 (2.0)&  8.6 (1.2)    &  14.7 &  10.6 &     145 &   47.3 (9.4)  &  3.1e+6 (6.2e+5)   \\
    H14 & 316.15 (0.01)& $-$0.68 (0.02)&  0.42 (0.02)& 0.31 (0.01)&\phn\phn 5  (5)  &  0.46 & $+$34.8 (1.9)&  8.8 (1.5)    &  14.8 &  10.5 &     120 &   43.8 (9.4)  &  2.0e+6 (4.3e+5)   \\
    H15 & 317.83 (0.02)& $-$0.64 (0.01)&  1.07 (0.01)& 0.54 (0.01)&\phn    91  (1)  &  1.13 & $+$35.8 (3.0)&  8.0 (1.3)    &  15.3 &  10.7 &     301 &   51.1 (11.8) &  1.5e+7 (3.5e+6)   \\
    H16 & 321.23 (0.02)& $-$0.44 (0.01)&  0.84 (0.01)& 0.52 (0.01)&\phn    89  (2)  &  0.94 & $+$29.1 (2.5)& 10.0 (3.4)\phn&  15.4 &  10.3 &     251 &   47.4 (9.1)  &  9.4e+6 (1.8e+6)   \\
    H17 & 323.04 (0.02)& $-$0.71 (0.01)&  0.33 (0.01)& 0.22 (0.01)&\phn    89  (6)  &  0.36 & $+$24.7 (1.7)&  7.5 (0.9)    &  15.4 &  10.0 & \phn 97 &   40.0 (7.1)  &  1.2e+6 (2.1e+5)   \\
    H18 & 324.60 (0.01)& $-$0.43 (0.01)&  0.59 (0.01)& 0.50 (0.01)&\phn    83  (7)  &  0.77 & $+$25.9 (2.2)&  8.3 (1.3)    &  15.8 &  10.2 &     213 &   46.6 (11.0) &  6.6e+6 (1.6e+6)   \\
    H19 & 325.91 (0.01)& $-$0.58 (0.01)&  0.44 (0.01)& 0.43 (0.01)&\phn    66 (131) &  0.63 & $+$27.7 (2.4)&  8.1 (1.4)    &  16.3 &  10.5 &     178 &   45.2 (10.2) &  4.5e+6 (1.0e+6)   \\
    H20 & 328.32 (0.02)& $-$0.56 (0.01)&  0.42 (0.01)& 0.24 (0.01)&\phn    88  (3)  &  0.47 & $+$32.0 (1.9)&  7.5 (0.9)    &  17.4 &  11.2 &     142 &   39.1 (7.9)  &  2.5e+6 (5.1e+5)   \\
    H21 & 329.71 (0.02)& $-$0.62 (0.01)&  0.61 (0.02)& 0.32 (0.01)&\phn    95  (2)  &  0.60 & $+$30.9 (1.6)& 10.2 (3.0)\phn&  17.5 &  11.2 &     182 &   42.7 (7.2)  &  4.5e+6 (7.6e+5)   \\
    H22 & 331.13 (0.03)& $-$0.83 (0.01)&  0.39 (0.02)& 0.16 (0.01)&\phn    85  (4)  &  0.34 & $+$29.9 (2.3)& 10.1 (3.0)\phn&  17.8 &  11.2 &     105 &   44.4 (10.9) &  1.5e+6 (3.7e+5)   \\
    H23 & 332.67 (0.01)& $-$0.21 (0.02)&  0.22 (0.01)& 0.15 (0.01)&       172  (9)  &  0.27 & $+$20.7 (2.1)&  6.7 (1.1)    &  16.9 &  10.3 & \phn 80 &   21.3 (5.2)  &  4.2e+5 (1.0e+5)   \\
    H24 & 333.77 (0.02)& $-$0.45 (0.01)&  0.25 (0.01)& 0.16 (0.01)&\phn    87  (8)  &  0.29 & $+$20.6 (2.5)&  7.4 (2.8)    &  17.1 &  10.3 & \phn 86 &   26.7 (7.9)  &  6.3e+5 (1.9e+5)   \\
    H25 & 334.81 (0.01)& $-$0.79 (0.02)&  0.41 (0.01)& 0.35 (0.01)&       158  (12) &  0.54 & $+$19.1 (1.8)&  6.1 (0.9)    &  17.1 &  10.2 &     162 &   29.2 (7.2)  &  2.4e+6 (5.9e+5)   \\
    H26 & 335.92 (0.02)& $-$0.62 (0.01)&  0.53 (0.01)& 0.40 (0.01)&       109  (6)  &  0.66 & $+$19.5 (1.2)&  6.1 (1.0)    &  17.4 &  10.4 &     201 &   28.7 (6.4)  &  3.6e+6 (8.1e+5)   \\
    H27 & 337.49 (0.02)& $-$0.71 (0.01)&  0.60 (0.01)& 0.33 (0.01)&\phn    93  (2)  &  0.66 & $+$20.6 (2.2)&  6.3 (1.3)    &  17.9 &  10.7 &     206 &   27.6 (8.8)  &  3.7e+6 (1.2e+6)   \\
    H28 & 338.96 (0.01)& $-$0.62 (0.01)&  0.44 (0.01)& 0.40 (0.01)&       169  (12) &  0.60 & $+$21.8 (1.6)&  6.0 (1.3)    &  18.5 &  11.1 &     193 &   34.1 (9.3)  &  4.0e+6 (1.1e+6)   \\
    H29 & 340.12 (0.02)& $-$0.62 (0.01)&  0.53 (0.01)& 0.39 (0.01)&       100  (4)  &  0.67 & $+$22.3 (1.7)&  5.7 (0.5)    &  18.9 &  11.4 &     220 &   33.0 (6.2)  &  5.0e+6 (9.4e+5)   \\
\enddata
\tablecomments{(1) Number assigned to a \schi\ cloud; 
(2--3) central position in Galactic coordinates;
(4--6) semi-major/minor axes and position angle (PA) of a best-fit ellipse; 
(7) angular geometrical mean radius;
(8) intensity-weighted mean LSR velocity;
(9) intensity-weighted velocity dispersion;
(10) Heliocentric distance;
(11) Galactocentric distance;
(12) linear geometrical mean radius;
(13) surface density;
(14) \schi\ cloud mass.
}
\tablenotetext{*}{These parameters were derived by an ellipse-fitting. 
                  The value in parentheses is $1\sigma$ uncertainty of each derived ellipse parameter. 
}
\tablenotetext{\dag}{The standard deviation of the outputs from 100 iterations.
}
\tablenotetext{\ddag}{The given uncertainty considers only that of the surface density (relevant to integral of decomposed Gaussian components), not including that of distance.
}
\end{deluxetable*}
\begin{deluxetable*}{ccccc clcll |crrlc c}
\tabletypesize{\scriptsize}
\setlength{\tabcolsep}{1mm}
\tablewidth{0pt}
\tablecaption{Molecular Clouds in the Carina Arm ($288\arcdeg \lesssim l \lesssim 340\arcdeg$) \label{tab:coclouds}}
\tablehead{
\colhead{} & \colhead{} &
 \colhead{$\lc$\tablenotemark{\scriptsize *}} &  \colhead{$\bc$\tablenotemark{\scriptsize *}} & \colhead{$\rmajor$\tablenotemark{\scriptsize *}} & \colhead{$\rminor$\tablenotemark{\scriptsize *}} & \colhead{PA\tablenotemark{\scriptsize *}} & 
\colhead{$\rgeo$} & 
\colhead{$\vc$\tablenotemark{\scriptsize \dag}} & \colhead{$\vsigma$\tablenotemark{\scriptsize \dag}} &
\colhead{$d$} & \colhead{$\Rgc$} & 
\colhead{$\rgeo$} & \colhead{$\SDht$\tablenotemark{\scriptsize \dag}} & \colhead{$\Mht$\tablenotemark{\scriptsize \ddag}}   \vspace{-2mm}\\
\colhead{\#$_{\rm CO}$} & \colhead{\#$_{\rm HI}$} & 
\colhead{(\arcdeg)} & \colhead{(\arcdeg)} & \colhead{(\arcdeg)} & \colhead{(\arcdeg)} & \colhead{(\arcdeg)} &
\colhead{(\arcdeg)} & 
\colhead{($\kms$)} & \colhead{($\kms$)} &
\colhead{(kpc)} & \colhead{(kpc)} &
 \colhead{(pc)} & \colhead{($\msol \rm pc^{-2}$)} & \colhead{($\msol$)} \vspace{-2mm}\\
\colhead{(1)} & \colhead{(2)} &
\colhead{(3)} & \colhead{(4)} & \colhead{(5)} & \colhead{(6)} & \colhead{(7)} &
\colhead{(8)} & 
\colhead{(9)} & \colhead{(10)} & 
\colhead{(11)} & \colhead{(12)} & 
\colhead{(13)} & \colhead{(14)} & \colhead{(15)}
}
\startdata
  \phn M1                              & \phn H1 &  288.93 (0.02)& $-$0.48 (0.02)& 0.47 (0.02) & 0.42 (0.02) &\phn   29 (29)  & 0.62 & $+$21.1 (0.3)    & 7.2 (0.5)    &  7.4 &  9.2 &  80 &  80.9 (1.7)   &  1.6e+6 (3.3e+4) \\   
  \phn M2                              & \phn H1 &  289.07 (0.04)& $-$1.36 (0.03)& 0.39 (0.04) & 0.29 (0.03) &\phn   37 (40)  & 0.37 & $+$21.4 (0.7)    & 6.9 (0.9)    &  7.5 &  9.2 &  49 &  28.8 (1.7)   &  2.2e+5 (1.3e+4) \\   
  \phn M3                              & \phn H1 &  289.37 (0.02)& $+$0.24 (0.02)& 0.26 (0.02) & 0.19 (0.01) &      174 (10)  & 0.32 & $+$19.8 (1.6)    & 4.3 (1.8)    &  7.4 &  9.1 &  42 &  19.0 (4.5)   &  1.0e+5 (2.4e+4) \\   
  \phn M4                              & \phn H1 &  289.66 (0.02)& $-$1.29 (0.03)& 0.26 (0.02) & 0.15 (0.01) &      100 (13)  & 0.28 & $+$19.3 (0.6)    & 9.5 (0.5)    &  7.4 &  9.1 &  36 &  41.8 (1.9)   &  1.7e+5 (7.8e+3) \\   
  \phn M5                              & \phn H1 &  289.87 (0.02)& $-$0.76 (0.02)& 0.33 (0.02) & 0.31 (0.02) &      126 (36)  & 0.43 & $+$17.8 (1.8)    & 6.6 (0.8)    &  7.3 &  9.0 &  55 &  38.9 (7.2)   &  3.7e+5 (6.8e+4) \\   
  \phn M6                              & \phn H1 &  290.09 (0.02)& $-$1.40 (0.02)& 0.21 (0.02) & 0.17 (0.02) &\phn   79 (25)  & 0.28 & $+$15.2 (1.6)    & 6.1 (1.4)    &  7.2 &  8.9 &  35 &  24.4 (3.1)   &  9.6e+4 (1.2e+4) \\   
  \phn M7                              & \phn H2 &  290.55 (0.03)& $-$0.23 (0.02)& 0.33 (0.02) & 0.25 (0.02) &\phn   97 (32)  & 0.40 & $+$22.9 (1.5)    & 7.7 (0.9)    &  7.9 &  9.3 &  55 &  26.8 (3.3)   &  2.5e+5 (3.0e+4) \\   
  \phn M8                              & \phn H2 &  290.85 (0.02)& $-$0.88 (0.02)& 0.29 (0.02) & 0.27 (0.01) &\phn   41 (***) & 0.39 & $+$22.4 (2.2)    & 7.7 (1.6)    &  8.0 &  9.3 &  55 &  19.7 (3.1)   &  1.9e+5 (3.0e+4) \\   
  \phn M9                              & \phn H2 &  291.25 (0.02)& $-$0.22 (0.03)& 0.27 (0.02) & 0.24 (0.02) &      155 (21)  & 0.32 & $+$20.9 (1.9)    & 7.5 (1.9)    &  7.9 &  9.2 &  43 &  16.0 (3.5)   &  9.5e+4 (2.0e+4) \\   
      M10                              & \phn H2 &  291.78 (0.02)& $-$0.72 (0.03)& 0.37 (0.02) & 0.33 (0.01) &\phn   27 (60)  & 0.46 & $+$22.2 (0.3)    & 8.0 (0.3)    &  8.2 &  9.3 &  66 &  38.8 (0.8)   &  5.3e+5 (1.1e+4) \\   
      M11                              & \phn H2 &  292.07 (0.02)& $-$0.14 (0.02)& 0.23 (0.02) & 0.22 (0.02) &      110 (39)  & 0.33 & $+$24.4 ($<$0.1) & 5.5 ($<$0.1) &  8.4 &  9.4 &  48 &  18.0 ($<$0.1) &  1.3e+5 (1.0e+2) \\   
      M12                              & \phn H2 &  292.50 (0.02)& $+$0.14 (0.02)& 0.21 (0.02) & 0.18 (0.01) &\phn   77 (18)  & 0.27 & $+$19.7 (0.5)    & 2.7 (0.5)    &  8.1 &  9.1 &  39 &  13.0 (1.0)   &  6.1e+4 (4.4e+3) \\   
      M13                              & \phn H3 &  293.69 (0.02)& $+$0.12 (0.03)& 0.17 (0.04) & 0.06 (0.03) &\phn   68 (28)  & 0.17 & $+$25.5 (0.1)    & 2.7 ($<$0.1) &  8.9 &  9.4 &  27 &  15.9 (0.2)   &  3.6e+4 (5.2e+2) \\   
      M14                              & \phn H3 &  293.81 (0.03)& $-$0.77 (0.03)& 0.31 (0.02) & 0.30 (0.02) &\phn   82 (52)  & 0.39 & $+$29.5 (0.2)    & 5.4 (0.2)    &  9.2 &  9.6 &  63 &  41.1 (1.7)   &  5.1e+5 (2.1e+4) \\   
      M15                              & \phn H3 &  293.81 (0.04)& $-$1.77 (0.02)& 0.28 (0.04) & 0.12 (0.02) &\phn   23 (163) & 0.24 & $+$21.4 (0.1)    & 2.6 (0.1)    &  8.6 &  9.2 &  37 &  16.2 (0.2)   &  6.9e+4 (1.0e+3) \\   
      M16                              & \phn H3 &  294.43 (0.02)& $-$0.96 (0.03)& 0.22 (0.03) & 0.19 (0.03) &      180 (14)  & 0.29 & $+$28.6 (0.9)    & 9.9 (1.6)    &  9.3 &  9.6 &  47 &  27.0 (3.0)   &  1.9e+5 (2.1e+4) \\   
      M17                              & \phn H3 &  294.55 (0.03)& $-$1.33 (0.02)& 0.21 (0.02) & 0.16 (0.02) &      113 (55)  & 0.26 & $+$23.3 ($<$0.1) & 3.0 ($<$0.1) &  8.9 &  9.3 &  41 &  17.5 (0.2)   &  9.2e+4 (1.2e+3) \\   
      M18                              & \phn H3 &  295.15 (0.02)& $-$0.77 (0.02)& 0.24 (0.02) & 0.19 (0.01) &      106 (19)  & 0.31 & $+$24.3 (2.3)    & 7.7 (1.5)    &  9.1 &  9.4 &  49 &  29.1 (5.6)   &  2.2e+5 (4.2e+4) \\   
      M19                              & \phn H5 &  296.04 (0.03)& $-$0.20 (0.03)& 0.17 (0.02) & 0.16 (0.04) &\phn   83 (6)   & 0.21 & $+$15.4 (1.6)    & 3.9 (0.9)    &  8.6 &  9.0 &  32 &  15.8 (1.6)   &  5.0e+4 (5.0e+3) \\   
      M20                              & \phn H5 &  296.34 (0.03)& $-$1.03 (0.03)& 0.39 (0.03) & 0.37 (0.03) &\phn   80 (145) & 0.49 & $+$21.9 (2.2)    & 6.5 (2.2)    &  9.2 &  9.3 &  78 &  18.3 ($<$0.1) &  3.5e+5 (5.8e$-$1) \\   
      M21                              & \phn H5 &  297.02 (0.02)& $-$1.50 (0.02)& 0.47 (0.02) & 0.31 (0.01) &      119 (***) & 0.54 & $+$21.7 ($<$0.1) & 5.3 ($<$0.1) &  9.3 &  9.3 &  88 &  23.5 ($<$0.1) &  5.7e+5 (9.8e+1) \\   
      M22                              & \phn H5 &  297.34 (0.02)& $-$0.70 (0.02)& 0.30 (0.02) & 0.25 (0.02) &      176 (9)   & 0.40 & $+$23.4 (0.3)    & 7.0 (0.2)    &  9.5 &  9.4 &  66 &  53.8 (1.2)   &  7.4e+5 (1.6e+4) \\   
      M23                              & \phn H5 &  297.56 (0.02)& $-$0.14 (0.02)& 0.28 (0.02) & 0.26 (0.01) &\phn   98 (8)   & 0.40 & $+$19.5 (0.2)    & 5.2 (0.1)    &  9.3 &  9.2 &  65 &  56.3 (1.3)   &  7.4e+5 (1.7e+4) \\   
      M24                              & \phn H6 &  298.15 (0.03)& $-$0.99 (0.02)& 0.34 (0.02) & 0.25 (0.02) &      132 (***) & 0.38 & $+$23.0 (0.1)    & 6.0 (0.1)    &  9.7 &  9.4 &  64 &  21.2 (0.1)   &  2.5e+5 (1.5e+3) \\   
      M25                              & \phn H6 &  298.22 (0.02)& $-$0.40 (0.02)& 0.26 (0.02) & 0.23 (0.02) &\phn   84 (31)  & 0.36 & $+$24.5 ($<$0.1) & 7.8 ($<$0.1) &  9.8 &  9.4 &  62 &  47.9 ($<$0.1) &  6.4e+5 (1.7e+2) \\   
      M26                              & \phn H6 &  298.84 (0.02)& $+$0.11 (0.02)& 0.28 (0.02) & 0.28 (0.02) &\phn   52 (***) & 0.41 & $+$23.8 (0.2)    & 5.5 (0.2)    &  9.9 &  9.4 &  70 &  45.3 (0.9)   &  7.0e+5 (1.4e+4) \\   
      M27                              & \phn H6 &  299.05 (0.03)& $-$0.57 (0.02)& 0.36 (0.03) & 0.24 (0.02) &\phn   90 (63)  & 0.41 & $+$25.0 (0.2)    & 6.7 (0.1)    & 10.1 &  9.5 &  71 &  40.1 (0.9)   &  6.4e+5 (1.4e+4) \\   
      M28                              & \phn H6 &  299.60 (0.02)& $-$0.08 (0.03)& 0.42 (0.02) & 0.29 (0.01) &\phn   83 (5)   & 0.46 & $+$23.3 ($<$0.1) & 7.7 ($<$0.1) & 10.1 &  9.4 &  81 &  28.4 ($<$0.1) &  5.8e+5 (8.1e+2) \\   
      M29                              & \phn H6 &  300.20 (0.03)& $-$0.33 (0.03)& 0.33 (0.02) & 0.29 (0.02) &\phn   86 (16)  & 0.42 & $+$30.6 ($<$0.1) & 4.9 ($<$0.1) & 10.8 &  9.8 &  79 &  28.7 ($<$0.1) &  5.6e+5 (2.8e+2) \\   
      M30                              & \phn H7 &  301.84 (0.02)& $-$0.07 (0.02)& 0.30 (0.02) & 0.24 (0.02) &      174 (13)  & 0.37 & $+$23.9 (0.1)    & 4.0 (0.1)    & 10.6 &  9.4 &  69 &  24.0 (0.5)   &  3.4e+5 (7.3e+3) \\   
      M31                              & \phn H7 &  302.26 (0.03)& $-$0.78 (0.02)& 0.56 (0.02) & 0.31 (0.01) &\phn   77 (7)   & 0.58 & $+$30.5 ($<$0.1) & 4.7 ($<$0.1) & 11.3 &  9.8 & 115 &  47.7 (0.4)   &  2.0e+6 (1.8e+4) \\   
      M32                              & \phn H7 &  303.49 (0.02)& $-$0.65 (0.02)& 0.39 (0.02) & 0.37 (0.02) &\phn   93 (18)  & 0.54 & $+$27.9 ($<$0.1) & 6.0 ($<$0.1) & 11.3 &  9.7 & 106 &  37.7 ($<$0.1) &  1.3e+6 (1.4e+3) \\   
      M33                              & \phn H7 &  304.29 (0.03)& $-$0.34 (0.03)& 0.52 (0.02) & 0.51 (0.02) &\phn\phn2 (8)   & 0.67 & $+$29.5 (0.1)    & 5.6 (0.2  )  & 11.6 &  9.8 & 136 &  31.9 (1.0)   &  1.9e+6 (6.1e+4) \\   
      M34                              & \phn H9 &  306.84 (0.02)& $-$0.41 (0.02)& 0.27 (0.02) & 0.19 (0.01) &\phn  114 (37)  & 0.32 & $+$24.8 (0.1)    & 3.2 (0.1)    & 11.9 &  9.6 &  67 &  26.2 (0.6)   &  3.7e+5 (8.4e+3) \\   
      M35                              &     H10 &  307.54 (0.02)& $-$0.79 (0.02)& 0.33 (0.02) & 0.21 (0.01) &\phn\phn0 (32)  & 0.39 & $+$32.7 (0.1)    & 3.8 ($<$0.1) & 12.7 & 10.1 &  86 &  27.9 (0.4)   &  6.4e+5 (9.0e+3) \\   
      M36                              &     H10 &  308.14 (0.02)& $-$0.50 (0.02)& 0.26 (0.02) & 0.22 (0.02) &\phn  135 (62)  & 0.35 & $+$32.1 (0.1)    & 6.6 ($<$0.1) & 12.7 & 10.0 &  77 &  31.8 (0.2)   &  5.9e+5 (4.1e+3) \\   
      M37                              &     H10 &  308.23 (0.03)& $-$0.99 (0.02)& 0.36 (0.03) & 0.20 (0.02) &\phn\phn0 (10)  & 0.38 & $+$37.2 (0.3)    & 4.0 (0.9)    & 13.2 & 10.4 &  87 &  19.7 (3.5)   &  4.7e+5 (8.4e+4) \\   
      M38                              &     H11 &  310.73 (0.02)& $-$0.47 (0.02)& 0.45 (0.02) & 0.42 (0.02) &\phn   92 (7)   & 0.63 & $+$28.2 (0.7)    & 7.9 (0.6)    & 13.0 &  9.9 & 142 &  41.9 (4.8)   &  2.7e+6 (3.1e+5) \\   
      M39                              &     H11 &  311.74 (0.02)& $-$0.34 (0.02)& 0.44 (0.02) & 0.42 (0.01) &      171 (***) & 0.62 & $+$28.6 (0.5)    & 9.2 (1.2)    & 13.3 &  9.9 & 144 &  57.5 (4.4)   &  3.7e+6 (2.8e+5) \\
      M40                              &     H12 &  313.23 (0.02)& $-$0.28 (0.03)& 0.33 (0.02) & 0.21 (0.01) &\phn   93 (2)   & 0.37 & $+$43.0 ($<$0.1) & 3.9 ($<$0.1) & 14.9 & 11.0 &  95 &  29.3 (0.2)   &  8.4e+5 (5.1e+3) \\
      M41                              &     H12 &  313.93 (0.03)& $-$0.14 (0.03)& 0.26 (0.03) & 0.24 (0.02) &\phn   77 (149) & 0.32 & $+$32.6 ($<$0.1) & 6.0 ($<$0.1) & 14.1 & 10.3 &  80 &  21.0 (0.1)   &  4.2e+5 (2.5e+3) \\
      M42                              &     H15 &  317.49 (0.04)& $-$0.51 (0.01)& 0.53 (0.04) & 0.14 (0.01) &\phn   90 (8)   & 0.40 & $+$30.1 (0.2)    & 5.8 (0.2)    & 14.7 & 10.2 & 102 &  22.3 (1.2)   &  7.3e+5 (4.1e+4) \\
      M43                              &     H15 &  317.90 (0.03)& $-$0.13 (0.03)& 0.23 (0.03) & 0.21 (0.02) &      180 (14)  & 0.27 & $+$23.7 (4.3)    & 8.0 (1.9)    & 14.2 &  9.8 &  68 &  34.4 (11)    &  5.0e+5 (1.7e+5) \\   
      M44                              &     H15 &  318.78 (0.03)& $-$0.44 (0.02)& 0.37 (0.02) & 0.23 (0.01) &      163 (38)  & 0.39 & $+$35.1 ($<$0.1) & 4.1 ($<$0.1) & 15.5 & 10.7 & 104 &  19.4 (0.3)   &  6.6e+5 (8.6e+3) \\   
 \phn M45\tablenotemark{\scriptsize \S}&     H16 &  320.49 (0.04)& $-$0.39 (0.04)& 0.22 (0.03) & 0.18 (0.05) &      167 (19)  & 0.27 & $+$22.7 (1.1)    & 4.8 (2.2)    & 14.7 &  9.8 &  70 &  17.4 (6.7)   &  2.7e+5 (1.0e+5) \\   
 \phn M46\tablenotemark{\scriptsize \S}&     H17 &  322.63 (0.02)& $-$0.72 (0.04)& 0.22 (0.03) & 0.10 (0.02) &\phn   89 (62)  & 0.21 & $+$19.0 (0.9)    & 6.2 (0.5)    & 14.8 &  9.6 &  55 &  20.2 (2.2)   &  1.9e+5 (2.1e+4) \\   
 \phn M47\tablenotemark{\scriptsize \S}&     H19 &  325.81 (0.02)& $-$0.12 (0.03)& 0.24 (0.02) & 0.20 (0.02) &\phn   86 (14)  & 0.29 & $+$29.2 (0.8)    & 5.0 (1.0)    & 16.4 & 10.6 &  83 &  20.8 (2.1)   &  4.5e+5 (4.4e+4) \\   
 \phn M48\tablenotemark{\scriptsize \S}&     H20 &  328.44 (0.02)& $-$0.12 (0.03)& 0.17 (0.04) & 0.06 (0.03) &      175 (16)  & 0.17 & $+$31.0 ($<$0.1) & 3.2 ($<$0.1) & 17.3 & 11.0 &  52 &  23.7 ($<$0.1) &  2.0e+5 (8.2e+0) \\   
      M49                              &     H28 &  339.23 (0.06)& $-$0.69 (0.02)& 0.28 (0.08) & 0.06 (0.04) &\phn   95 (7)   & 0.17 & $+$23.4 (1.4)    & 7.5 (1.2)    & 18.9 & 11.5 &  57 &  17.6 (2.2)   &  1.8e+5 (2.2e+4) \\   
\enddata

\tablecomments{
(1) Number assigned to a MC; (2) Number of associated \schi\ cloud;
(3--4) central position in Galactic coordinates;
(5--7) semi-major/minor axes and PA of a best-fit ellipse;
(8) angular geometrical mean radius;
(9) intensity-weighted mean LSR velocity;
(10) intensity-weighted velocity dispersion;
(11) Heliocentric distance;
(12) Galactocentric distance;
(13) linear geometrical mean radius; 
(14) surface density;
(15) $\rmht$ mass.
}
\tablenotetext{*}{These parameters were derived by an ellipse-fitting. 
                  The value in parentheses is $1\sigma$ uncertainty of each derived ellipse parameter 
                  but is replaced by the symbol `***' if it is too large (larger than 180\arcdeg).
}
\tablenotetext{\dag}{The standard deviation of the outputs from 100 iterations.
}
\tablenotetext{\ddag}{The given uncertainty considers only that of the surface density (relevant to the integral of decomposed Gaussian components), not including that of distance.
}
\tablenotetext{\S}{This MC is on the edge of the \schi\ cloud when overlaying the $v$-integrated CO map onto the $v$-integrated \schi\ map.
}

\end{deluxetable*}

It is noteworthy that all CO-bright \schi\ clouds observed in this study have masses greater than $1 \times 10^6~\msol$, with the majority exceeding $5 \times 10^6~\msol$. 
These clouds also exhibit \schi\ surface densities of $\SDhi \gtrsim$ 34~$\msol \rm pc^{-2}$, while some CO-dark \schi\ clouds have \schi\ surface densities ranging from 34~$\msol \rm pc^{-2}$ to 47~$\msol \rm pc^{-2}$.
In addition, with respect to the surface density of total gas (see Equation~\ref{eq:sdgas}), the majority of CO-bright \schi\ clouds have values of $\SDgas \gtrsim$ 50~$\msol \rm pc^{-2}$, while the minimum value for the \schi\ clouds in this study is 21.3~$\msol \rm pc^{-2}$.

Observational studies of extragalaxies and the Milky Way suggest that a minimum \schi\ surface density of $\sim 10~\msol\,{\rm pc^{-2}}$ at solar metallicity is required for $\rmht$ formation \citep[e.g.,][]{wong2002, bigiel2008, m.lee2012, m.lee2015}.
The Analytical study by \citet{krumholz2009} has shown that $\rmht$ formation is closely related to metallicity and total gas surface density.
Based on the Galactic metallicity gradient shown in \citep{pedicelli2009} and the fact that our \schi\ clouds are located at $\Rgc \sim$ 9--11~kpc, a rough estimate suggests that the metallicity has decreased to about 0.8 times the solar value, while the most extreme case indicated in their Figure~3 suggests a value of about 0.3 times the solar value.
For the 11 out of 26 \schi\ clouds in our sample that do not show detectable CO emission according to our criteria, the absence of CO may be due to the need for higher \schi\ surface densities resulting from their lower metallicity, although the possibility of CO emission with column densities below our detection limit cannot be entirely excluded.

However, it should be noted that the previous studies of the minimum \schi\ surface density  required for $\rmht$ formation were derived from pixel-to-pixel analysis, while this study is based on the total area of each cloud.
Therefore, a direct comparison between the two values should be made with caution.

The average geometrical mean radius of 48 MCs, excluding M34, is 70~pc, and their $\Mht$ ranges from $3.6 \times 10^4~\msol$ to $3.7 \times 10^6~\msol$, with a mean value of $6.0 \times 10^5~\msol$.
Of the 48 MCs, 41 have $\Mht$ values that equal or exceed $10^5~\msol$, which implies that either giant molecular clouds or groups of MCs, namely, molecular cloud complexes.

In terms of \schi\ clouds, we investigated the relationship between $\Rgc$ and mass or radius using the Spearman's rank correlation test, which examines if two variables are monotonically related.
We found a weak negative correlation between $\Rgc$ and the \schi\ mass of \schi\ clouds (Spearman's correlation coefficient ($\rho$) = $-0.36$ and p-value = 0.071).
This means that as $\Rgc$ increases, the \schi\ mass tends to decrease.
However, the correlation is not statistically significant, so it is unclear whether the two variables have a real relationship.
On the other hand, we found that there is a strong negative correlation between $\Rgc$ and $\rmht$ mass assigned to the corresponding \schi\ clouds ($\rho = -0.52$ and p-value = 0.048). 
This means that as $\Rgc$ increases, the $\rmht$ mass tends to decrease.
The p-value is very small, so the correlation is statistically significant.
Finally, we found that there is no correlation between $\Rgc$ and $\rgeo$ of the \schi\ clouds ($\rho = -0.085$ and p-value = 0.68).
This means that there is no clear relationship between the two variables.

\subsection{Spatial Distribution of \schi\ Superclouds} \label{sec:spatial_hi}

Most of the identified \schi\ superclouds and MCs are below $b=0\arcdeg$.
The mean deviation from $b=0\arcdeg$ is $\Delta b \simeq -0\fdg6$, corresponding to a linear distance of $-110$~pc in the $\Rgc$ range of $\sim9$~kpc to $\sim11$~kpc, based on the distances provided in Tables~\ref{tab:hiclouds} and \ref{tab:coclouds}.
This spatial distribution is likely due to the warped Galactic plane \citep{burton1988,levine2006apj,koo2017}.
Generally, the MCs appear to be situated within their corresponding \schi\ superclouds, although their integrated-intensity peak positions often differ (see Section~\ref{sec:HInCO_spatial} for details).

\begin{figure}[h]
\centering
\epsscale{1.2}
\includegraphics[width=0.45\textwidth]{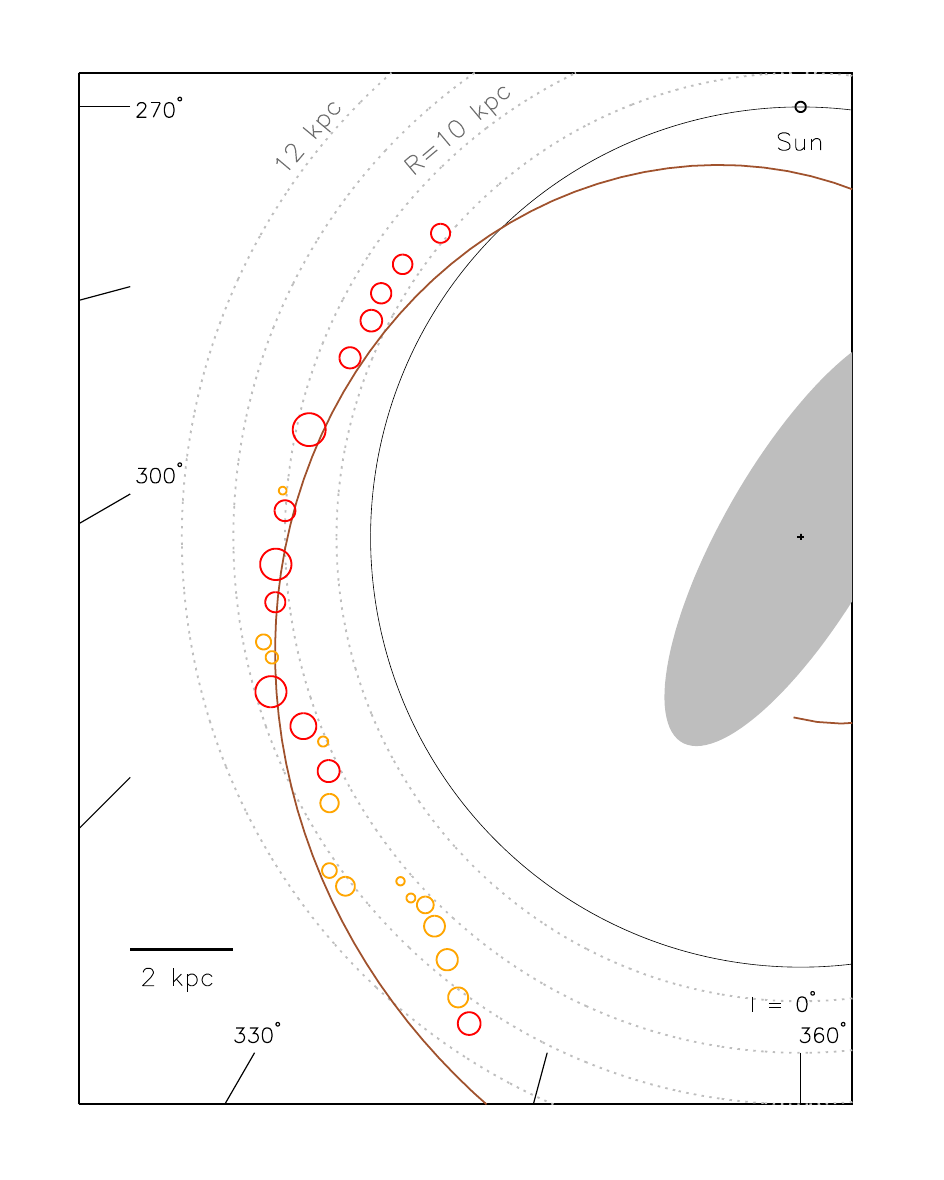}
\caption{
Schematic face-on view of the \schi\ clouds analyzed in this paper.
Circles indicate clouds of mass $\geq 5\times10^6~\msol$ (in red) or $< 5\times10^6~\msol$ (in orange).
The size of each circle corresponds to the geometrical mean radius listed in Table~\ref{tab:hiclouds}.
The brown curve denotes the (Sagittarius-)Carina arm spiral model, as derived by Paper~I.
The filled ellipse located near the Galactic center represents the central bar, as reported by \citep{wegg2015}.
\label{fig:hicloud_xyplane}}
\end{figure}

The spatial distribution of the 26 \schi\ clouds is illustrated in a face-on view in Figure~\ref{fig:hicloud_xyplane}.
Even a cursory glance shows that the clouds are regularly distributed.
The average separation between cloud ($i$) and its immediate neighboring clouds ($i-1$ and $i+1$) can be determined using the cosine law, which is given by the following equation:
\begin{equation}
Separation = 0.5\,\sqrt{ d_{i+1}^2 + d_{i-1}^2– 2\,d_{i+1}\,d_{i-1}\,cos(\Delta l)},
\label{eq:sep}
\end{equation}
where $\Delta l = l_{i+1}-l_{i-1}$.

\begin{figure*}
\centering
\epsscale{1.}
\includegraphics[width=0.45\textwidth]{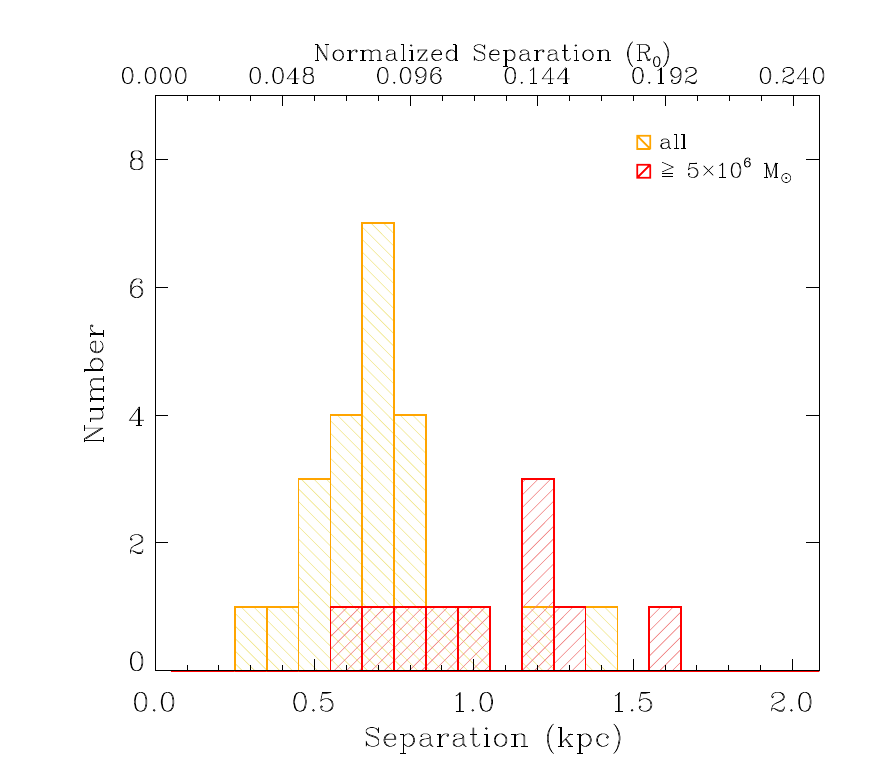}
\includegraphics[width=0.45\textwidth]{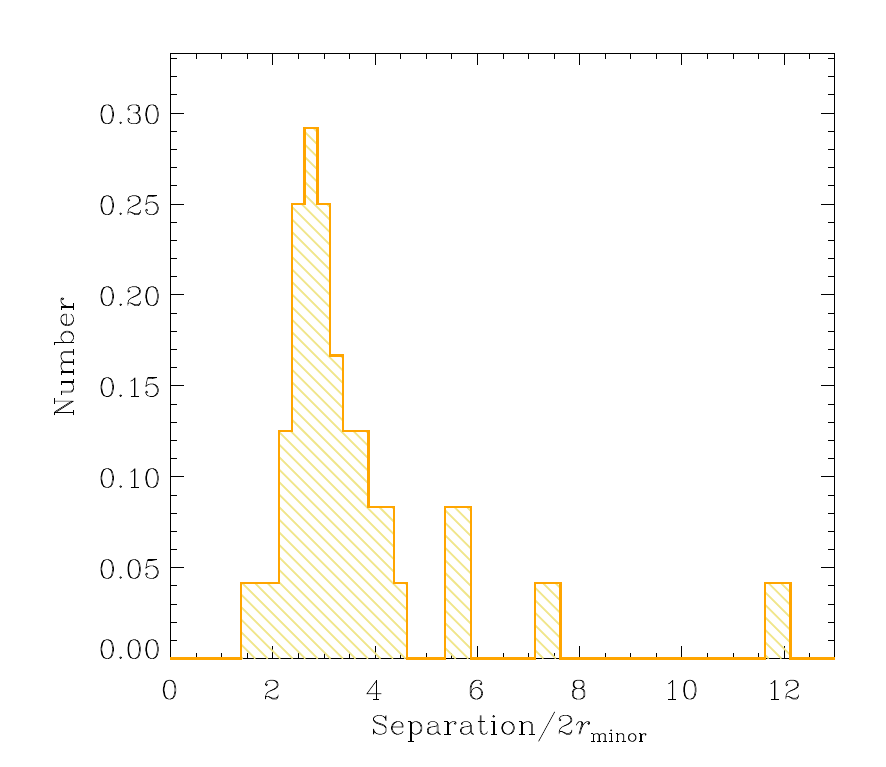}
\caption{
Histogram of the distribution of spacings between neighboring \schi\ clouds (left), and the ratio of the separation to filament diameter (right).
The filament diameter is determined as the cloud minor axis ($2\rminor$).
The yellow sticks with left diagonals represent all twenty-six \schi\ clouds, while the red sticks with right diagonals represent massive clouds with $\Mhi \geq 5 \times 10^{6}~\msol$.
The bin size is $0.012\,\Rsun$, which corresponds to around 100~pc.
The separation of 3.3~kpc between the last two massive clouds (H18 and H29) with $\Mhi \geq 5 \times 10^{6}~\msol$ is not included in this plot.
\label{fig:hicloud_spacing_histo}}
\end{figure*}

The left-hand panel of Figure~\ref{fig:hicloud_spacing_histo} presents a histogram depicting this average spacing between adjacent \schi\ clouds along the Carina arm, covering a distance of approximately 20~kpc.
The bin size used for this analysis is $0.012\,\Rsun$, which is equivalent to 100~pc, assuming $\Rsun =$ 8.34~kpc.
The yellow histogram indicates the distribution of all \schi\ clouds, which displays a prominent peak at $0.084\,\Rsun$ ($\sim 700$~pc).
On the other hand, the red histogram, which represents neighboring \schi\ superclouds with $\Mhi \geq 5\times10^6~\msol$, shows a peak at $0.144\,\Rsun$ ($\sim 1.2$~kpc).

The separation between \schi\ superclouds along the Carina arm has been investigated in previous studies
\citep[e.g.,][]{efremov1998, efremov2009},
which corroborate our finding.
It is worth noting that EE87 reported a spacing of approximately 1500~pc between \schi\ superclouds in the first quadrant.
Although this value is somewhat larger than ours, it will be still comparable, considering the use of different assumptions, such as the Sun's location and the Galactic rotation model.

Regular segmentation in the spiral arms of nearby galaxies \citep{elmegreen1983} has been observed recently in M31 \citep{efremov2010}, M100 \citep{elmegreen2018}, 15 other spirals \citep{elmegreen2019}, and 3 more \citep{gusev2022}. 
It has been attributed to gravitational or combined gravitational and magnetic (Parker-Jeans) instabilities in compressed spiral-arm gas \citep[e.g,][]{elmegreen1979,elmegreen1982,kim2002,dobbs08a,renaud2013,s.lee2011,inoue2019} or wiggle instabilities \citep{wada2004,mandowara2022}. 
The three-dimensional MHD simulation conducted by \citet{s.lee2011} demonstrated that when a perturbation wavelength greater than the Jeans critical wavelength acts on a self-gravitating disk of magnetized isothermal gas, the cooperation of the Parker and Jeans instabilities suppresses convection and generates a dense, large-scale structure of mass and size corresponding to the observed \schi\ superclouds.

We introduce the concept of the ratio of {\it lengths} as an indicator of filament regularity and the underlying instability processes.
The ratio of lengths, predicted to be around 3.9 in \citet{elmegreen2018}, represents the ratio of separation to the effective filament diameter for a non-magnetic cylinder, derived using equation~4.2 of \citet{nagasawa1987}.
While our observations do not directly measure the effective diameter or exhibit all the ideal conditions assumed in the theoretical derivation, the essence lies in capturing the regularity of the clouds and the presence of an instability mechanism rather than random cloud agglomeration.
Building upon this understanding, we have considered an alternative approach to assess the regularity of the clouds by focusing on the ratio of separation to the minor axis.
We believe the minor axis, being a representative measure of the filament diameter, offers a closer approximation to the relevant diameter compared to the major axis. The right-hand panel of Figure~\ref{fig:hicloud_spacing_histo} shows a histogram of the ratio of separation to cloud minor axis. The peak is at a ratio of around 3. 

In addition, the relative separation difference (RSD), defined as the difference between two adjacent separations divided by their average, provides further insight into the regularity of cloud spacing along the filament.
As described in \citet{elmegreen2018}, the relative difference in the separations between three adjacent clumps, i.e., $i-1$, $i$, and $i+1$, is
\begin{equation}
RSD = 2\,(S_{i,i-1} - S_{i,i+1}) / (S_{i,i-1} + S_{i,i+1}).
\label{eq:relative_sep}
\end{equation}
Here, $S$ is the separation between two adjacent clouds and not the average separation between a cloud and its neighbors, as in equation \ref{eq:sep}. 
This RSD quantifies the deviation from equally spaced separations along the filament.
This value ranges from 0 to a maximum value of 2, where a value of 0 indicates equally spaced separations and a larger value signifies a deviation from equal spacing.
A peak in the histogram of these RSDs at a small value would indicate a higher level of regularity in cloud spacing, providing further evidence for the presence of instability that shapes the filamentary structures.

\begin{figure}
\centering
\epsscale{1}
\includegraphics[width=0.45\textwidth]{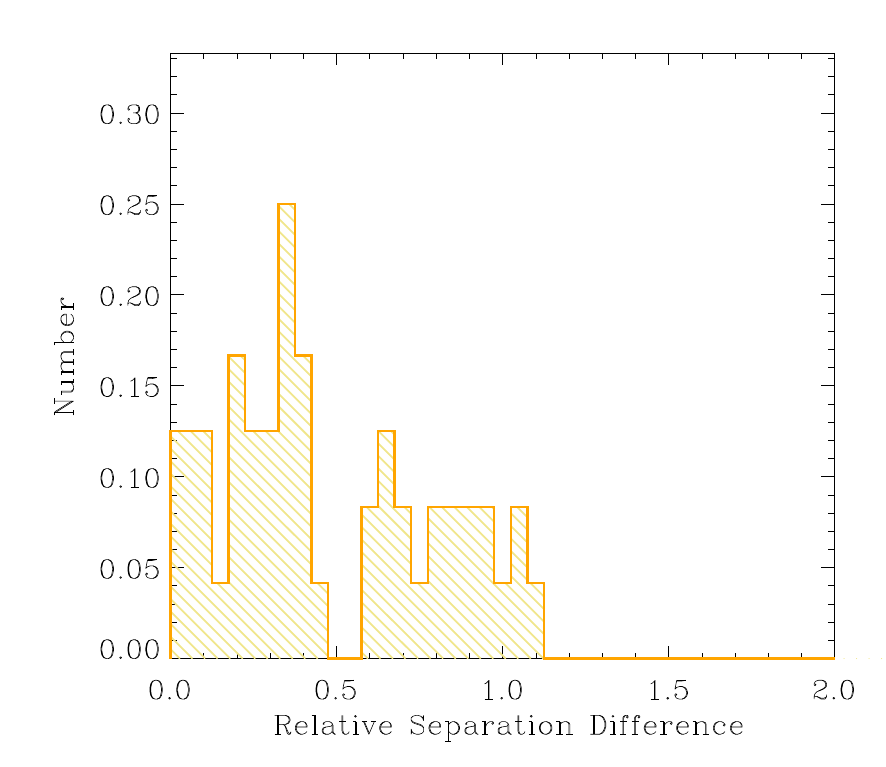}
\caption{
Histogram of the relative separation difference between clouds. 
}
\label{fig:hicloud_relspacing_histo}
\end{figure}

The RSD histogram is shown in Figure \ref{fig:hicloud_relspacing_histo}. 
The RSD peaks at $\sim 0.35$.
Although the observed clouds may not exhibit the same level of regularity as seen in other cases, such as the well-studied M100 \citep{elmegreen2018}, this does not diminish the significance of our findings.
We can confidently state that the ratio of separation to the minor axis closely aligns with theoretical expectations, and the modest regularity observed in the RSD histogram suggests the involvement of a gravity-driven process in shaping these structures.

\begin{figure*}
\centering
\epsscale{.9}
\includegraphics[width=0.8\textwidth]{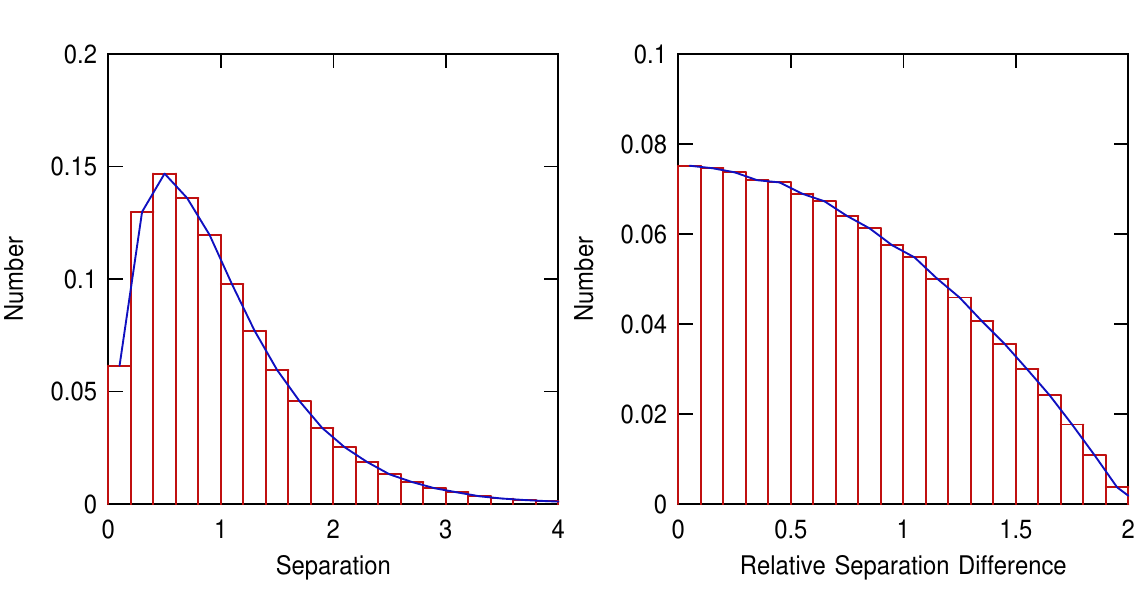}
\caption{Theoretical model for the separation distribution (left) and the distribution of relative separation differences (right) for points uniformly distributed on a line. 
The units of separation in the theoretical distribution are arbitrary.
The theoretical relative separation difference is non-zero above 1.1 where the observed distribution has no examples, suggesting that the observed distribution is not random. A selection effect related to cloud blending could prevent these high values. 
}
\label{fig:hicloud_relspacing_theory}
\end{figure*}

Figure \ref{fig:hicloud_relspacing_theory} shows a theoretical distribution of separation and RSD for $N=10^6$ points randomly placed on a line. The placement is made by assigning the point position to be a random number uniformly distributed between 0 and 1. These positions are then ordered by increasing value and their separations determined. To account for the trend of decreasing average separation with increasing $N$, the separations are multiplied by $N$; then the average is 1, and the separations range from 0 to some large number up to $N$. The separation distribution on the left of Figure \ref{fig:hicloud_relspacing_theory} peaks at 0.5 with a dip toward zero separation because it is increasingly unlikely for two points to be much closer than the average separation.  The RSD distribution on the right has a slowly decreasing trend toward a value of 2. This maximum occurs when all the points are at one end of the line and one point is at the other end, making $S_{i,i-1}=0$ and $S_{i,i+1}=1$. 

Figure \ref{fig:hicloud_relspacing_histo} should be compared to Figure \ref{fig:hicloud_relspacing_theory} (right) because both are on relative scales. The theoretical result has 26\% of the separation differences larger than 1.1 whereas the observations have none. The average RSD from theory is 0.75. 

We consider the Kolmogorov–Smirnov test to determine how likely it is that the observed RSD comes from the same model as the random distribution. This test uses the normalized cumulative distributions. For the RSD, this cumulative distribution (not shown) reaches unity at the maximum difference of $\sim1.1$, from Figure \ref{fig:hicloud_relspacing_histo}.  At this same value, the cumulative distribution for the theoretical case equals only 0.55. The difference is 0.45, which we take to be the Kolmogorov–Smirnov statistic when considering the hypothesis that the observed and theoretical distributions are from the same model. This statistic should be compared with $c(\alpha)\left[(N_{\rm clouds}+N)/(N_{\rm clouds}N)\right]^{0.5}$ for number of clouds in Figure \ref{fig:hicloud_relspacing_histo} equal to $N_{\rm clouds}=26$ and $N=10^6$ in the model. Here $c(\alpha)=\left[-0.5\ln(\alpha/2)\right]^{0.5}$ for level $1-\alpha$ at which the hypothesis is rejected. Setting $0.45=c(\alpha)/\left(N_{\rm clouds}\right)^{0.5}$ for $N>>N_{\rm clouds}$, we derive $\alpha=5.3\times10^{-5}$, meaning that the hypothesis the observed distribution is from the same model as the random distribution is rejected at a confidence level of 99.995\%. 

The lack of RSDs larger than 1.1 in the observations could be the result of a selection effect, where clouds that are too close to each other are called a single cloud. If we consider two nearby clouds at a separation $a$ with another cloud a distance $b$ from one of them, then the RSD is $2(b-a)/(b+a)$. Setting this less than 1.1 as in the observed limit gives $a>0.29b$. If the minimum separation between two adjacent clouds is never less than $0.29$ times their near-neighbor distance, we can explain the lack of high values. This type of selection effect is possible because the mean separation is $\sim700$ pc from Figure \ref{fig:hicloud_spacing_histo} (left) and 0.29 times this is $\sim200$ pc, which is comparable to the average major axis length of the \schi\ clouds, $\sim300$ pc. Thus, the upper limit in Figure \ref{fig:hicloud_relspacing_histo} could be because ``touching'' clouds have become confused with single clouds, eliminating the short spacings.

\subsection{Virial Equilibrium} \label{sec:virial_equil}

Determining whether \schi\ superclouds or MCs are gravitationally bound is a fundamental question in understanding star formation.
To examine the degree of self-gravitational bounding, one can compare the one-dimensional virial theorem velocity dispersion ($v_\textrm{VT}$) to the measured velocity dispersion ($\vsigma$) of the \schi\ cloud.
Assuming a spherical cloud with an isothermal mass distribution, the virial dispersion is given by 
\begin{equation}
  v_\textrm{VT}^2 = \frac{GM}{4R},
\label{eq:vir}
\end{equation}
where $M$ and $R$ are the total mass and the radius of the complex, respectively (EE87).
The factor $\beta$ in Equation~18 of EE87 indicates the ratio of magnetic to turbulent pressures, and they adopted $\beta =1$ based on the understanding that magnetism plays an important role in contributing to cloud support \citep[see Appendix~A in][]{elmegreen1985}. 
We adopt this value, resulting in the constant in the denominator of Equation~\ref{eq:vir} being 4. 
$M$ is the sum of $\Mhi$ in Table~\ref{tab:hiclouds} and $\Mht$ in Table~\ref{tab:coclouds},
and $R$ is the geometrical mean radius of the \schi\ cloud in Table~\ref{tab:hiclouds}.
If a cloud is in virial equilibrium, the velocity ratio $v_\textrm{VT}/\vsigma = 1$ or virial parameter (defined as $\alphav$ $= 4R\vsigma^2/GM$) is equal to 1.

\begin{figure}[!h]
\centering
\epsscale{1.1}
\includegraphics[width=0.45\textwidth]{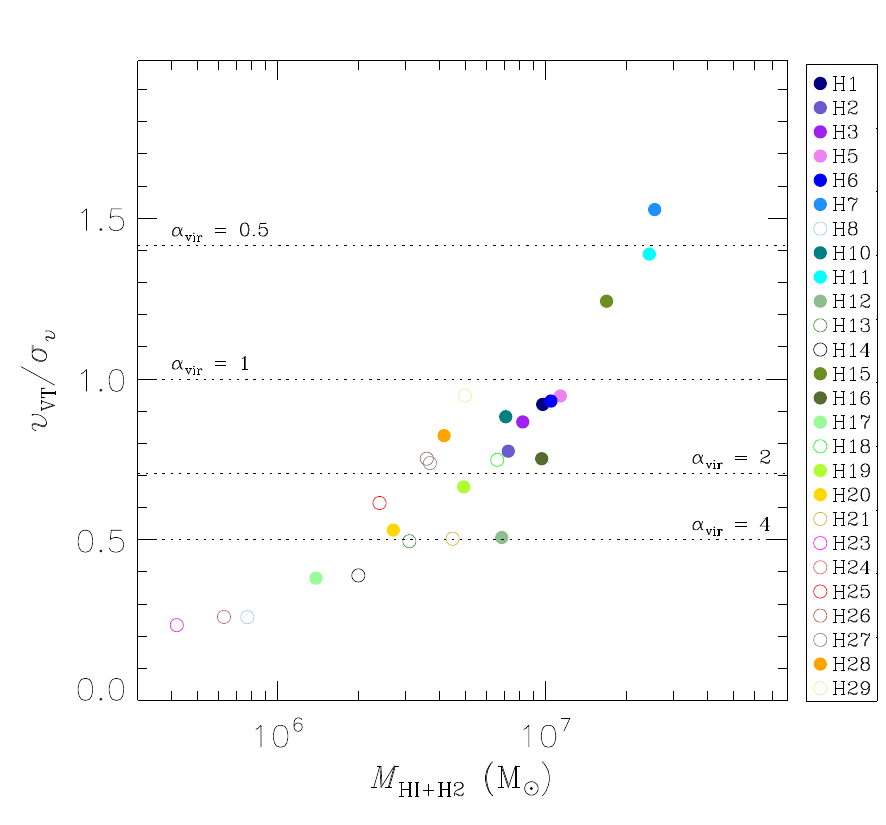}
\caption{
Relationship between the total cloud mass ($\Mgas$) and the velocity ratio ($\vVT/\vsigma$).
Filled circles represent CO-bright \schi\ clouds, while unfilled circles indicate other \schi\ clouds.
The lines mark the loci for virial parameters ($\alphav$) of 0.5, 1, 2, and 4.
\label{fig:hicloud_vir}}
\end{figure}

Figure~\ref{fig:hicloud_vir} displays the velocity ratio as a function of the total cloud mass for the 26 \schi\ clouds identified in this study.
The total cloud mass ($\Mgas$) is defined as
\begin{equation}
 \Mgas = \Mhi+\Mhtall,
\label{eq:mgas}
\end{equation}
where $\Mhtall$ is the sum of masses of MCs associated with an \schi\ cloud.
The velocity ratios fall in the range of approximately 0.3--1.5, which is close to unity, indicating that some of the clouds may be gravitationally bound.
For reference, if we assume a uniform density profile for the cloud in Equation~\ref{eq:vir}, the constant 4 changes to 5, which results in a slightly downward of the data points shown in this figure.
Comparison of CO-bright and CO-dark \schi\ clouds reveals that CO-dark clouds generally have lower total gas mass, but the velocity ratio shows no significant difference between the two types of clouds in the mass range where both types coexist. 
Moreover, the velocity ratio appears to increase with increasing total gas mass, with lower-mass clouds having lower velocity ratios and higher-mass clouds having higher velocity ratios. 
Most of the CO-bright \schi\ clouds have velocity ratios $\gtrsim 0.7$ ($\alphav \lesssim 2$), suggesting  that they are gravitationally bound or marginally bound.
Conversely, we infer that clouds with velocity ratios $\lesssim 0.5$ ($\alphav \gtrsim 4$) are likely to be gravitationally unbound and will either expand or dissipate in the absence of confinement from external pressure.

In addition to examining the virial equilibrium of the \schi\ clouds, we further investigate the relationship between the critical velocity dispersion ($\vsigmac$) and the observed velocity dispersion ($\vsigma$).
The critical velocity dispersion is denoted as the minimum velocity dispersion required for achieving equilibrium within filamentary structure \citep[e.g.,][]{ostriker1964, inutsuka1992, chandrasekhar1953, stodolkiewicz1963, nagasawa1987, fiege2000}.
It provides a measure of the velocity dispersion necessary for self-gravitational stability of the spiral arm \schi\ to determine whether it might have collapsed into the observed clouds from a more uniform filamentary structure in the long spiral shock.

In the context of our study, we consider the critical velocity dispersion as a key parameter for assessing the gravitational stability of the \schi\ in the spiral arm before it formed the observed superclouds.
The critical velocity dispersion is determined based on the equilibrium mass ($\mu$) per unit length.
\begin{equation}
  \vsigmac^2 = G\mu/2.
\label{eq:critical_sigmav}
\end{equation}
We estimate $\mu$ as the total gas mass ($\Mgas$) divided by the spacing between the clouds which is derived using Equation~\ref{eq:sep}.
This approach allows us to identify the minimum velocity dispersion required to maintain a state of equilibrium within the Carina arm.

\begin{figure*}
\centering
\epsscale{.5}
\includegraphics[width=0.45\textwidth]{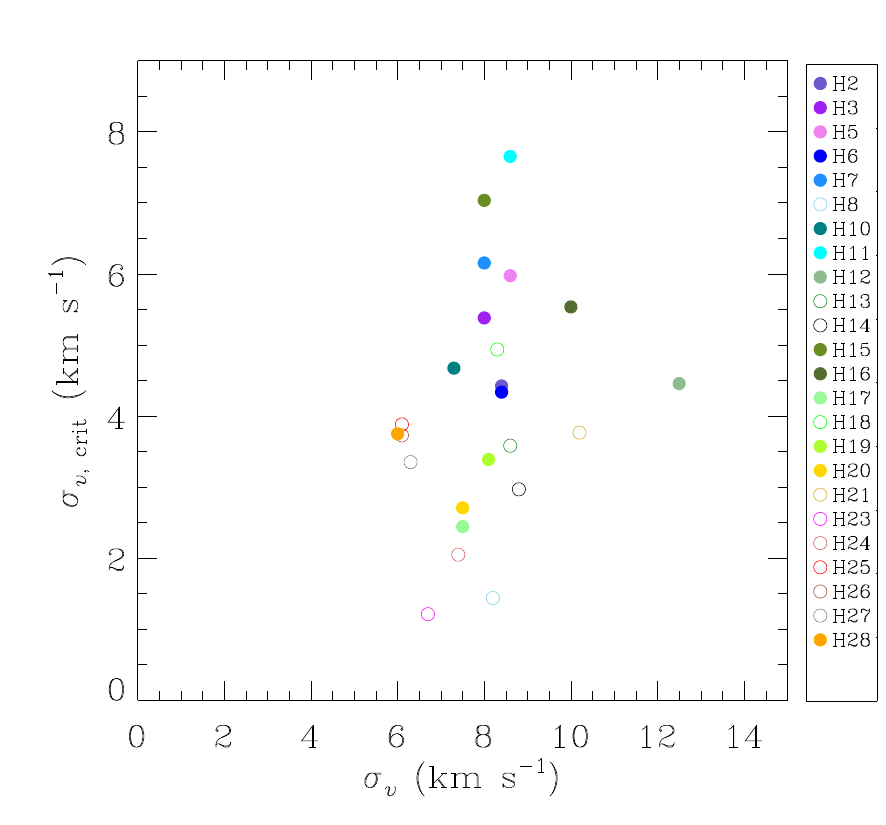}
\epsscale{1}
\includegraphics[width=0.8\textwidth]{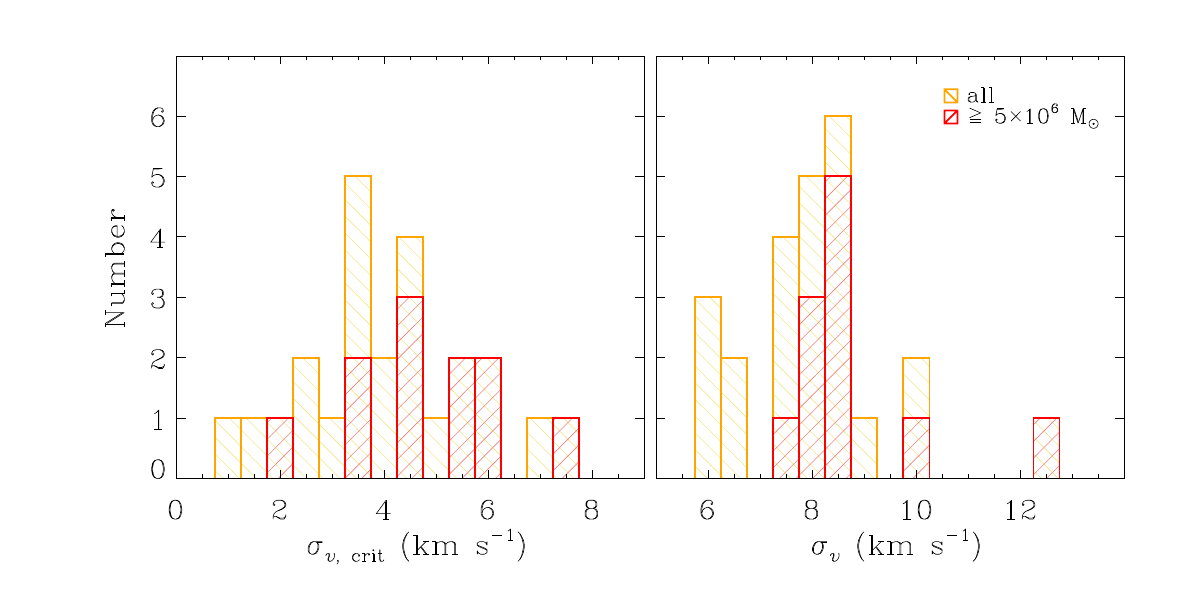}
\caption{
{
Scatter plot of critical velocity dispersion ($\sigma_{v,\mathrm{crit}}$) and observed velocity dispersion ($\vsigma$), with histograms of each.
The left histogram displays the distribution of $\sigma_{v,\mathrm{crit}},$ while the right histogram shows the distribution of $\vsigma$.
In the histograms, the yellow sticks with left diagonals represent all \schi\ clouds, whereas the red sticks with right diagonals represent massive clouds with $\Mhi \geq 5 \times 10^{6}~\msol$.
The bin size for both histograms is set to 0.5~$\kms$.
}
\label{fig:hicloud_critical_sigmav}}
\end{figure*}

Figure~\ref{fig:hicloud_critical_sigmav} presents a scatter plot of $\sigma_{v,\mathrm{crit}}$ versus the $\vsigma$ for the \schi\ clouds analyzed in this work.
The corresponding histograms illustrate the distributions of the critical and observed velocity dispersions.
Notably, the observed velocity dispersions, calculated from brightness temperature-weighted measurements, exhibit a relatively constant behavior.
The critical dispersion averages around 4 km s$^{-1}$. If the observed clouds formed by gravitational instabilities in a previously more uniform spiral arm gas, then the velocity dispersion in this gas had to be less than  or equal to $\sim4$ km s$^{-1}$ for this process to be rapid. Such a decrease is expected theoretically \citep{cowie1981,dobbs2008b} and there are indications of a low dispersion for the cool component \schi\ in observations of spiral galaxies by \cite{ianjamasimanana2012}. 
The higher observed velocity dispersions in present-day \schi\ clouds suggest the presence of additional energy sources, such as gravitational energy from their formation or young stellar feedback after stars appeared.

\section{\schi\ Superclouds and Molecular clouds} \label{sec:HInCO}

\subsection{Spatial Relation between \schi\ Superclouds and MCs} \label{sec:HInCO_spatial}

The formation of molecular clouds from \schi\ gas is a complex process that can be driven by a combination of physical processes, such as turbulent compression and fragmentation (which create density fluctuations), radiative compression (which triggers thermal instability), gravitational instability (which enhances self-gravity), and magnetic instability (which can contribute to cloud collapse and fragmentation) \citep[e.g.,][]{kwan1987, elmegreen1996, hennebelle2000, ostriker2004, dobbs2014}.
While the classical understanding of the spatial distribution of \schi\ and molecular gas is a layered structure with a giant molecular cloud surrounded by atomic gas \citep{blitz1993}, the resulting spatial distribution of molecular clouds relative to their parent \schi\ clouds is expected to depend on a variety of factors, including the specific physical conditions of the gas and the relative importance of these formation mechanisms.
However, the ubiquity of \schi\ gas in the Galactic disk and the difficulty in determining its exact kinematic distance along the  line of sight makes it challenging to establish a definitive association between \schi\ and molecular clouds and to provide observational insight.
Nevertheless, it is worthwhile to investigate the spatial distribution of \schi\ superclouds and MCs, particularly in the outer Carina arm, which is relatively well-identified.

A simple question arises as to whether the detected MCs are centered on the associated \schi\ superclouds or matched with local \schi\ emission peaks.
Extragalactic observational studies of the spatial distribution and kinematics of molecular clouds relative to their associated \schi\ gas have been conducted using a variety of tracers, such as CO and \schi\ emission \citep[e.g.,][]{wong2009, tosaki2011}.
It is found that CO emission was typically associated with high-intensity \schi\ gas, but not all regions of high \schi\ intensity were found to have CO emission.

\begin{figure*}
\centering
\epsscale{1.}
\includegraphics[width=\textwidth]{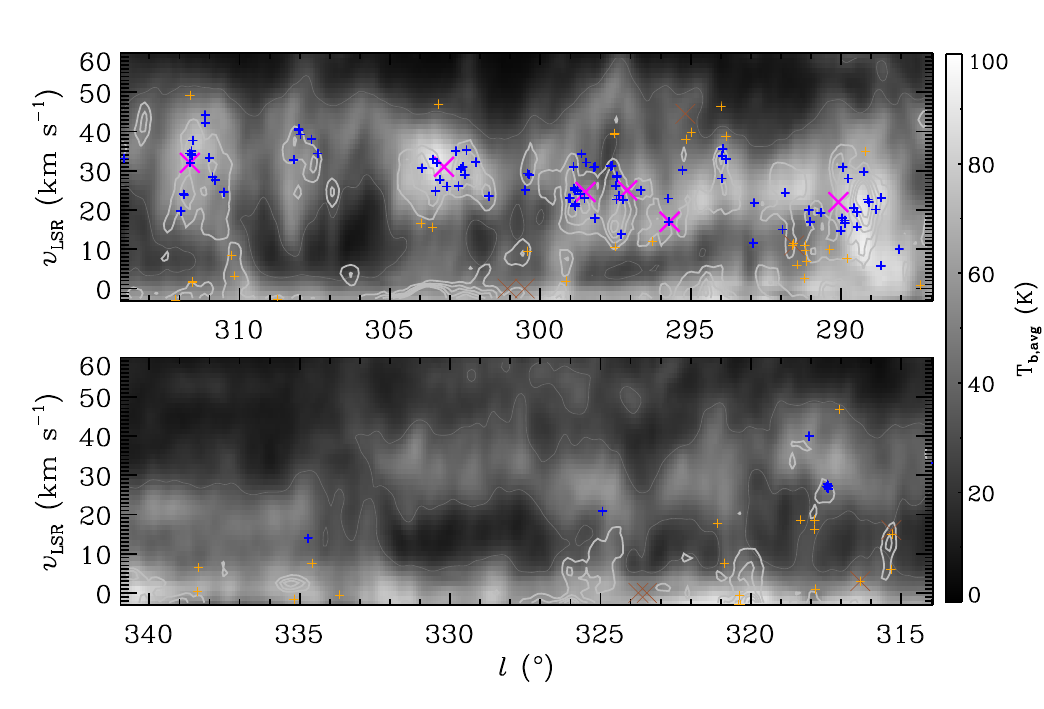}
\caption{
$b$-averaged ($l$, $\vlsr$) \schi\ map, overlaid with CO contours and star-forming regions.
The integrated latitude range for the map is from -2\arcdeg\ to $+1$\arcdeg.
The background image with a thin, dark gray contour represents \schi, and the contour level is set at 30~K.
CO contours, displayed as thick, bright gray lines, indicate levels of 0.1, 0.2, 0.3, 0.4, and 0.5~K.
\wmap\ sources,  likely related to the outer Carina arm, are marked with thick, blue pluses, whereas thin, orange pluses represent those that are unlikely to be related.
Similarly,  thick, magenta crosses represent \wise\ \schii\ regions with a single measured velocity, likely associated with the outer Carina arm, while thin, brown crosses indicate those that are unlikely to be related.
\label{fig:comp_hinco_lv}}
\end{figure*}

Figure~\ref{fig:comp_hinco_lv} displays a position-velocity diagram of \schi\ and CO emission features, which is averaged over latitudes.
The \schi\ cloud emission features defined in this work are generally visible within the gray contour of 30~K (see Section~\ref{sec:clf_step1} and Figure~\ref{fig:gas_lv} for details).
The CO emission features, displayed with bright gray contours, are commonly placed in the velocity range where the \schi\ cloud features are visible, although some small and weak CO emission features at $l \gtrsim 320\arcdeg$ do not appear in this averaged diagram.
Extended, an arc-like feature in \schi\ emission is observed at $l \sim$ 296\arcdeg\ to 300\arcdeg\ and $\vlsr \sim$ 35 to 60~$\kms$, which appear as a ring-like structure when lower velocity \schi\ gas is included.
However, this feature is not within the scope of this study.
This may indicate activity such as a supershell or chimney, and is very likely to be related to GSH~298$-$01$+$35, proposed as a chimney structure, identified by \citet{mcclure2002}.
A detailed 3D kinematic analysis would provide further insight into the structures of the \schi\ and molecular gas.

\begin{figure*}
\centering
\epsscale{1.}
\includegraphics[width=\textwidth]{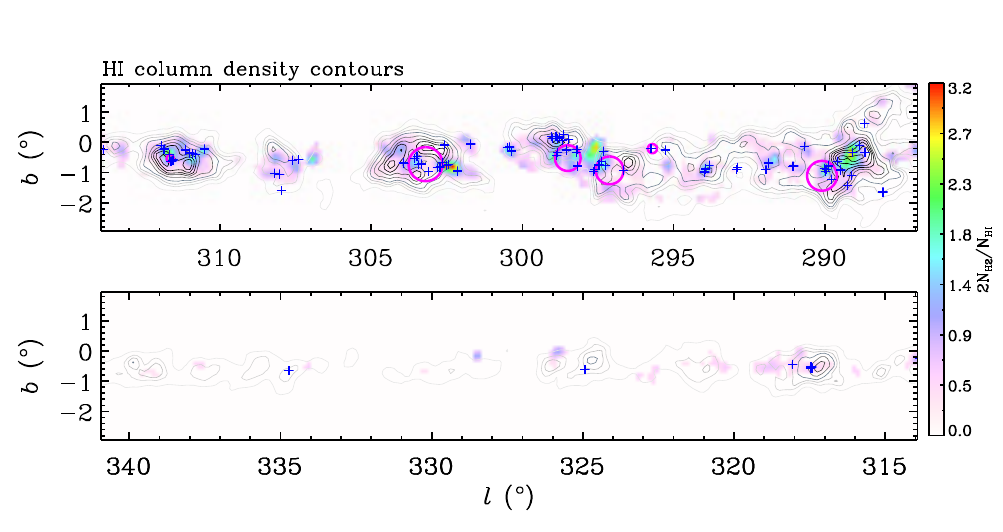}
\includegraphics[width=\textwidth]{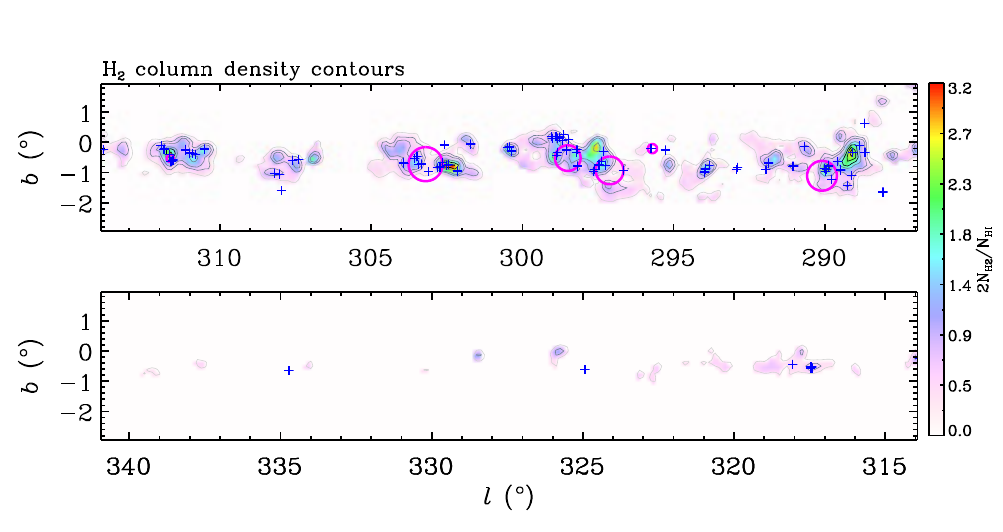}
\caption{
Ratio map of $\rmht$ column density ($\Nht$) to \schi\ column density ($\Nhi$), specifically $2\Nht/\Nhi$, overlaid with the \schi\ column density contours (top) or $\rmht$ column density contours (bottom).
The white areas in both the top and bottom panels indicate regions where $\Nht < 5 \times 10^{20}~\mathrm{cm}^{-2}$,
which corresponds to the minimum threshold for MC identification, and have been masked. 
We also masked the area of $l > 291\arcdeg$ and $b < -2\arcdeg$ or $b > 0.8\arcdeg$.
Some weak CO emission features are visible in the region, but their validity is uncertain as they were obtained by different surveys with relatively low sensitivity and a large sampling value. 
The contours are color-graded from gray to black to indicate increasing levels.
In the top panel, the $\Nhi$ levels are 2.2, 2.7, 3.0, 3.3, 3.7, 4.0, 4.2, 4.4, 4.6, $4.8 \times 10^{21}~\mathrm{cm}^{-2}$.
In the bottom panel, the $\Nht$ levels are 5, 10, 15, 30, 40, 50, $60 \times 10^{20}~\mathrm{cm}^{-2}$.
The star-forming regions, which are likely related to the outer Carina arm, are represented by magenta circles and blue pluses for \wmap\ sources and \wise\ \schii\ regions, respectively, as shown in Figure~\ref{fig:comp_hinco_lv}.
The size of each circle for the \wmap\ sources corresponds to their respective effective radii.
\label{fig:comp_hinco_cN}}
\end{figure*}

Figure~\ref{fig:comp_hinco_cN} presents a comparison of the \schi\ and CO emission features in a position-position diagram. 
The figure shows the ratio of column densities between $\rmht$ column density($\Nht$) and \schi\ column density ($\Nhi$) obtained by using Figure~\ref{fig:gas_lb}.
The column density ratio, $2\Nht/\Nhi$, is beneficial for studies of molecular clouds and star formation, as it reflects the degree of molecular gas enrichment in a given region, which refers to the proportion of molecular gas present relative to atomic gas.
This parameter is important for understanding the conditions necessary for star formation.
The color image displays the column density ratio with contours indicating $\Nhi$ from 2.2 to 4.8 $\times 10^{21}~\textrm{cm}^{-2}$ for the top panel, and $\Nht$ from 5 to 60 $\times 10^{20}~\textrm{cm}^{-2}$ for the bottom panel.
The column density ratios were calculated only where $\Nht$ is greater than or equal to $5 \times 10^{20}~\textrm{cm}^{-2}$, which is equal to the minimum limit for the identification of MCs in the work.
The $v$-integrated positions of all CO emissions of the identified MCs are equal to where column density ratios are expressed.
The column density ratios displayed in this figure range from approximately 0.2 to 3.2, with the highest value observed at M31 in association with H7.

The minimum value for \schi\ column density used to identify \schi\ clouds in this study may provide insight into areas where $\rmht$ formation occurs more efficiently through self-shielding.
Numerous previous studies have demonstrated that the saturation of \schi\ column density at $\approx 10^{21} \mathrm{cm}^{-2}$ in the correlation between \schi\ column density and tracers of total gas column density, such as infrared surface brightness and optical reddening, indicates the presence of $\rmht$ \citep[e.g.,][]{reach1994, meyerdierks1996, barriault2010, liszt2014, park2018}.
However, it is important to note that the conversion from atomic to molecular gas can also be influenced by other factors, such as metallicity \citep[and references therein]{bolatto2013}.

Most of the MCs appear to be spatially associated within the corresponding \schi\ superclouds, but a majority of them are not situated in the central regions of these superclouds.
Notably, many peaks in CO emission appear somewhat distant from the nearby \schi\ emission peaks.
It should be noted that some MCs (e.g., H47 at $l\simeq326\arcdeg$ and H48 at $l\simeq328\arcdeg$) are found to be situated at or beyond the periphery of the relevant \schi\ supercloud.
Although the cloud boundary, particularly for \schi\ superclouds, was arbitrarily delineated based on bright, dense regions,
the physical associations of the aforementioned cases may be less clear than those of other cases.
Nonetheless, we have included those MCs in the catalog.

As seen in the bottom panel in Figure~\ref{fig:comp_hinco_cN}, $\Nht$ is positively correlated with the column density ratio. 
That is, the CO emissions peak in regions with high column density ratios, indicating that the conversion of \schi\ to $\rmht$ is more efficient in those regions.
This may be due to certain physical conditions that favor the formation of molecular gas, such as high gas density or low gas temperature.
However, as seen in the top panel in Figure~\ref{fig:comp_hinco_cN}, the correlation between column density ratios and local \schi\ peaks seems lacking.
This may be due to various factors.
One possible explanation is that the conversion of \schi\ to $\rmht$ is not solely determined by the local \schi\ column density but also by other physical conditions such as gas density, temperature, radiation field, and turbulence or magnetic fields.
In regions where the physical conditions are favorable for forming molecular gas, the conversion may be more efficient even if the local $\Nhi$ is not high.
On the other hand, in regions where the physical conditions are unfavorable, the conversion may be less efficient even if the local $\Nhi$ is high.
It is possible that triggered molecular cloud formation could play a role in the lack of correlation between the column density ratios and local \schi\ peaks.
When molecular clouds form due to external triggering mechanisms, such as shock compression or radiation from massive stars, the conversion from \schi\ to $\rmht$ can happen more efficiently in certain regions that are influenced by these triggering mechanisms \citep[e.g.,][]{ballesteros1999, hartmann2001, inoue2009, inutsuka2015}.
This can lead to molecular gas being present in regions with lower \schi\ column densities or in regions where the \schi\ gas has been dispersed or compressed by the triggering event.

\begin{figure*}
\centering
\epsscale{.8}
\includegraphics[width=0.35\textwidth]{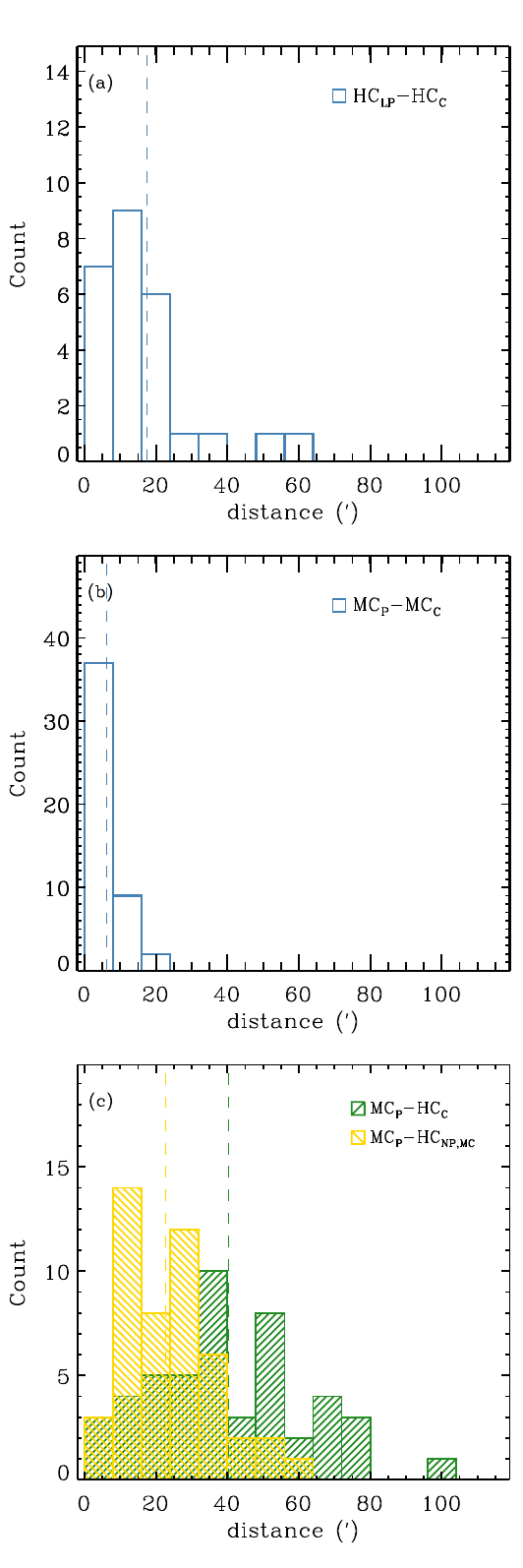}
\includegraphics[width=0.35\textwidth]{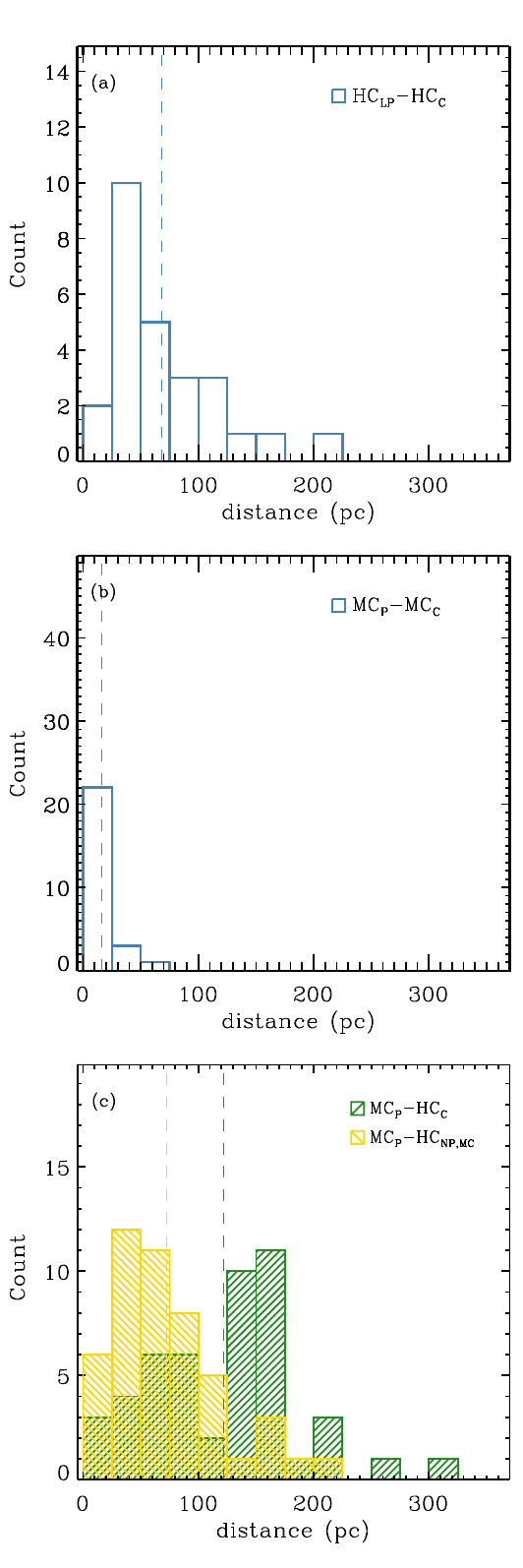}
\caption{
Histogram of the mutual distances between different positions, including:
(a) the local emission peaks of \schi\ clouds (${\rm HC_{LP}}$) and the geometric centers of the \schi\ clouds (${\rm HC_{C}}$) (${\rm HC_{LP}}$--${\rm HC_{C}}$),
(b) the emission peaks of MCs (${\rm MC_{P}}$) and the geometric centers of the MCs (${\rm MC_{C}}$) (${\rm MC_{P}}$--${\rm MC_{C}}$), and
(c) the emission peaks of MCs (${\rm MC_{P}}$) and either the geometric centers of MC-related \schi\ clouds (${\rm HC_{C}}$) (${\rm MC_{P}}$--(${\rm HC_{C}}$; green with right diagonals) or the local emission peaks of \schi\ clouds nearest the MCs (${\rm HC_{NP, MC}}$) (${\rm MC_{P}}$--${\rm HC_{NP, MC}}$; yellow with left diagonals).
The left group of four panels shows the distances projected in the sky in angular scale, and the right group shows the distances calculated assuming the heliocentric distance of relevant \schi\ cloud, provided in Table~\ref{tab:hiclouds}, in linear scale.
These result in the minimum values, which are not the actual separation in 3D.
The emission peaks were derived from $\vlsr$-integrated maps (e.g., Figure~\ref{fig:gas_lb}).
For the left panels, we chose a bin size is 8\arcmin, which is approximately half of the angular resolution of the data used.
The right panels, on the other hand, were binned with an arbitrary bin size of 25~pc.
A vertical dotted line marks the mean value of each case (the parentheses refer to the right panel):
(a) 18\arcmin\ (68 pc),
(b)  6\arcmin\ (16 pc),
(c) 23\arcmin\ (72 pc); 40\arcmin\ (122 pc).
The two mean values for the left and right panels do not correspond directly to each other since the distances to the \schi\ clouds are different.
\label{fig:comp_hinco_peaks}}
\end{figure*}

Figure~\ref{fig:comp_hinco_peaks} presents a quantitative comparison between the positions of \schi\ clouds and MCs.
The left group of three panels shows the projected distances on the sky in angular scale, with an adopted angular bin size of 8\arcmin, which is approximately half of the angular resolution of the data used.
The right group of three panels displays the distances computed using the heliocentric distance of the corresponding \schi\ cloud provided in Table~\ref{tab:hiclouds}, presented in linear scale with an arbitrarily selected linear bin size of 25~pc.
According to the cloud definition method in this paper, the emission peaks of MCs (${\rm MC_{P}}$) are typically located at or near the centers of the MCs (${\rm MC_{C}}$), given the angular resolution (panel (b)). 
In contrast, the local emission peaks of \schi\ clouds (${\rm HC_{LP}}$) are not always situated at their centers (${\rm HC_{C}}$) (panel (a)), and some clouds even exhibit multiple peaks (e.g., H7 and H11).
When comparing ${\rm MC_{P}}$ with ${\rm HC_{C}}$, we observe that most ${\rm MC_{P}}$ are significantly distant from ${\rm HC_{C}}$ (green in panel (c)).
Additionally, we identified the nearest \schi\ peak for a given MC and labeled it as ${\rm HC_{NP, MC}}$, then measured the distance between them (yellow in panel (c)). 
The distribution of ${\rm MC_{P}}$--${\rm HC_{NP, MC}}$ exhibits a relatively stronger correlation than that of the ${\rm MC_{P}}$--${\rm HC_{C}}$ and displays two prominent peaks with a valley around the mean mutual distances of $\sim 23\arcmin$. 
Specifically, 17 pairs on the left side of the distribution exhibit a strong positional correspondence or marginal overlap, while the remaining 31 pairs show a clear positional discrepancy.

\subsection{Molecular Mass Fraction of \schi\ Superclouds} \label{sec:mmf}

\begin{figure*}
\centering
\epsscale{.8}
\includegraphics[width=0.8\textwidth]{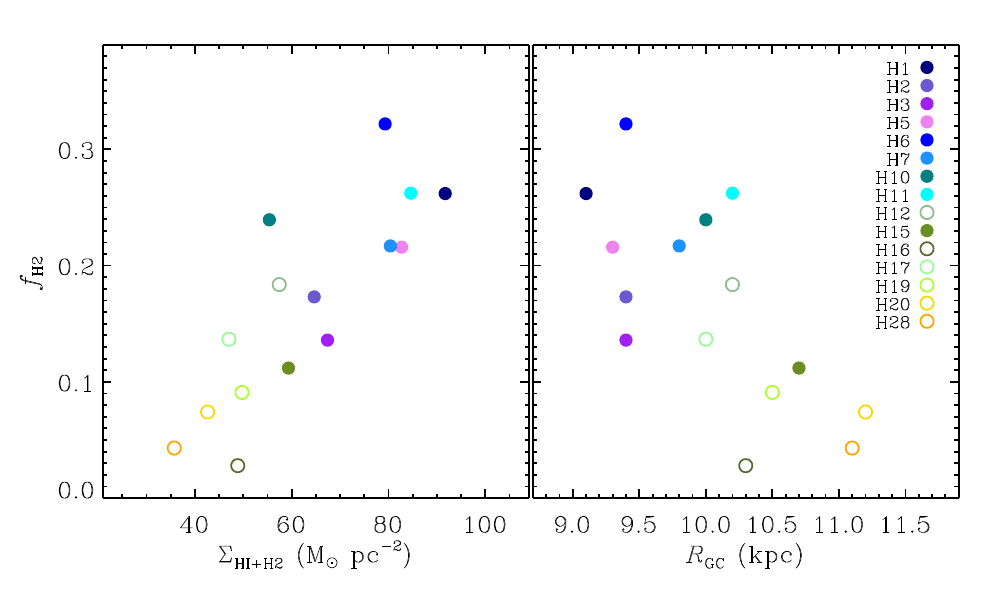}
\caption{
Relation of the molecular mass fractions ($\fht$) and the total gas surface densities ($\SDgas$) of \schi\ superclouds (left) or their Galactocentric distances (right).
Filled symbols indicate \schi\ superclouds associated with \schii\ region(s), while unfilled symbols represent \schi\ superclouds without \schii\ region(s).
\label{fig:hinco_relation1}}
\end{figure*}

In Figure~\ref{fig:hinco_relation1}, we present two distributions showing the molecular mass fraction as a function of Galactocentric distance (left panel) or total gas surface density (right panel) of \schi\ superclouds.
The molecular mass fraction,$\fht$, is defined as
\begin{equation}
 \fht = \frac{\Mhtall}{\Mhi + \Mhtall}.
\label{eq:fht}
\end{equation}
The total gas surface density, $\SDgas$, is defined as
\begin{equation}
 \SDgas = \frac{\Mhi + \Mhtall}{A_\mathrm{HI}},
\label{eq:sdgas}
\end{equation}
where $A_\mathrm{HI}$ is the area of a \schi\ supercloud (see Section~\ref{sec:size}).

As anticipated, a larger mass of an \schi\ cloud is generally associated with a larger mass of an MC.
The relation between $\fht$ and $\SDgas$ exhibits a clear positive correlation.
In regions where the $\SDgas$ is higher, there is generally a higher pressure environment, which facilitates the conversion of atomic gas into molecular gas.
The positive correlation between $\fht$ and $\SDgas$ can be understood within the context of interconnection processes that drive the conversion of atomic hydrogen into molecular hydrogen and ultimately lead to star formation.

Furthermore, previous studies, such as \citet{elmegreen1987}, have observed a trend where the value of $\fht$ decreases as the Galactocentric distance increases.
In our study, we observe this trend in the outer Carina arm clouds spanning $l = 288\arcdeg$ to $340\arcdeg$, which extends over the Galactocentric distance range of $\Rgc \sim$ 9--11~kpc.
We find that the decrease in $\fht$ smoothly declines with increasing $\Rgc$, with few deviations.
Additionally, the majority of \schi\ superclouds with $\fht > 0.1$ or $\SDgas \gtrsim$ $60~\msol \mathrm{pc^{-2}}$ exhibit evidence of star-forming activity, as indicated by the presence of \schii\ region(s).
The $\fht$ values of \schi\ superclouds between $\Rgc \sim$ 9--10~kpc exhibit a relatively wide distribution ranging from $\sim0.1$ to 0.3, suggesting that variations in $\fht$ may be influenced by local environmental factors or other processes affecting the conversion of atomic to molecular gas.
For example, higher gas densities in some \schi\ superclouds might lead to increased molecular mass fractions and consequently promote star formation.
In contrast, the range of 0.1--0.3 at $\Rgc \sim$ 9--10~kpc is somewhat larger than the $\fht$ values ($< 0.1$) of solar-neighboring \schi\ superclouds reported by \citet{elmegreen1987},
It is worth noting, however, that the mean $\fht$ value over the entire gas disk, integrated over $b$ from $-30\arcdeg$ to $+30\arcdeg$, at the solar radius is 0.1--0.2, while it increases to $\sim$~0.5 at the Galactic midplane when integrated over $b = -1\fdg5$ to $+1\fdg5$ \citep{koda2016}.
This observation underscores the importance of considering the relationship between molecular mass fraction, gas surface density, and the presence of \schii\ regions in understanding star formation processes within these superclouds.

\section{Star Formation} \label{sec:sf}

\subsection{Spatial Relation of \schii\ regions with \schi\ Superclouds and MCs} \label{sec:SFnHInCO}

The presence of \schii\ regions is a strong indicator of the existence of high-mass star-forming regions.
To investigate the relationship between \schi\ superclouds and \schii\ regions in the outer Carina arm, We overlaid the \schii\ regions, obtained from the \schi\ catalog, in the outer Galaxy onto $b$-averaged ($l$, $v$) or $v$-integrated ($l$, $b$) diagrams of Figures~\ref{fig:comp_hinco_lv} and \ref{fig:comp_hinco_cN}.
We used star-forming regions including those cataloged by \citet{e.lee2012} using Wilkinson Microwave Anisotropy Probe (\wmap) data and the \schii\ regions identified in the outer Galaxy from the all-sky Wide-field Infrared Survey Explorer (\wise) survey \citep{anderson2014}\footnote{We used Version 2.0}.
Specifically, we focused on \schii\ regions with a single observed velocity.
In Figure~\ref{fig:comp_hinco_lv}, the associations of two objects, namely \wise\ \schii\ regions and \wmap\ sources, with the outer Carina arm are indicated through different symbols and line thicknesses.
The associations are determined based on a simple comparison of observed LSR velocities.
\wise\ \schii\ regions are represented by pluses, marked in blue or orange, with varying line thickness.
The thicker blue pluses correspond to likely associations with the outer Carina arm, while the thinner orange pluses indicate likely no association.
Similarly, \wmap\ sources are depicted as crosses in magenta or brown, with different line thicknesses.
The magenta crosses with thicker lines indicate likely associations, while the thinner brown crosses do not exhibit a clear association.
Figure~\ref{fig:comp_hinco_cN} specifically highlights the blue pluses for \wise\ \schii\ regions and marks \wmap\ sources likely related to the outer Carina arm with magenta circles.
The circle size of each \wmap\ source represents its effective radius.
Most of the \schii\ regions appear to be associated with the outer Carina arm, suggesting a clear positional correlation between the three components: \schi\ superclouds, GMCs, \schii\ regions.

Upon closer examination, we found that eleven \schi\ superclouds (H1--3, H5--7, H10--11, H15, H18, and H25) are associated with \wise\ \schii\ regions, indicating ongoing high-mass star formation.
However, we were unable to detect any CO emission, which is commonly used as a tracer of molecular gas, associated with two of them (H18 and H25).
This could imply that either the molecular gas in these regions is not abundant enough to be traced by CO emission or that the CO emission is too weak to be detected with our current observations.

The spatial distribution of star-forming regions relative to their associated \schi\ supercloud is also not centralized.
However, these regions are typically located within the range of the \schi\ supercloud and distributed closer to the CO emission peaks. 
This is a natural phenomenon as stars form in dense regions of MCs.
\citet{murray2010} proposed that the \wmap\ sources are bubbles generated by massive star clusters, representing the extended low-density \schii\ regions previously described by \citet{mezger1978}. 
Furthermore, they found that classical giant \schii\ regions are distributed in the bubble walls, which can be explained by triggered star formation.
H7 provides a good example of such a scenario, where a \wmap\ source is located in the center of H7, which exhibits low $\Nhi$ and $\Nht$. 
Several \wise\ \schii\ regions are located near the effective radius boundary of the \wmap\ source and are closer to the CO emission peaks. 
Based on these facts, we propose that the earlier star formation at the cloud center produced the \wmap\ bubble, which then triggered the next star formation event in the \wise\ \schii\ regions.
Additionally, it is likely that the atomic and molecular gases in the cloud center were blown away during this process.

\begin{figure*}
\centering
\epsscale{1.}
\includegraphics[width=0.45\textwidth]{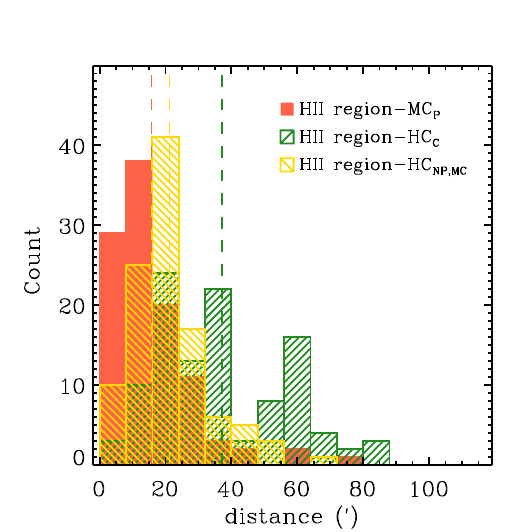}
\includegraphics[width=0.45\textwidth]{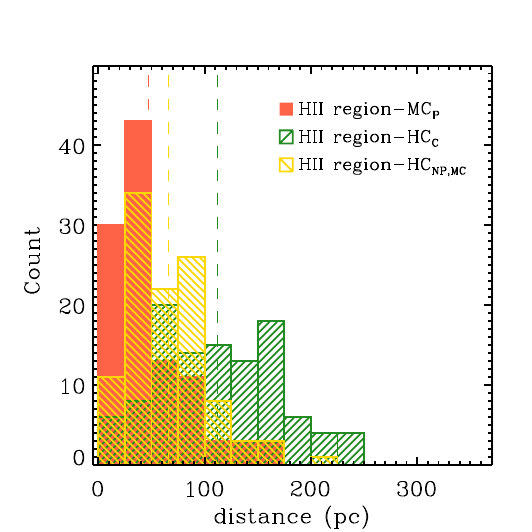}
\caption{
Histogram similar to Figure~\ref{fig:comp_hinco_peaks}, but for the mutual distances between different positions, the centers of \wise\ \schii\ regions and either ${\rm MC_{P}}$ (\schii\ region--${\rm MC_{P}}$; red filled) or ${\rm HC_{C}}$ (\schii\ region--${\rm HC_{C}}$; green with right diagonals) or ${\rm HC_{NP, MC}}$ (\schii\ region--${\rm HC_{NP, MC}}$; yellow with left diagonals).
Vertical dotted lines mark mean values of 16\arcmin\ (47 pc); 21\arcmin\ (66 pc); 37\arcmin\ (112 pc), with the values in parentheses corresponding to those for the right panel. 
\label{fig:comp_hinco_hii_peaks}}
\end{figure*}

The results of the comparison between the centers of \wise\ \schii\ regions and  ${\rm MC_{P}}$, ${\rm HC_{C}}$, or ${\rm HC_{NP, MC}}$ are shown in Figure~\ref{fig:comp_hinco_hii_peaks}.
The mean mutual distances increase from ${\rm MC_{P}}$ to ${\rm HC_{NP, MC}}$, and their distributions exhibit some differences (red, yellow, and green, respectively).
We found that the degree of correlation with the central positions of the \schii\ regions follows the order of ${\rm HC_{NP, MC}}$ $< {\rm HC_{C}}$ $< {\rm MC_{P}}$.
Notably, the distribution of \schii\ region--${\rm MC_{P}}$ is positively skewed by numerous pairs with lower values than their mean, suggesting a strong positional correspondence or marginal overlap for many pairs. 
These results indicate that \schii\ regions are more closely related to emission peaks of MCs than those of \schi\ clouds.
This is consistent with the fact that stars form within the densest cores within MCs, and the ionizing radiation from young massive stars creates \schii\ regions within or near the MCs.
On the other hand, as stars form and evolve, stellar feedback, such as radiation pressure, stellar winds, and supernova explosions,
can remove the surrounding gas and dust, potentially influencing the correlation between the central positions of \schii\ regions and emission peaks of MCs.
However, our angular resolution might not be high enough to resolve the detailed effects of stellar feedback, especially for targets at distances over 7~kpc.
Also, there is a limitation in the examination of correlation in 2D space integrated along velocity.
Despite these limitations, our findings are in line with the general understanding that \schii\ regions are more closely related to MCs than large, diffuse \schii\ clouds. 
The fact that 67 out of 106 pairs ($\sim 63\%$) in the \schii\ region--${\rm MC_{P}}$ distribution show good matching supports the notion that \schii\ regions and MCs are physically associated with each other. 

\subsection{Star Formation Rate} \label{sec:sfr}

The star formation rate (SFR) in the Galactic star-forming regions has been extensively studied in the literature, often using free-free emission flux measured from \wmap\ \citep{murray2010, e.lee2012}.
In this work, we use physical information of star-forming complexes(SFCs) cataloged by \citet{e.lee2012}.
We match seven SFCs to six \schi\ superclouds, and their catalog numbers are listed in Table~\ref{tab:sfr}.
We derive SFRs from the total ionizing photon luminosity ($Q$) using the following equations, which are similar to Equations (3), (9)--(10) of \citeauthor{e.lee2012}:
 \begin{equation}
  \frac{\rm SFR}{(\msol\,\mathrm{yr}^{-1})} = 4.1 \times 10^{-54} \, Q,
 \end{equation}
 \begin{equation}
  \frac{Q}{(s^{-1})} = 1.34\times 10^{26} \times \left(\frac{4\pi d^2 F_{\nu}}{\mathrm{erg\,s^{-1}}}\right) \times 1.37,
 \end{equation}
where 1.37 is a correction factor accounting for the absorption of ionizing photons by dust grains, and the term of $4\pi d^2 F_{\nu}$ describes the specific free-free luminosity at 94~GHz of a given \schii\ region, which is calculated with the 94~GHz free-free flux cataloged in Table~1 of \citeauthor{e.lee2012} and its recalculated distance. 
The distances were derived using the velocities given in \citeauthor{e.lee2012} and the flat Galactic rotation model adopted in this work.
Table~\ref{tab:sfr} lists $Q$, SFR, and surface densities of SFR, molecular gas, and total gas ($\SDSFR$, $\SDht$, and $\SDgas$) derived for a given \schi\ supercloud.
The surface densities are based on the area of a matched \schi\ supercloud, e.g., $\mathrm{SFR} \times \mathrm{A}_\mathrm{HI}^{-1}$ as similar to Equation~\ref{eq:sdgas}.

\begin{deluxetable*}{ccccc cl}
\tabletypesize{\scriptsize}
\tablewidth{0pt}
\tablecaption{Star Formation Properties of \schi\ Superclouds \label{tab:sfr}}
\tablehead{
\colhead{} & \colhead{$Q$} & \colhead{SFR} & 
\colhead{$\SDSFR$\tablenotemark{\scriptsize a}} & \colhead{$\SDht$\tablenotemark{\scriptsize a}} & 
\colhead{$\SDgas$\tablenotemark{\scriptsize a}} & \\
\colhead{\#$_{\rm HI}$} & \colhead{(s$^{-1}$)} & \colhead{($\msol \mathrm{yr}^{-1}$)} & 
\colhead{($\msol \mathrm{pc}^{-2} \mathrm{Myr}^{-1}$)} & \colhead{($\msol \mathrm{pc}^{-2}$)} &
\colhead{($\msol \mathrm{pc}^{-2}$)} & \colhead{Lee SFC\tablenotemark{\scriptsize b}}
}
\startdata
\phn H1& 1.3e+51& 5.4e-03 & 1.050 &    24.0 & 91.7  & 167 \\
\phn H3& 4.6e+50& 1.9e-03 & 0.016 & \phn9.2 & 67.4  & 171 \\
\phn H5& 2.4e+51& 9.9e-03 & 0.071 &    17.9 & 82.7  & 172, 173 \\
\phn H6& 4.3e+51& 1.8e-02 & 0.134 &    25.5 & 79.3  & 174 \\
\phn H7& 4.4e+51& 1.8e-02 & 0.057 &    17.4 & 80.4  & 181 \\
    H11& 1.2e+51& 4.9e-03 & 0.017 &    22.2 & 84.6  & 192 \\
\enddata
\tablenotetext{a}{All surface densities are based on the area of a matched \schi\ supercloud.}
\tablenotetext{b}{SFC(s) cataloged in \citet{e.lee2012} which is/are matched to a given \schi\ supercloud.}
\end{deluxetable*}

The empirical relationship well-known as the Kennicutt-Schimidt (K-S) law \citep{schmidt1959, kennicutt1989, kennicutt1998} describes a power-law relation between gas surface density and SFR surface density.
The original K-S relation uses the total gas surface density, which includes both atomic and molecular gas. 
It appears still prominent in the global galaxy scale, that is, disk-averaged \citep[][reference therein]{kennicutt2012}.
However, in some cases, variations of the K-S relation focus on molecular gas only, as molecular gas is more directly related to star formation.
For example, \citet{bigiel2008} studied the relationship between $\SDSFR$ and $\SDhi$ or $\SDht$ at sub-kpc scales within nearby spiral galaxies.
They found that $\SDSFR$ correlates well with $\SDht$, but shows little or no correlation with $\SDhi$.
Similar results known as the resolved K-S relation have been reported by other extragalactic studies \citep[e.g.,][]{leroy2008, schruba2011, williams2018}.

\begin{figure*}
\centering
\epsscale{1.1}
\includegraphics[width=0.9\textwidth]{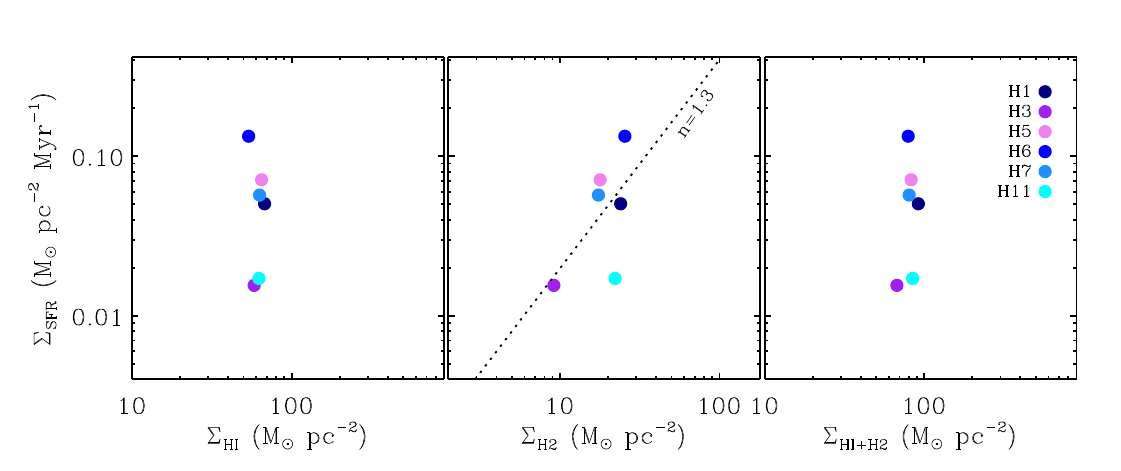}
\caption{
Relation between the SFR surface density ($\SDSFR$) and the surface densities of \schi\ ($\SDhi$; left), $\rmht$ ($\SDht$; middle), or total gas ($\SDgas$; right) in six \schi\ superclouds.
All surface densities in this plot were calculated based on the area of a matched \schi\ supercloud.
In the middle panel, the best single power-law fit with a power-law index of $n \simeq 1.3$ is marked with a dotted line.
\label{fig:relation_with-star}}
\end{figure*}

Figure~\ref{fig:relation_with-star} presents the relation between the SFR surface density ($\SDSFR$) and the surface densities of \schi\ ($\SDhi$), \schii\ ($\SDht$), or total gas ($\SDgas$) in \schi\ superclouds (referred to as $\SDSFR$--$\SDhi$, $\SDSFR$--$\SDht$, and $\SDSFR$--$\SDgas$, respectively).
As shown in the previous studies, We also find that $\SDSFR$ appears independent of $\SDhi$ and $\SDgas$ in \schi\ superclouds.
The correlations are weak, with $\rho$ of $-0.029$ and p-value of 0.96 for both.
Although our findings are limited to a very small number of samples, they are still meaningful to compare to others.
Since the \schi\ component dominates the total gas content for our \schi\ superclouds, the $\SDSFR$--$\SDgas$ relation follows the $\SDSFR$--$\SDhi$ trend closely.
Our finding of no-correlation in $\SDSFR$--$\SDhi$ and $\SDSFR$--$\SDgas$ is consistent with those obtained by \citet{lada2013} for local GMCs in the Milky Way.
They explained their findings as a characteristic of constant gas surface density for Galactic GMCs \citep[e.g.,][]{lombardi2010, heyer2009}, which might also be acceptable to our result of \schi\ superclouds.

On the other hand, we do observe a positive correlation between $\SDSFR$ and $\SDht$, similar to the resolved K-S relation, although this correlation can be less clear due to a small number of data points.
Spearman's rank correlation coefficient of 0.54 and p-value of 0.27 also indicate a moderate positive correlation.
The power-law index of $\sim 1.3$ we found is in good agreement with the well-known value of $\sim 1.4$ for the global K-S relation \citet{kennicutt1998}.
This consistency implies that the star formation process in our study region follows a similar scaling relation as observed on global scales in other galaxies.
Our results are also consistent with the findings of \citet{bigiel2008} and \citet{leroy2013}, who demonstrated that molecular gas plays a dominant role in regulating star formation across a wide range of spatial scales and environments.
This reinforces the importance of molecular gas in regulating star formation and provides additional support for the K-S relation as a fundamental scaling law governing the conversions of gas into stars.

\section{Summary} \label{sec:summary}

In this study, we investigated \schi\ clouds, MCs, and star formation in the Carina spiral arm of the outer Galaxy.

Using HI4PI survey and CfA CO survey data, we identified \schi\ clouds and MCs based on ($l$, $\vlsr$) locations of the Carina arm obtained from Paper I.
We derived physical parameters, including size, mass, and distance, for 29 \schi\ clouds and 49 MCs, but conducted further analysis on 26 \schi\ clouds and 48 MCs (see Section~\ref{sec:clf_step1}).

Most of the identified \schi\ clouds are more massive than $10^6~\msol$ and are referred to as \schi\ superclouds. 
Fifteen of the 26 \schi\ clouds have associated MC(s) with masses exceeding $10^6~\msol$ with $\SDgas \gtrsim$ 50~$\msol \rm pc^{-2}$.
Our virial equilibrium analysis suggests that these CO-bright \schi\ clouds are gravitationally bound or marginally bound.

We also found an anti-correlation between molecular mass fractions and Galactocentric distances, as well as a correlation between molecular mass fractions and total gas surface densities.
Among the CO-bright \schi\ superclouds, nine are associated with \schii\ regions, indicating ongoing star formation.
The star-forming \schi\ superclouds have molecular mass fractions larger than 0.1, although some \schi\ superclouds that meet this criterion do not match with \schii\ regions.

Consistent with previous studies, we showed that \schi\ superclouds are regularly spaced along the spiral arm, with a typical spacing of $0.084\,\Rsun$ ($\sim 700$~pc); this could be partly a selection effect from blending at short spacings.
The presence of a peak in the ratio of separation to the minor axis, which closely aligns with the theoretical expectation of 3.9, suggests that the observed regular spacing is not a result of chance but rather an outcome of an underlying physical process.
We observed a strong spatial correlation between \schii\ regions and molecular clouds, with some offsets between molecular clouds and local \schi\ column density peaks.

Lastly, we examined the relationship between the SFR surface density and surface densities of \schi, $\rmht$, and total gas.
In agreement with extragalactic studies of the resolved K-S relation and local GMCs study by \citet{lada2013}, our results indicate that $\SDSFR$ is independent of $\SDhi$ and $\SDgas$, but show a positive correlation with $\SDht$ with a power-law index of $\sim 1.3$.
This consistency with the well-established value of $\sim 1.4$ for the global K-S relation highlights the significant role molecular gas plays in controlling star formation and lends further credence to the K-S relation as a key scaling law that governs the transformation of gas into stars.

\begin{acknowledgments}
G.P. was supported by the National Research Foundation of Korea through grants NRF-2020R1A6A3A01100208 and RS-2023-00242652.
This work was partly supported by the Korea Astronomy and Space Science Institute grant funded by the Korea government(MSIT) (Project No. 2022-1-840-05).
\end{acknowledgments}

%

\vspace{5mm}


\software{
         ASTROML \citep{VanderPlas2012, ivezic2014},
         The IDL Astronomy User’s Library \citep{landsman1993}
         }

\bibliography{main}{}

\begin{thebibliography}{}
\expandafter\ifx\csname natexlab\endcsname\relax\def\natexlab#1{#1}\fi
\providecommand{\url}[1]{\href{#1}{#1}}
\providecommand{\dodoi}[1]{doi:~\href{http://doi.org/#1}{\nolinkurl{#1}}}
\providecommand{\doeprint}[1]{\href{http://ascl.net/#1}{\nolinkurl{http://ascl.net/#1}}}
\providecommand{\doarXiv}[1]{\href{https://arxiv.org/abs/#1}{\nolinkurl{https://arxiv.org/abs/#1}}}

\bibitem[{{Abdo} {et~al.}(2010){Abdo}, {Ackermann}, {Ajello}, {Baldini},
  {Ballet}, {Barbiellini}, {Bastieri}, {Baughman}, {Bechtol}, {Bellazzini},
  {Berenji}, {Bloom}, {Bonamente}, {Borgland}, {Bregeon}, {Brez}, {Brigida},
  {Bruel}, {Burnett}, {Buson}, {Caliandro}, {Cameron}, {Caraveo}, {Casandjian},
  {Cecchi}, {{\c{C}}elik}, {Chekhtman}, {Cheung}, {Chiang}, {Ciprini}, {Claus},
  {Cohen-Tanugi}, {Cominsky}, {Conrad}, {Dermer}, {de Palma}, {Digel}, {Silva},
  {Drell}, {Dubois}, {Dumora}, {Farnier}, {Favuzzi}, {Fegan}, {Focke},
  {Fortin}, {Frailis}, {Fukazawa}, {Funk}, {Fusco}, {Gargano}, {Gehrels},
  {Germani}, {Giavitto}, {Giebels}, {Giglietto}, {Giordano}, {Glanzman},
  {Godfrey}, {Grenier}, {Grondin}, {Grove}, {Guillemot}, {Guiriec}, {Harding},
  {Hayashida}, {Horan}, {Hughes}, {Jackson}, {J{\'o}hannesson}, {Johnson},
  {Johnson}, {Kamae}, {Katagiri}, {Kataoka}, {Kawai}, {Kerr}, {Kn{\"o}dlseder},
  {Kuss}, {Lande}, {Latronico}, {Lemoine-Goumard}, {Longo}, {Loparco}, {Lott},
  {Lovellette}, {Lubrano}, {Makeev}, {Mazziotta}, {McEnery}, {Meurer},
  {Michelson}, {Mitthumsiri}, {Mizuno}, {Monte}, {Monzani}, {Morselli},
  {Moskalenko}, {Murgia}, {Nolan}, {Norris}, {Nuss}, {Ohsugi}, {Okumura},
  {Omodei}, {Orlando}, {Ormes}, {Paneque}, {Pelassa}, {Pepe}, {Pesce-Rollins},
  {Piron}, {Porter}, {Rain{\`o}}, {Rando}, {Razzano}, {Reimer}, {Reimer},
  {Reposeur}, {Rodriguez}, {Ryde}, {Sadrozinski}, {Sanchez}, {Sander}, {Saz
  Parkinson}, {Sgr{\`o}}, {Siskind}, {Smith}, {Spandre}, {Spinelli}, {Starck},
  {Strickman}, {Strong}, {Suson}, {Takahashi}, {Tanaka}, {Thayer}, {Thayer},
  {Thompson}, {Tibaldo}, {Torres}, {Tosti}, {Tramacere}, {Uchiyama}, {Usher},
  {Vasileiou}, {Vilchez}, {Vitale}, {Waite}, {Wang}, {Winer}, {Wood}, {Ylinen},
  {Ziegler}, \& {Fermi/LAT Collaboration}}]{abdo2010}
{Abdo}, A.~A., {Ackermann}, M., {Ajello}, M., {et~al.} 2010, \apj, 710, 133,
  \dodoi{10.1088/0004-637X/710/1/133}

\bibitem[{{Allen}(1973)}]{allen1973}
{Allen}, C.~W. 1973, {Astrophysical quantities}

\bibitem[{{Anderson} {et~al.}(2014){Anderson}, {Bania}, {Balser}, {Cunningham},
  {Wenger}, {Johnstone}, \& {Armentrout}}]{anderson2014}
{Anderson}, L.~D., {Bania}, T.~M., {Balser}, D.~S., {et~al.} 2014, \apjs, 212,
  1, \dodoi{10.1088/0067-0049/212/1/1}

\bibitem[{{Ballesteros-Paredes} {et~al.}(1999){Ballesteros-Paredes},
  {Hartmann}, \& {V{\'a}zquez-Semadeni}}]{ballesteros1999}
{Ballesteros-Paredes}, J., {Hartmann}, L., \& {V{\'a}zquez-Semadeni}, E. 1999,
  \apj, 527, 285, \dodoi{10.1086/308076}

\bibitem[{{Barriault} {et~al.}(2010){Barriault}, {Joncas}, {Lockman}, \&
  {Martin}}]{barriault2010}
{Barriault}, L., {Joncas}, G., {Lockman}, F.~J., \& {Martin}, P.~G. 2010,
  \mnras, 407, 2645, \dodoi{10.1111/j.1365-2966.2010.17105.x}

\bibitem[{{Bigiel} {et~al.}(2008){Bigiel}, {Leroy}, {Walter}, {Brinks}, {de
  Blok}, {Madore}, \& {Thornley}}]{bigiel2008}
{Bigiel}, F., {Leroy}, A., {Walter}, F., {et~al.} 2008, \aj, 136, 2846,
  \dodoi{10.1088/0004-6256/136/6/2846}

\bibitem[{{Blitz}(1993)}]{blitz1993}
{Blitz}, L. 1993, in Protostars and Planets III, ed. E.~H. {Levy} \& J.~I.
  {Lunine}, 125

\bibitem[{{Bolatto} {et~al.}(2013){Bolatto}, {Wolfire}, \&
  {Leroy}}]{bolatto2013}
{Bolatto}, A.~D., {Wolfire}, M., \& {Leroy}, A.~K. 2013, \araa, 51, 207,
  \dodoi{10.1146/annurev-astro-082812-140944}

\bibitem[{{Boulanger} \& {Viallefond}(1992)}]{boulanger1992}
{Boulanger}, F., \& {Viallefond}, F. 1992, \aap, 266, 37

\bibitem[{{Burton}(1988)}]{burton1988}
{Burton}, W.~B. 1988, in Galactic and Extragalactic Radio Astronomy, ed. K.~I.
  {Kellermann} \& G.~L. {Verschuur}, 295--358

\bibitem[{{Chandrasekhar} \& {Fermi}(1953)}]{chandrasekhar1953}
{Chandrasekhar}, S., \& {Fermi}, E. 1953, \apj, 118, 116,
  \dodoi{10.1086/145732}

\bibitem[{{Cowie}(1981)}]{cowie1981}
{Cowie}, L. 1981, \apj, 245, 66, \dodoi{10.1086/158786}

\bibitem[{{Dame} {et~al.}(2001){Dame}, {Hartmann}, \& {Thaddeus}}]{dame2001}
{Dame}, T.~M., {Hartmann}, D., \& {Thaddeus}, P. 2001, \apj, 547, 792,
  \dodoi{10.1086/318388}

\bibitem[{{Dobbs}(2008)}]{dobbs08a}
{Dobbs}, C. 2008, \mnras, 391, 844–858,
  \dodoi{10.1111/j.1365-2966.2008.13939.x}

\bibitem[{{Dobbs} {et~al.}(2008){Dobbs}, {Glover}, {Clark}, \&
  {Klessen}}]{dobbs2008b}
{Dobbs}, C., {Glover}, S., {Clark}, P., \& {Klessen}, R.~S. 2008, \mnras, 389,
  1097, \dodoi{10.1111/j.1365-2966.2008.13646.x}

\bibitem[{{Dobbs} {et~al.}(2014){Dobbs}, {Krumholz}, {Ballesteros-Paredes},
  {Bolatto}, {Fukui}, {Heyer}, {Low}, {Ostriker}, \&
  {V{\'a}zquez-Semadeni}}]{dobbs2014}
{Dobbs}, C.~L., {Krumholz}, M.~R., {Ballesteros-Paredes}, J., {et~al.} 2014, in
  Protostars and Planets VI, ed. H.~{Beuther}, R.~S. {Klessen}, C.~P.
  {Dullemond}, \& T.~{Henning}, 3--26,
  \dodoi{10.2458/azu_uapress_9780816531240-ch001}

\bibitem[{{Efremov}(1998)}]{efremov1998}
{Efremov}, Y.~N. 1998, Astronomical and Astrophysical Transactions, 15, 3,
  \dodoi{10.1080/10556799808201745}

\bibitem[{{Efremov}(2009)}]{efremov2009}
---. 2009, Astronomy Letters, 35, 507, \dodoi{10.1134/S1063773709080015}

\bibitem[{{Efremov}(2010)}]{efremov2010}
---. 2010, \mnras, 405, 1531, \dodoi{10.1111/j.1365-2966.2010.16578.x}

\bibitem[{{Elmegreen}(1979)}]{elmegreen1979}
{Elmegreen}, B.~G. 1979, \apj, 231, 372

\bibitem[{{Elmegreen}(1982)}]{elmegreen1982}
---. 1982, \apj, 253, 655, \dodoi{10.1086/159666}

\bibitem[{{Elmegreen}(1996)}]{elmegreen1996}
{Elmegreen}, B.~G. 1996, in Unsolved Problems of the Milky Way, ed. L.~{Blitz}
  \& P.~J. {Teuben}, Vol. 169, 551

\bibitem[{{Elmegreen}(2004)}]{elmegreen2004}
---. 2004, arXiv e-prints, astro, \dodoi{10.48550/arXiv.astro-ph/0411282}

\bibitem[{{Elmegreen} \& {Clemens}(1985)}]{elmegreen1985}
{Elmegreen}, B.~G., \& {Clemens}, C. 1985, \apj, 294, 523,
  \dodoi{10.1086/163320}

\bibitem[{{Elmegreen} \& {Elmegreen}(1983)}]{elmegreen1983}
{Elmegreen}, B.~G., \& {Elmegreen}, D.~M. 1983, \mnras, 203, 31,
  \dodoi{10.1093/mnras/203.1.31}

\bibitem[{{Elmegreen} \& {Elmegreen}(1987)}]{elmegreen1987}
---. 1987, \apj, 320, 182, \dodoi{10.1086/165534}

\bibitem[{{Elmegreen} \& {Elmegreen}(2019)}]{elmegreen2019}
---. 2019, \apjs, 245, 14, \dodoi{10.3847/1538-4365/ab4903}

\bibitem[{{Elmegreen} {et~al.}(2018){Elmegreen}, {Elmegreen}, \&
  {Efremov}}]{elmegreen2018}
{Elmegreen}, B.~G., {Elmegreen}, D.~M., \& {Efremov}, Y.~N. 2018, \apj, 863,
  59, \dodoi{10.3847/1538-4357/aacf9a}

\bibitem[{{Engargiola} {et~al.}(2003){Engargiola}, {Plambeck}, {Rosolowsky}, \&
  {Blitz}}]{engargiola2003}
{Engargiola}, G., {Plambeck}, R.~L., {Rosolowsky}, E., \& {Blitz}, L. 2003,
  \apjs, 149, 343, \dodoi{10.1086/379165}

\bibitem[{{Fiege} \& {Pudritz}(2000)}]{fiege2000}
{Fiege}, J.~D., \& {Pudritz}, R.~E. 2000, \mnras, 311, 105,
  \dodoi{10.1046/j.1365-8711.2000.03067.x}

\bibitem[{{Fukui} \& {Kawamura}(2010)}]{fukui2010}
{Fukui}, Y., \& {Kawamura}, A. 2010, \araa, 48, 547,
  \dodoi{10.1146/annurev-astro-081309-130854}

\bibitem[{{Grabelsky} {et~al.}(1987){Grabelsky}, {Cohen}, {Bronfman},
  {Thaddeus}, \& {May}}]{grabelsky1987}
{Grabelsky}, D.~A., {Cohen}, R.~S., {Bronfman}, L., {Thaddeus}, P., \& {May},
  J. 1987, \apj, 315, 122, \dodoi{10.1086/165118}

\bibitem[{{Grenier} {et~al.}(2005){Grenier}, {Casandjian}, \&
  {Terrier}}]{grenier2005}
{Grenier}, I.~A., {Casandjian}, J.-M., \& {Terrier}, R. 2005, Science, 307,
  1292, \dodoi{10.1126/science.1106924}

\bibitem[{{Gusev} {et~al.}(2022){Gusev}, {Shimanovskaya}, \&
  {Zaitseva}}]{gusev2022}
{Gusev}, A., {Shimanovskaya}, E., \& {Zaitseva}, N. 2022, \mnras, 514,
  3953–3964, \dodoi{doi.org/10.1093/mnras/stac1592}

\bibitem[{{Hartmann} {et~al.}(2001){Hartmann}, {Ballesteros-Paredes}, \&
  {Bergin}}]{hartmann2001}
{Hartmann}, L., {Ballesteros-Paredes}, J., \& {Bergin}, E.~A. 2001, \apj, 562,
  852, \dodoi{10.1086/323863}

\bibitem[{{Hennebelle} \& {P{\'e}rault}(2000)}]{hennebelle2000}
{Hennebelle}, P., \& {P{\'e}rault}, M. 2000, \aap, 359, 1124

\bibitem[{{Heyer} {et~al.}(2009){Heyer}, {Krawczyk}, {Duval}, \&
  {Jackson}}]{heyer2009}
{Heyer}, M., {Krawczyk}, C., {Duval}, J., \& {Jackson}, J.~M. 2009, \apj, 699,
  1092, \dodoi{10.1088/0004-637X/699/2/1092}

\bibitem[{{HI4PI Collaboration} {et~al.}(2016){HI4PI Collaboration}, {Ben
  Bekhti}, {Fl{\"o}er}, {Keller}, {Kerp}, {Lenz}, {Winkel}, {Bailin},
  {Calabretta}, {Dedes}, {Ford}, {Gibson}, {Haud}, {Janowiecki}, {Kalberla},
  {Lockman}, {McClure-Griffiths}, {Murphy}, {Nakanishi}, {Pisano}, \&
  {Staveley-Smith}}]{hi4pi2016}
{HI4PI Collaboration}, {Ben Bekhti}, N., {Fl{\"o}er}, L., {et~al.} 2016, \aap,
  594, A116, \dodoi{10.1051/0004-6361/201629178}

\bibitem[{{Ianjamasimanana} {et~al.}(2012){Ianjamasimanana}, de~Blok, {Walter},
  \& {Heald}}]{ianjamasimanana2012}
{Ianjamasimanana}, R., de~Blok, W., {Walter}, F., \& {Heald}, G. 2012, \apj,
  144, 96 (25pp), \dodoi{10.1088/0004-6256/144/4/96}

\bibitem[{{Inoue} \& {Yoshida}(2019)}]{inoue2019}
{Inoue}, S., \& {Yoshida}, N. 2019, \mnras, 485, 3024,
  \dodoi{10.1093/mnras/stz584}

\bibitem[{{Inoue} \& {Inutsuka}(2009)}]{inoue2009}
{Inoue}, T., \& {Inutsuka}, S.-i. 2009, \apj, 704, 161,
  \dodoi{10.1088/0004-637X/704/1/161}

\bibitem[{{Inutsuka} {et~al.}(2015){Inutsuka}, {Inoue}, {Iwasaki}, \&
  {Hosokawa}}]{inutsuka2015}
{Inutsuka}, S.-i., {Inoue}, T., {Iwasaki}, K., \& {Hosokawa}, T. 2015, \aap,
  580, A49, \dodoi{10.1051/0004-6361/201425584}

\bibitem[{{Inutsuka} \& {Miyama}(1992)}]{inutsuka1992}
{Inutsuka}, S.-I., \& {Miyama}, S.~M. 1992, \apj, 388, 392,
  \dodoi{10.1086/171162}

\bibitem[{{Ivezi{\'c}} {et~al.}(2014){Ivezi{\'c}}, {Connolly}, {VanderPlas}, \&
  {Gray}}]{ivezic2014}
{Ivezi{\'c}}, {\v{Z}}., {Connolly}, A.~J., {VanderPlas}, J.~T., \& {Gray}, A.
  2014, {Statistics, Data Mining, and Machine Learning in Astronomy: A
  Practical Python Guide for the Analysis of Survey Data},
  \dodoi{10.1515/9781400848911}

\bibitem[{{Kalberla} {et~al.}(2005){Kalberla}, {Burton}, {Hartmann}, {Arnal},
  {Bajaja}, {Morras}, \& {P{\"o}ppel}}]{kalberla2005}
{Kalberla}, P.~M.~W., {Burton}, W.~B., {Hartmann}, D., {et~al.} 2005, \aap,
  440, 775, \dodoi{10.1051/0004-6361:20041864}

\bibitem[{{Kalberla} \& {Haud}(2015)}]{kalberla2015}
{Kalberla}, P.~M.~W., \& {Haud}, U. 2015, \aap, 578, A78,
  \dodoi{10.1051/0004-6361/201525859}

\bibitem[{{Kalberla} {et~al.}(2010){Kalberla}, {McClure-Griffiths}, {Pisano},
  {Calabretta}, {Ford}, {Lockman}, {Staveley-Smith}, {Kerp}, {Winkel},
  {Murphy}, \& {Newton-McGee}}]{kalberla2010}
{Kalberla}, P.~M.~W., {McClure-Griffiths}, N.~M., {Pisano}, D.~J., {et~al.}
  2010, \aap, 521, A17, \dodoi{10.1051/0004-6361/200913979}

\bibitem[{{Kennicutt}(1989)}]{kennicutt1989}
{Kennicutt}, Robert~C., J. 1989, \apj, 344, 685, \dodoi{10.1086/167834}

\bibitem[{{Kennicutt}(1998)}]{kennicutt1998}
---. 1998, \apj, 498, 541, \dodoi{10.1086/305588}

\bibitem[{{Kennicutt} \& {Evans}(2012)}]{kennicutt2012}
{Kennicutt}, R.~C., \& {Evans}, N.~J. 2012, \araa, 50, 531,
  \dodoi{10.1146/annurev-astro-081811-125610}

\bibitem[{{Kerp} {et~al.}(2011){Kerp}, {Winkel}, {Ben Bekhti}, {Fl{\"o}er}, \&
  {Kalberla}}]{kerp2011}
{Kerp}, J., {Winkel}, B., {Ben Bekhti}, N., {Fl{\"o}er}, L., \& {Kalberla},
  P.~M.~W. 2011, Astronomische Nachrichten, 332, 637,
  \dodoi{10.1002/asna.201011548}

\bibitem[{{Kim} {et~al.}(2002){Kim}, {Ostriker}, \& {Stone}}]{kim2002}
{Kim}, W.-T., {Ostriker}, E.~C., \& {Stone}, J.~M. 2002, \apj, 581, 1080

\bibitem[{{Koda} {et~al.}(2016){Koda}, {Scoville}, \& {Heyer}}]{koda2016}
{Koda}, J., {Scoville}, N., \& {Heyer}, M. 2016, \apj, 823, 76,
  \dodoi{10.3847/0004-637X/823/2/76}

\bibitem[{{Koo} {et~al.}(2017){Koo}, {Park}, {Kim}, {Lee}, {Balser}, \&
  {Wenger}}]{koo2017}
{Koo}, B.-C., {Park}, G., {Kim}, W.-T., {et~al.} 2017, \pasp, 129, 094102
  (Paper~I), \dodoi{10.1088/1538-3873/aa7c08}

\bibitem[{{Krumholz} {et~al.}(2009){Krumholz}, {McKee}, \&
  {Tumlinson}}]{krumholz2009}
{Krumholz}, M.~R., {McKee}, C.~F., \& {Tumlinson}, J. 2009, \apj, 693, 216,
  \dodoi{10.1088/0004-637X/693/1/216}

\bibitem[{{Kwan} \& {Valdes}(1987)}]{kwan1987}
{Kwan}, J., \& {Valdes}, F. 1987, \apj, 315, 92, \dodoi{10.1086/165116}

\bibitem[{{Lada} {et~al.}(2013){Lada}, {Lombardi}, {Roman-Zuniga}, {Forbrich},
  \& {Alves}}]{lada2013}
{Lada}, C.~J., {Lombardi}, M., {Roman-Zuniga}, C., {Forbrich}, J., \& {Alves},
  J.~F. 2013, \apj, 778, 133, \dodoi{10.1088/0004-637X/778/2/133}

\bibitem[{{Lada} {et~al.}(1988){Lada}, {Margulis}, {Sofue}, {Nakai}, \&
  {Handa}}]{lada1988}
{Lada}, C.~J., {Margulis}, M., {Sofue}, Y., {Nakai}, N., \& {Handa}, T. 1988,
  \apj, 328, 143, \dodoi{10.1086/166275}

\bibitem[{{Landsman}(1993)}]{landsman1993}
{Landsman}, W.~B. 1993, in Astronomical Society of the Pacific Conference
  Series, Vol.~52, Astronomical Data Analysis Software and Systems II, ed.
  R.~J. {Hanisch}, R.~J.~V. {Brissenden}, \& J.~{Barnes}, 246

\bibitem[{{Lee} {et~al.}(2012{\natexlab{a}}){Lee}, {Murray}, \&
  {Rahman}}]{e.lee2012}
{Lee}, E.~J., {Murray}, N., \& {Rahman}, M. 2012{\natexlab{a}}, \apj, 752, 146,
  \dodoi{10.1088/0004-637X/752/2/146}

\bibitem[{{Lee} {et~al.}(2015){Lee}, {Stanimirovi{\'c}}, {Murray}, {Heiles}, \&
  {Miller}}]{m.lee2015}
{Lee}, M.-Y., {Stanimirovi{\'c}}, S., {Murray}, C.~E., {Heiles}, C., \&
  {Miller}, J. 2015, \apj, 809, 56, \dodoi{10.1088/0004-637X/809/1/56}

\bibitem[{{Lee} {et~al.}(2012{\natexlab{b}}){Lee}, {Stanimirovi{\'c}},
  {Douglas}, {Knee}, {Di Francesco}, {Gibson}, {Begum}, {Grcevich}, {Heiles},
  {Korpela}, {Leroy}, {Peek}, {Pingel}, {Putman}, \& {Saul}}]{m.lee2012}
{Lee}, M.-Y., {Stanimirovi{\'c}}, S., {Douglas}, K.~A., {et~al.}
  2012{\natexlab{b}}, \apj, 748, 75, \dodoi{10.1088/0004-637X/748/2/75}

\bibitem[{{Lee} \& {Hong}(2011)}]{s.lee2011}
{Lee}, S.~M., \& {Hong}, S.~S. 2011, \apj, 734, 101,
  \dodoi{10.1088/0004-637X/734/2/101}

\bibitem[{{Leroy} {et~al.}(2008){Leroy}, {Walter}, {Brinks}, {Bigiel}, {de
  Blok}, {Madore}, \& {Thornley}}]{leroy2008}
{Leroy}, A.~K., {Walter}, F., {Brinks}, E., {et~al.} 2008, \aj, 136, 2782,
  \dodoi{10.1088/0004-6256/136/6/2782}

\bibitem[{{Leroy} {et~al.}(2013){Leroy}, {Walter}, {Sandstrom}, {Schruba},
  {Munoz-Mateos}, {Bigiel}, {Bolatto}, {Brinks}, {de Blok}, {Meidt}, {Rix},
  {Rosolowsky}, {Schinnerer}, {Schuster}, \& {Usero}}]{leroy2013}
{Leroy}, A.~K., {Walter}, F., {Sandstrom}, K., {et~al.} 2013, \aj, 146, 19,
  \dodoi{10.1088/0004-6256/146/2/19}

\bibitem[{{Levine} {et~al.}(2006){Levine}, {Blitz}, \&
  {Heiles}}]{levine2006apj}
{Levine}, E.~S., {Blitz}, L., \& {Heiles}, C. 2006, \apj, 643, 881,
  \dodoi{10.1086/503091}

\bibitem[{{Liszt}(2014)}]{liszt2014}
{Liszt}, H. 2014, \apj, 783, 17, \dodoi{10.1088/0004-637X/783/1/17}

\bibitem[{{Liszt} \& {Lucas}(1996)}]{liszt1996}
{Liszt}, H., \& {Lucas}, R. 1996, \aap, 314, 917

\bibitem[{{Lombardi} {et~al.}(2010){Lombardi}, {Alves}, \&
  {Lada}}]{lombardi2010}
{Lombardi}, M., {Alves}, J., \& {Lada}, C.~J. 2010, \aap, 519, L7,
  \dodoi{10.1051/0004-6361/201015282}

\bibitem[{{Lucas} \& {Liszt}(1996)}]{lucas1996}
{Lucas}, R., \& {Liszt}, H. 1996, \aap, 307, 237

\bibitem[{{Mandowara} {et~al.}(2022){Mandowara}, {Sormani}, {Sobacchi}, \&
  {Klessen}}]{mandowara2022}
{Mandowara}, Y., {Sormani}, M.~C., {Sobacchi}, E., \& {Klessen}, R. 2022,
  \mnras, 513, 5052, \dodoi{10.1093/mnras/stac1214}

\bibitem[{{Markwardt}(2009)}]{markwardt2009}
{Markwardt}, C.~B. 2009, in Astronomical Society of the Pacific Conference
  Series, Vol. 411, Astronomical Data Analysis Software and Systems XVIII, ed.
  D.~A. {Bohlender}, D.~{Durand}, \& P.~{Dowler}, 251,
  \dodoi{10.48550/arXiv.0902.2850}

\bibitem[{{McClure-Griffiths} {et~al.}(2002){McClure-Griffiths}, {Dickey},
  {Gaensler}, \& {Green}}]{mcclure2002}
{McClure-Griffiths}, N.~M., {Dickey}, J.~M., {Gaensler}, B.~M., \& {Green},
  A.~J. 2002, \apj, 578, 176, \dodoi{10.1086/342470}

\bibitem[{{McClure-Griffiths} {et~al.}(2005){McClure-Griffiths}, {Dickey},
  {Gaensler}, {Green}, {Haverkorn}, \& {Strasser}}]{mcclure2005}
{McClure-Griffiths}, N.~M., {Dickey}, J.~M., {Gaensler}, B.~M., {et~al.} 2005,
  \apjs, 158, 178, \dodoi{10.1086/430114}

\bibitem[{{McClure-Griffiths} {et~al.}(2009){McClure-Griffiths}, {Pisano},
  {Calabretta}, {Ford}, {Lockman}, {Staveley-Smith}, {Kalberla}, {Bailin},
  {Dedes}, {Janowiecki}, {Gibson}, {Murphy}, {Nakanishi}, \&
  {Newton-McGee}}]{mcclure2009}
{McClure-Griffiths}, N.~M., {Pisano}, D.~J., {Calabretta}, M.~R., {et~al.}
  2009, \apjs, 181, 398, \dodoi{10.1088/0067-0049/181/2/398}

\bibitem[{{McGee} \& {Milton}(1964)}]{mcgee1964}
{McGee}, R.~X., \& {Milton}, J.~A. 1964, Australian Journal of Physics, 17,
  128, \dodoi{10.1071/PH640128}

\bibitem[{{Meyerdierks} \& {Heithausen}(1996)}]{meyerdierks1996}
{Meyerdierks}, H., \& {Heithausen}, A. 1996, \aap, 313, 929

\bibitem[{{Mezger}(1978)}]{mezger1978}
{Mezger}, P.~O. 1978, \aap, 70, 565

\bibitem[{{Murray} \& {Rahman}(2010)}]{murray2010}
{Murray}, N., \& {Rahman}, M. 2010, \apj, 709, 424,
  \dodoi{10.1088/0004-637X/709/1/424}

\bibitem[{{Nagasawa}(1987)}]{nagasawa1987}
{Nagasawa}, M. 1987, Progress of Theoretical Physics, 77, 635,
  \dodoi{10.1143/PTP.77.635}

\bibitem[{{Ostriker} \& {Kim}(2004)}]{ostriker2004}
{Ostriker}, E.~C., \& {Kim}, W.~T. 2004, in Astronomical Society of the Pacific
  Conference Series, Vol. 317, Milky Way Surveys: The Structure and Evolution
  of our Galaxy, ed. D.~{Clemens}, R.~{Shah}, \& T.~{Brainerd}, 248

\bibitem[{{Ostriker}(1964)}]{ostriker1964}
{Ostriker}, J. 1964, \apj, 140, 1056, \dodoi{10.1086/148005}

\bibitem[{{Park} {et~al.}(2018){Park}, {Koo}, {Kim}, {Byun}, \&
  {Heiles}}]{park2018}
{Park}, G., {Koo}, B.-C., {Kim}, K.-T., {Byun}, D.-Y., \& {Heiles}, C.~E. 2018,
  \apss, 363, 150, \dodoi{10.1007/s10509-018-3372-4}

\bibitem[{{Pedicelli} {et~al.}(2009){Pedicelli}, {Bono}, {Lemasle},
  {Fran{\c{c}}ois}, {Groenewegen}, {Lub}, {Pel}, {Laney}, {Piersimoni},
  {Romaniello}, {Buonanno}, {Caputo}, {Cassisi}, {Castelli}, {Leurini},
  {Pietrinferni}, {Primas}, \& {Pritchard}}]{pedicelli2009}
{Pedicelli}, S., {Bono}, G., {Lemasle}, B., {et~al.} 2009, \aap, 504, 81,
  \dodoi{10.1051/0004-6361/200912504}

\bibitem[{{Pineda} {et~al.}(2013){Pineda}, {Langer}, {Velusamy}, \&
  {Goldsmith}}]{pineda2013}
{Pineda}, J.~L., {Langer}, W.~D., {Velusamy}, T., \& {Goldsmith}, P.~F. 2013,
  \aap, 554, A103, \dodoi{10.1051/0004-6361/201321188}

\bibitem[{{Planck Collaboration} {et~al.}(2011{\natexlab{a}}){Planck
  Collaboration}, {Abergel}, {Ade}, {Aghanim}, {Arnaud}, {Ashdown}, {Aumont},
  {Baccigalupi}, {Balbi}, {Banday}, {Barreiro}, {Bartlett}, {Battaner},
  {Benabed}, {Beno{\^\i}t}, {Bernard}, {Bersanelli}, {Bhatia}, {Blagrave},
  {Bock}, {Bonaldi}, {Bond}, {Borrill}, {Bouchet}, {Boulanger}, {Bucher},
  {Burigana}, {Cabella}, {Cantalupo}, {Cardoso}, {Catalano}, {Cay{\'o}n},
  {Challinor}, {Chamballu}, {Chiang}, {Chiang}, {Christensen}, {Clements},
  {Colombi}, {Couchot}, {Coulais}, {Crill}, {Cuttaia}, {Danese}, {Davies},
  {Davis}, {de Bernardis}, {de Gasperis}, {de Rosa}, {de Zotti},
  {Delabrouille}, {Delouis}, {D{\'e}sert}, {Dickinson}, {Donzelli}, {Dor{\'e}},
  {D{\"o}rl}, {Douspis}, {Dupac}, {Efstathiou}, {En{\ss}lin}, {Eriksen},
  {Finelli}, {Forni}, {Frailis}, {Franceschi}, {Galeotta}, {Ganga}, {Giard},
  {Giardino}, {Giraud-H{\'e}raud}, {Gonz{\'a}lez-Nuevo}, {G{\'o}rski},
  {Gratton}, {Gregorio}, {Gruppuso}, {Hansen}, {Harrison}, {Helou},
  {Henrot-Versill{\'e}}, {Herranz}, {Hildebrandt}, {Hivon}, {Hobson}, {Holmes},
  {Hovest}, {Hoyland}, {Huffenberger}, {Jaffe}, {Joncas}, {Jones}, {Jones},
  {Juvela}, {Keih{\"a}nen}, {Keskitalo}, {Kisner}, {Kneissl}, {Knox},
  {Kurki-Suonio}, {Lagache}, {Lamarre}, {Lasenby}, {Laureijs}, {Lawrence},
  {Leach}, {Leonardi}, {Leroy}, {Linden-V{\o}rnle}, {Lockman},
  {L{\'o}pez-Caniego}, {Lubin}, {Mac{\'\i}as-P{\'e}rez}, {MacTavish}, {Maffei},
  {Maino}, {Mandolesi}, {Mann}, {Maris}, {Marshall}, {Martin},
  {Mart{\'\i}nez-Gonz{\'a}lez}, {Masi}, {Matarrese}, {Matthai}, {Mazzotta},
  {McGehee}, {Meinhold}, {Melchiorri}, {Mendes}, {Mennella},
  {Miville-Desch{\^e}nes}, {Moneti}, {Montier}, {Morgante}, {Mortlock},
  {Munshi}, {Murphy}, {Naselsky}, {Nati}, {Natoli}, {Netterfield},
  {N{\o}rgaard-Nielsen}, {Noviello}, {Novikov}, {Novikov}, {O'Dwyer},
  {Osborne}, {Pajot}, {Paladini}, {Pasian}, {Patanchon}, {Perdereau},
  {Perotto}, {Perrotta}, {Piacentini}, {Piat}, {Pinheiro Gon{\c{c}}alves},
  {Plaszczynski}, {Pointecouteau}, {Polenta}, {Ponthieu}, {Poutanen},
  {Pr{\'e}zeau}, {Prunet}, {Puget}, {Rachen}, {Reach}, {Reinecke}, {Renault},
  {Ricciardi}, {Riller}, {Ristorcelli}, {Rocha}, {Rosset}, {Rowan-Robinson},
  {Rubi{\~n}o-Mart{\'\i}n}, {Rusholme}, {Sandri}, {Santos}, {Savini}, {Scott},
  {Seiffert}, {Shellard}, {Smoot}, {Starck}, {Stivoli}, {Stolyarov}, {Stompor},
  {Sudiwala}, {Sygnet}, {Tauber}, {Terenzi}, {Toffolatti}, {Tomasi}, {Torre},
  {Tristram}, {Tuovinen}, {Umana}, {Valenziano}, {Vielva}, {Villa}, {Vittorio},
  {Wade}, {Wandelt}, {Wilkinson}, {Yvon}, {Zacchei}, \&
  {Zonca}}]{planck2011a24}
{Planck Collaboration}, {Abergel}, A., {Ade}, P.~A.~R., {et~al.}
  2011{\natexlab{a}}, \aap, 536, A24, \dodoi{10.1051/0004-6361/201116485}

\bibitem[{{Planck Collaboration} {et~al.}(2011{\natexlab{b}}){Planck
  Collaboration}, {Ade}, {Aghanim}, {Arnaud}, {Ashdown}, {Aumont},
  {Baccigalupi}, {Balbi}, {Banday}, {Barreiro}, {Bartlett}, {Battaner},
  {Benabed}, {Beno{\^\i}t}, {Bernard}, {Bersanelli}, {Bhatia}, {Bock},
  {Bonaldi}, {Bond}, {Borrill}, {Bouchet}, {Boulanger}, {Bucher}, {Burigana},
  {Cabella}, {Cardoso}, {Catalano}, {Cay{\'o}n}, {Challinor}, {Chamballu},
  {Chiang}, {Chiang}, {Christensen}, {Clements}, {Colombi}, {Couchot},
  {Coulais}, {Crill}, {Cuttaia}, {Dame}, {Danese}, {Davies}, {Davis}, {de
  Bernardis}, {de Gasperis}, {de Rosa}, {de Zotti}, {Delabrouille}, {Delouis},
  {D{\'e}sert}, {Dickinson}, {Dobashi}, {Donzelli}, {Dor{\'e}}, {D{\"o}rl},
  {Douspis}, {Dupac}, {Efstathiou}, {En{\ss}lin}, {Eriksen}, {Falgarone},
  {Finelli}, {Forni}, {Fosalba}, {Frailis}, {Franceschi}, {Fukui}, {Galeotta},
  {Ganga}, {Giard}, {Giardino}, {Giraud-H{\'e}raud}, {Gonz{\'a}lez-Nuevo},
  {G{\'o}rski}, {Gratton}, {Gregorio}, {Grenier}, {Gruppuso}, {Hansen},
  {Harrison}, {Helou}, {Henrot-Versill{\'e}}, {Herranz}, {Hildebrandt},
  {Hivon}, {Hobson}, {Holmes}, {Hovest}, {Hoyland}, {Huffenberger}, {Jaffe},
  {Jones}, {Juvela}, {Kawamura}, {Keih{\"a}nen}, {Keskitalo}, {Kisner},
  {Kneissl}, {Knox}, {Kurki-Suonio}, {Lagache}, {Lamarre}, {Lasenby},
  {Laureijs}, {Lawrence}, {Leach}, {Leonardi}, {Leroy}, {Lilje},
  {Linden-V{\o}rnle}, {L{\'o}pez-Caniego}, {Lubin}, {Mac{\'\i}as-P{\'e}rez},
  {MacTavish}, {Maffei}, {Maino}, {Mandolesi}, {Mann}, {Maris}, {Martin},
  {Mart{\'\i}nez-Gonz{\'a}lez}, {Masi}, {Matarrese}, {Matthai}, {Mazzotta},
  {McGehee}, {Meinhold}, {Melchiorri}, {Mendes}, {Mennella},
  {Miville-Desch{\^e}nes}, {Moneti}, {Montier}, {Morgante}, {Mortlock},
  {Munshi}, {Murphy}, {Naselsky}, {Natoli}, {Netterfield},
  {N{\o}rgaard-Nielsen}, {Noviello}, {Novikov}, {Novikov}, {O'Dwyer}, {Onishi},
  {Osborne}, {Pajot}, {Paladini}, {Paradis}, {Pasian}, {Patanchon},
  {Perdereau}, {Perotto}, {Perrotta}, {Piacentini}, {Piat}, {Plaszczynski},
  {Pointecouteau}, {Polenta}, {Ponthieu}, {Poutanen}, {Pr{\'e}zeau}, {Prunet},
  {Puget}, {Reach}, {Reinecke}, {Renault}, {Ricciardi}, {Riller},
  {Ristorcelli}, {Rocha}, {Rosset}, {Rowan-Robinson}, {Rubi{\~n}o-Mart{\'\i}n},
  {Rusholme}, {Sandri}, {Santos}, {Savini}, {Scott}, {Seiffert}, {Shellard},
  {Smoot}, {Starck}, {Stivoli}, {Stolyarov}, {Stompor}, {Sudiwala}, {Sygnet},
  {Tauber}, {Terenzi}, {Toffolatti}, {Tomasi}, {Torre}, {Tristram}, {Tuovinen},
  {Umana}, {Valenziano}, {Vielva}, {Villa}, {Vittorio}, {Wade}, {Wandelt},
  {Wilkinson}, {Yvon}, {Zacchei}, \& {Zonca}}]{planck2011a19}
{Planck Collaboration}, {Ade}, P.~A.~R., {Aghanim}, N., {et~al.}
  2011{\natexlab{b}}, \aap, 536, A19, \dodoi{10.1051/0004-6361/201116479}

\bibitem[{{Reach} {et~al.}(1994){Reach}, {Koo}, \& {Heiles}}]{reach1994}
{Reach}, W.~T., {Koo}, B.-C., \& {Heiles}, C. 1994, \apj, 429, 672,
  \dodoi{10.1086/174353}

\bibitem[{{Reid} \& {Dame}(2016)}]{reid2016}
{Reid}, M.~J., \& {Dame}, T.~M. 2016, \apj, 832, 159,
  \dodoi{10.3847/0004-637X/832/2/159}

\bibitem[{{Reid} {et~al.}(2014){Reid}, {Menten}, {Brunthaler}, {Zheng}, {Dame},
  {Xu}, {Wu}, {Zhang}, {Sanna}, {Sato}, {Hachisuka}, {Choi}, {Immer},
  {Moscadelli}, {Rygl}, \& {Bartkiewicz}}]{reid2014}
{Reid}, M.~J., {Menten}, K.~M., {Brunthaler}, A., {et~al.} 2014, \apj, 783,
  130, \dodoi{10.1088/0004-637X/783/2/130}

\bibitem[{{Renaud} {et~al.}(2013){Renaud}, {Bournaud}, {Emsellem}, {Elmegreen},
  {Teyssier}, {Alves}, {Chapon}, {Combes}, {Dekel}, {Gabor}, {Hennebelle}, \&
  {Kraljic}}]{renaud2013}
{Renaud}, F., {Bournaud}, F., {Emsellem}, E., {et~al.} 2013, \mnras, 436, 1836,
  \dodoi{10.1093/mnras/stt1698}

\bibitem[{{Schmidt}(1959)}]{schmidt1959}
{Schmidt}, M. 1959, \apj, 129, 243, \dodoi{10.1086/146614}

\bibitem[{{Schruba} {et~al.}(2011){Schruba}, {Leroy}, {Walter}, {Bigiel},
  {Brinks}, {de Blok}, {Dumas}, {Kramer}, {Rosolowsky}, {Sandstrom},
  {Schuster}, {Usero}, {Weiss}, \& {Wiesemeyer}}]{schruba2011}
{Schruba}, A., {Leroy}, A.~K., {Walter}, F., {et~al.} 2011, \aj, 142, 37,
  \dodoi{10.1088/0004-6256/142/2/37}

\bibitem[{{Stod{\'o}lkiewicz}(1963)}]{stodolkiewicz1963}
{Stod{\'o}lkiewicz}, J.~S. 1963, \actaa, 13, 30

\bibitem[{{Tang} {et~al.}(2017){Tang}, {Li}, {Heiles}, {Yue}, {Dawson},
  {Goldsmith}, {Kr{\v{c}}o}, {McClure-Griffiths}, {Wang}, {Zuo}, {Pineda}, \&
  {Wang}}]{tang2017}
{Tang}, N., {Li}, D., {Heiles}, C., {et~al.} 2017, \apj, 839, 8,
  \dodoi{10.3847/1538-4357/aa67e9}

\bibitem[{{Tosaki} {et~al.}(2011){Tosaki}, {Kuno}, {Onodera}, {Sawada},
  {Muraoka}, {Nakanishi}, {Komugi}, {Nakanishi}, {Kaneko}, {Hirota}, {Kohno},
  \& {Kawabe}}]{tosaki2011}
{Tosaki}, T., {Kuno}, N., {Onodera}, Sachiko~Miura, R., {et~al.} 2011, \pasj,
  63, 1171, \dodoi{10.1093/pasj/63.6.1171}

\bibitem[{{van Dishoeck} \& {Black}(1988)}]{vanDishoeck1988}
{van Dishoeck}, E.~F., \& {Black}, J.~H. 1988, \apj, 334, 771,
  \dodoi{10.1086/166877}

\bibitem[{{VanderPlas} {et~al.}(2012){VanderPlas}, {Connolly}, {Ivezic}, \&
  {Gray}}]{VanderPlas2012}
{VanderPlas}, J., {Connolly}, A.~J., {Ivezic}, Z., \& {Gray}, A. 2012, in
  Proceedings of Conference on Intelligent Data Understanding (CIDU, 47--54,
  \dodoi{10.1109/CIDU.2012.6382200}

\bibitem[{{Wada} \& {Koda}(2004)}]{wada2004}
{Wada}, K., \& {Koda}, J. 2004, \mnras, 349, 270–280,
  \dodoi{110.1111/j.1365-2966.2004.07484.x}

\bibitem[{{Wegg} {et~al.}(2015){Wegg}, {Gerhard}, \& {Portail}}]{wegg2015}
{Wegg}, C., {Gerhard}, O., \& {Portail}, M. 2015, \mnras, 450, 4050,
  \dodoi{10.1093/mnras/stv745}

\bibitem[{{Williams} {et~al.}(1994){Williams}, {de Geus}, \&
  {Blitz}}]{williams1994}
{Williams}, J.~P., {de Geus}, E.~J., \& {Blitz}, L. 1994, \apj, 428, 693,
  \dodoi{10.1086/174279}

\bibitem[{{Williams} {et~al.}(2018){Williams}, {Gear}, \&
  {Smith}}]{williams2018}
{Williams}, T.~G., {Gear}, W.~K., \& {Smith}, M. W.~L. 2018, \mnras, 479, 297,
  \dodoi{10.1093/mnras/sty1476}

\bibitem[{{Winkel} {et~al.}(2016){Winkel}, {Kerp}, {Fl{\"o}er}, {Kalberla},
  {Ben Bekhti}, {Keller}, \& {Lenz}}]{winkel2016}
{Winkel}, B., {Kerp}, J., {Fl{\"o}er}, L., {et~al.} 2016, \aap, 585, A41,
  \dodoi{10.1051/0004-6361/201527007}

\bibitem[{{Wolfire} {et~al.}(2010){Wolfire}, {Hollenbach}, \&
  {McKee}}]{wolfire2010}
{Wolfire}, M.~G., {Hollenbach}, D., \& {McKee}, C.~F. 2010, \apj, 716, 1191,
  \dodoi{10.1088/0004-637X/716/2/1191}

\bibitem[{{Wong} \& {Blitz}(2002)}]{wong2002}
{Wong}, T., \& {Blitz}, L. 2002, \apj, 569, 157, \dodoi{10.1086/339287}

\bibitem[{{Wong} {et~al.}(2009){Wong}, {Hughes}, {Fukui}, {Kawamura}, {Mizuno},
  {Ott}, {Muller}, {Pineda}, {Welty}, {Kim}, {Mizuno}, {Murai}, \&
  {Onishi}}]{wong2009}
{Wong}, T., {Hughes}, A., {Fukui}, Y., {et~al.} 2009, \apj, 696, 370,
  \dodoi{10.1088/0004-637X/696/1/370}

\end{thebibliography}
\bibliographystyle{aasjournal}

\end{document}